\documentclass[graybox]{svmult}

\usepackage{mathptmx}       
\usepackage{helvet}         
\usepackage{courier}        
\usepackage{type1cm}        
\usepackage{amssymb,amsfonts}
\usepackage{makeidx}         
\usepackage{graphicx}        
\usepackage{multicol}        
\usepackage[bottom]{footmisc}

\usepackage{color}
\usepackage{enumerate}

\makeindex             

\begin{document}

\title*{The Smaller (SALI) and the Generalized (GALI)
Alignment Indices: Efficient Methods of Chaos Detection}
\titlerunning{The SALI and the GALI methods of chaos detection}
\author{Charalampos Skokos and Thanos Manos}
\institute{Charalampos Skokos \at Department of Mathematics and
  Applied Mathematics, University of Cape Town, Rondebosch, 7701,
  South Africa, \email{haris.skokos@uct.ac.za} \and Thanos Manos \at
  Center for Applied Mathematics and Theoretical Physics (CAMTP),
  University of Maribor, Krekova 2, SI-2000 Maribor, Slovenia, \at
  School of Applied Sciences, University of Nova Gorica - Vipavska
  11c, SI-5270 Ajdov\v{s}\v{c}ina, Slovenia, \at Institute of
  Neuroscience and Medicine Neuromodulation (INM-7), Research Center
  J\"{u}lich, 52425 J\"{u}lich, Germany,
  \email{t.manos@fz-juelich.de}}
\maketitle

\abstract{We provide a concise presentation of the Smaller (SALI) and
  the Generalized Alignment Index (GALI) methods of chaos
  detection. These are efficient chaos indicators based on the
  evolution of two or more, initially distinct, deviation vectors
  from the studied orbit. After explaining the motivation behind the
  introduction of these indices, we sum up the behaviors they
  exhibit for regular and chaotic motion, as well as for stable and
  unstable periodic orbits, focusing mainly on finite-dimensional
  conservative systems: autonomous Hamiltonian models and symplectic
  maps. We emphasize the advantages of these methods in studying the
  global dynamics of a system, as well as their ability to identify
  regular motion on low dimensional tori. Finally we discuss several
  applications of these indices to problems originating from different
  scientific fields like celestial mechanics, galactic dynamics,
  accelerator physics and condensed matter physics.}

\section{Introduction and Basic Concepts}
\label{sect:Intro}

A fundamental aspect in studies of dynamical systems is the identification of
chaotic behavior, both locally, i.e.~in the neighborhood of individual orbits,
and globally, i.e.~for large samples of initial conditions. The most commonly
used method to characterize chaos is the computation of the maximum Lyapunov
exponent (mLE) $\lambda_1$. In general, Lyapunov exponents (LEs) are asymptotic measures characterizing the average rate of growth or shrinking of small
perturbations to orbits of dynamical systems. They were introduced by Lyapunov
\cite{Lyapunov_1892} and they were applied to characterize chaotic motion by
Oseledec in \cite{O_68}, where the Multiplicative Ergodic Theorem (which
 provided the theoretical basis for the numerical computation of the
LEs) was stated and proved. For a recent review of the theory and the numerical evaluation of LEs the reader is referred to \cite{S_10}.  The numerical
evaluation of the mLE was achieved in the late 1970's \cite{BGS_76,NS_77,CGG_78} and allowed the discrimination between regular and
chaotic motion. This evaluation is performed through the time
evolution of an infinitesimal perturbation of the orbit's initial condition,
which is described by a deviation vector from the orbit itself. The evolution
of the deviation vector is governed by the so-called \emph{variational equations} \cite{CGG_78}.

In practice, $\lambda_1$ is evaluated as the limit for $t\rightarrow\infty$ of the \emph{finite time maximum Lyapunov exponent}
\begin{equation}\label{eq:ftLE}
    \Lambda_1(t)=\frac{1}{t} \ln \frac{\| \vec{w}(t) \|}{\| \vec{w}(0) \|},
\end{equation}
where $t$ denotes the time and $\| \vec{w}(0) \|$, $\| \vec{w}(t) \|$
are the Euclidean norms\footnote{We note that the value of $\lambda_1$ is independent of the used norm.} of the deviation vector $\vec{w}$
at times $t=0$ and $t>0$ respectively. Thus
\begin{equation}\label{eq:mLE}
    \lambda_1= \lim_{t\rightarrow\infty} \Lambda_1(t).
\end{equation}

The computation of the mLE was extensively used for studying chaos and it is still implemented nowadays for this purpose. Nevertheless, one of its major practical disadvantages is the slow convergence of the finite time Lyapunov exponent (\ref{eq:ftLE}) to its limit value (\ref{eq:mLE}). Since $\Lambda_1(t)$ is influenced by the whole evolution of the deviation vector, the time needed for it to converge to $\lambda_1$ is not known a priori, and in many cases it may become extremely long. This delay can result in CPU-time expensive computations, especially when the study of many orbits is required for the global investigation of a system. In order to overcome this
problem several other fast chaos detection techniques have been
developed over the years; some of which are presented in this volume.

Throughout this chapter we consider finite-dimensional conservative
dynamical systems and in particular, autonomous Hamiltonian models and
symplectic maps (except from Sect.~\ref{sect:time_dep} where a time dependent Hamiltonian system is studied). In these systems regular motion occurs on the surface of a torus in the system's phase space and is characterized by $\lambda_1=0$. Any deviation vector $\vec{w}(0)$ from a regular orbit eventually falls on the tangent space of this torus and its norm
will approximately grow  linearly in time, i.e. eventually becoming
proportional to $t$, $\| \vec{w}(t)\| \propto t $. Consequently, $\Lambda_1(t) \propto \ln t /t $,  which practically means that $\Lambda_1(t)$ tends
asymptotically to zero following the power law $t^{-1}$ because the
values of $\ln t$ change much slower than $t$ as time grows (see for example \cite{BGS_76,CCF_80} and Sect.~5.3 of \cite{S_10}). On the other hand, in the case of chaotic orbits the use of any initial deviation vector in
(\ref{eq:ftLE}) and (\ref{eq:mLE}) practically leads to the computation of the
mLE $\lambda_1>0$ because this vector eventually is stretched towards the
direction associated to the mLE, assuming of course that $\lambda_1 >
\lambda_2$, with $\lambda_2$ being the second largest LE. We note here that, from the first numerical attempts to evaluate the mLE \cite{BGS_76,CGG_78} it became apparent that a random choice of the initial deviation vector $\vec{w}(0)$ leads with probability one to the computation of $\lambda_1$. This means that, the choice of $\vec{w}(0)$ does not affect the limiting value of $\Lambda_1(t)$, but only the initial phases of its evolution.  This behavior introduces some difficulties when we want to evaluate the whole spectrum of LEs of chaotic orbits because any set of initially distinct deviation vectors eventually end up to vectors aligned along the direction defined by the mLE. It is worth-noting that even in cases where we could theoretically know the initial choice of deviation vectors which would lead to the evaluation of LEs other than the maximum one, the unavoidable numerical errors in the computational procedure will lead again to the computation of the mLE \cite{BGGS_80b}. This problem was bypassed by the development of a procedure based on repeated orthonormalizations of the evolved deviation vectors \cite{BGGS_78,BG_79,SN_79,BFS_79,BGGS_80a,BGGS_80b,WSSV_85}.

Although the eventual coincidence of distinct initial deviation vectors for chaotic orbits with $\lambda_1 > \lambda_2$ was well-known from the early 1980's, this property was not directly used to identify chaos for about two decades until the introduction of the Smaller Alignment Index (SALI) method in \cite{S_01}. In the 1990's some indirect consequences of the fact that two initially distinct deviation vectors eventually coincide for chaotic motion, while they will have different directions on the tangent space of the torus for regular ones, were used to determine the nature of orbits, but not the fact itself. In particular, in \cite{VCE_98} the spectra of what was named the `\emph{stretching number}', i.e.~the quantity
\begin{equation} \label{eq:strnum}
 \alpha=\frac{\ln \left(  \frac{\left\| \vec{w}(t+\Delta t)\right\|}
  {\left\| \vec{w}(t) \right\|} \right)}  {\Delta t},
\end{equation}
where $\Delta t$ is a small time step, were considered. The main outcome of that paper was that `\emph{the spectra for two different  initial deviations are the same for chaotic orbits, but different for ordered orbits}', as was stated in the abstract of \cite{VCE_98}. This feature was later quantified in \cite{VCE_99} by the introduction of a quantity measuring the `difference' of two spectra, the so-called `spectral distance'. In \cite{VCE_99} it was shown that this quantity attains constant, positive values for regular orbits, while it becomes zero for chaotic ones. It is worth noting that in \cite{VCE_98} it was explained that the observed behavior of the two spectra was due to the fact that the deviation vectors eventually coincide for chaotic orbits, producing the same sequences of stretching numbers, while they remain different for regular ones resulting in different spectra of stretching numbers. Nevertheless, instead of directly checking the matching (or not) of the two deviation vectors the method developed in \cite{VCE_98,VCE_99} requires unnecessary, additional computations as it goes through the construction of the two spectra and the evaluation of their `distance'. Naturally, this procedure is influenced by the whole time evolution of the deviation vectors, which in turn results in the delay of the matching of the two spectra with respect to the matching of the two deviation vectors.

Apparently, the direct determination of the possible coincidence (or not) of the deviation vectors is a much faster and more efficient approach to reveal the regular or chaotic nature of orbits than the evaluation of the spectral distance, as it requires less computations (see \cite{S_01} for a comparison between the two approaches). This observation led to the introduction in \cite{S_01} of the SALI method which actually checks the possible coincidence of deviation vectors, while the later introduced Generalized Alignment Index (GALI) \cite{SBA_07} extends this criterion to more deviation vectors. As we see in Sect.~\ref{sect:GALI} this extension allows the correct characterization of chaotic orbits also in the case where the spectrum of the LEs is degenerate and the second, or even more, largest LEs are equal to $\lambda_1$.

In order to illustrate the behaviors of both the SALI and the GALI methods for regular and chaotic motion we use in this chapter some simple models of Hamiltonian systems and symplectic maps.

In particular, as a two degrees of freedom (2D) Hamiltonian model we consider the well-known H\'{e}non-Heiles system \cite{HH_64}, described by the Hamiltonian
\begin{equation}
H_2 = \frac{1}{2} (p_1^2+p_2^2) + \frac{1}{2} (q_1^2+q_2^2) + q_1^2
q_2 - \frac{1}{3} q_2^3. \label{eq:2DHam}
\end{equation}
We also consider the 3D Hamiltonian system
\begin{equation}
H_3 = \sum_{i=1}^3 \frac{\omega_i}{2} (q_i^2+p_i^2) + q_1^2 q_2+q_1^2
q_3, \label{eq:3DHam}
\end{equation}
initially studied in \cite{CGG_78,BGGS_80b}. Note that $\omega_i$ in
(\ref{eq:3DHam}) are some constant coefficients. As a model of higher
dimensions we use the $N$D Hamiltonian
\begin{equation}
H_{N} = \frac{1}{2} \sum_{i=1}^{N} p_i^2 + \sum_{i=0}^{N}
\left[\frac{1}{2}(q_{i+1}-q_i)^2 + \frac{1}{4}\beta (q_{i+1}-q_i)^4
  \right], \label{eq:NDHam}
\end{equation}
which describes a chain of $N$ particles with quadratic and quartic
nearest neighbor interactions, known as the Fermi-Pasta-Ulam $\beta$
model (FPU-$\beta$) \cite{FPU_55}, where $q_0=q_{N+1}=0$. In all the
above-mentioned $N$D Hamiltonian models, $q_i$, $p_i$, $i=1,2,\ldots
N$ are respectively the generalized coordinates and the conjugate
momenta defining the 2$N$-dimensional (2$N$d) phase space of the
system.

As a symplectic map model we consider in our presentation the
2$M$-dimensional (2$M$d) system of coupled standard maps studied in
\cite{KG_88}
\begin{equation}\label{eq:Md_map}
\begin{array}{ccl}
   \displaystyle
   x'_j & \displaystyle = & \displaystyle
   x_j +y'_j \\ & & \\
   \displaystyle y'_j & \displaystyle =
   & \displaystyle y_j  +\frac{K_j}{2 \pi} \sin \left( 2 \pi
   x_j \right) - \frac{\gamma}{2 \pi} \left\{ \sin \left[ 2 \pi \left(
     x_{j+1}  -x_j  \right) \right] + \sin \left[ 2 \pi \left(
     x_{j-1}  -x_j  \right) \right] \right\},
\end{array}
\end{equation}
where $j=1,2,\ldots,M$ is the index of each standard map, $K_j$ and
$\gamma$ are the model's parameters and the prime ($'$) denotes the
new values of the variables after one iteration of the map.  We note
that each variable is given modulo 1, i.e.~$0\leq x_j <1$, $0\leq y_j
<1$ and also that the conventions $x_0=x_M$ and $x_{M+1}=x_1$ hold.

In order to make this chapter more focused and easier to read we
decided not to present any analytical proofs for the various
mathematical statements given in the text; we prefer to direct the
reader to the publications where these proofs can be
found. Nevertheless, we want to emphasize here that all the laws
describing the behavior of the SALI and the GALI have been obtained
theoretically and they are not numerical estimations or fits to
numerical data. Indeed, these laws succeed to accurately reproduce the
evolution of the indices in actual numerical simulations, some of
which are presented in the following sections.

The chapter is organized as follows. In Sect.~\ref{sect:SALI} the SALI method
is presented and the behavior of the index for regular and chaotic orbits is
discussed. Section \ref{sect:GALI} is devoted to the GALI method. After
explaining the motivation that led to the introduction of the GALI, the
definition of the index is given and its practical computation is discussed in Sect.~\ref{sect:GALI_compute}.  Then, in Sect.~\ref{sect:GALI_ch_reg} the behavior of the index for regular and chaotic motion is presented and several example orbits of Hamiltonian systems and symplectic maps of various dimensions are used to illustrate these behaviors. The ability of the GALI to identify motion on low dimensional tori is presented in Sect.~\ref{sect:low}, while Sect.~\ref{sect:GALI_po} is devoted to the behavior of the index for stable and unstable periodic orbits. In Sect.~\ref{sect:appl} several applications of the SALI and the GALI methods are presented. In particular, in Sect.~\ref{sect:global} we explain how the SALI and the GALI can be used for understanding the global dynamics of a system, while specific applications of the indices to various dynamical models are briefly discussed in Sect.~\ref{sect:appl_studies}. The particular case of time dependent Hamiltonians is considered in Sect.~\ref{sect:time_dep}. Finally, in Sect.~\ref{sect:disc} we summarize the advantages of the SALI and
the GALI methods and briefly discuss some recent comparative studies
of different chaos indicators.

\section{The Smaller Alignment Index (SALI)}
\label{sect:SALI}

The idea behind the SALI's introduction was the need for a simple, easily
computed quantity which could clearly identify the possible alignment of two
multidimensional vectors. As has been already explained, it was well-known that any two deviation vectors from a chaotic orbit with $\lambda_1 > \lambda_2$ are stretched towards the direction defined by the mLE, eventually becoming aligned having the same or opposite directions. Thus, it would be quite helpful to devise a quantity which could clearly indicate this alignment.

Since we are only interested in the direction of the two deviation
vectors and not in their actual size, we can normalize them before
checking their alignment. This process also eliminates the problem of
potential numerical overflow due to vectors' growth in size, which
appears especially in the case of chaotic orbits.  So in practice, we
let the two deviation vectors evolve under the system's dynamics
(according to the variational equations for Hamiltonian models, or the
so-called tangent map for symplectic maps) normalizing them after a
fixed number of evolution steps to a predefined norm value. For
simplicity in our presentation we consider the usual Euclidean norm
(denoted by $\|\, \cdot \,\|$) and renormalize the evolved vectors to
unity.

\begin{figure}
\includegraphics[width=\linewidth]{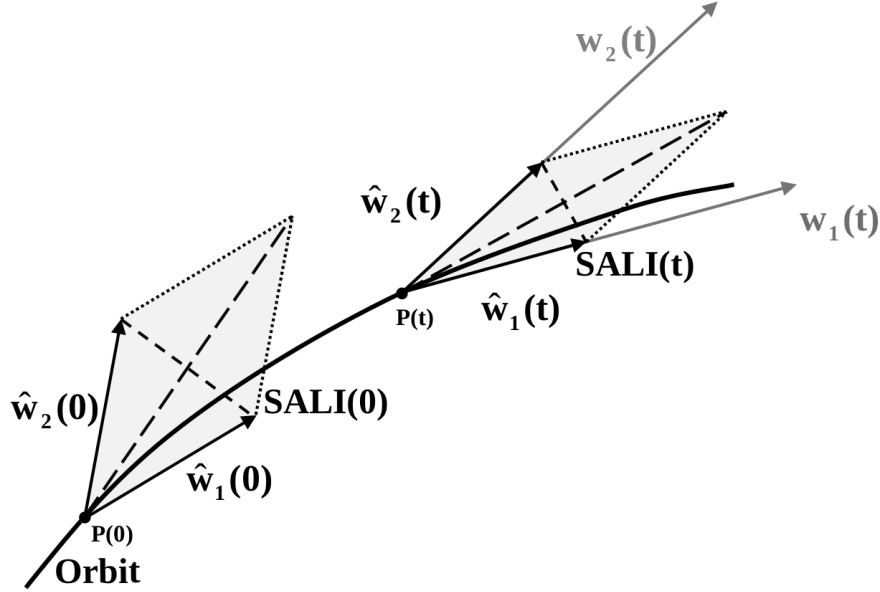}
\caption{Schematic representation of the evolution of two deviation
  vectors and of the corresponding SALI for a chaotic orbit. Two
  initially distinct unit deviation vectors $\hat{\vec{w}}_1(0)$,
  $\hat{\vec{w}}_2(0)$ from point $P(0)$ of a chaotic orbit become
  $\vec{w}_1(t)$, $\vec{w}_2(t)$ after some time $t>0$ when the orbit
  reaches point $P(t)$, with $\hat{\vec{w}}_1(t)$,
  $\hat{\vec{w}}_2(t)$ being the unit vectors along these
  directions. The length of the shortest diagonals of the grey-shaded
  parallelograms defined by $\hat{\vec{w}}_1(0)$, $\hat{\vec{w}}_2(0)$
  and $\hat{\vec{w}}_1(t)$, $\hat{\vec{w}}_2(t)$ are the values of the
  SALI$(0)$ and the SALI$(t)$ respectively}
\label{fig:SM_SALI_vects_ch}
\end{figure}

In the case of chaotic orbits this procedure is schematically shown in
Fig.~\ref{fig:SM_SALI_vects_ch} where the two initially distinct unit
deviation vectors\footnote{We note that throughout this chapter we use
  the hat symbol ($\, \hat{}\,$) to denote a unit vector.}
$\hat{\vec{w}}_1(0)$, $\hat{\vec{w}}_2(0)$ converge to the same direction.
We emphasize that Fig.~\ref{fig:SM_SALI_vects_ch} is just a schematic
representation on the plane of the real deviation vectors which are objects
evolving in multidimensional spaces. Since the mLE $\lambda_1>0$ denotes the
mean exponential rate of each vector's stretching, they are elongated
at some later time $t>0$\footnote{For Hamiltonian systems the time is a
continuous variable, while for maps it is a discrete one counting the map's
iterations.}, becoming $\vec{w}_1(t)$, $\vec{w}_2(t)$, while the
corresponding unit vectors are $\hat{\vec{w}}_1(t)$,
$\hat{\vec{w}}_2(t)$. Then the diagonals of the
parallelograms defined by $\hat{\vec{w}}_1(t)$, $\hat{\vec{w}}_2(t)$,
both for $t=0$ and $t>0$, depict the sum and the difference of the two
unit vectors.

In the particular case shown in Fig.~\ref{fig:SM_SALI_vects_ch} the
two unit vectors tend to align by becoming equal. This means that $\|
\hat{\vec{w}}_1(t) - \hat{\vec{w}}_2(t) \| \rightarrow 0$ and $\|
\hat{\vec{w}}_1(t) + \hat{\vec{w}}_2(t) \| \rightarrow 2$. Of course
the dynamics could have led the vectors to become opposite. In that
case we get $\| \hat{\vec{w}}_1(t) - \hat{\vec{w}}_2(t) \| \rightarrow
2$ and $\| \hat{\vec{w}}_1(t) + \hat{\vec{w}}_2(t) \| \rightarrow
0$. Since we are not interested in the particular orientation of the
deviation vectors, i.e.~whether they become equal or opposite to each
other, when we check their possible alignment, a rather natural choice
is to define the minimum of norms $\| \hat{\vec{w}}_1(t) +
\hat{\vec{w}}_2(t) \| $, $\| \hat{\vec{w}}_1(t) - \hat{\vec{w}}_2(t)
\| $ as an indicator of the vectors' alignment. This is the reason of
the appellation, as well as of the definition of the SALI in
\cite{S_01} as
\begin{equation} \label{eq:SALI} \mbox{SALI}(t)=\min \left\{\|
    \hat{\vec{w}}_1(t) + \hat{\vec{w}}_2(t) \|, \| \hat{\vec{w}}_1(t)
    - \hat{\vec{w}}_2(t) \| \right\},
\end{equation}
with $\hat{\vec{w}}_i(t) = \frac{\vec{w}_i(t)}{\| \vec{w}_i(t) \|}$,
$i=1,2$ being unit vectors.

\begin{figure}
\includegraphics[width=\linewidth]{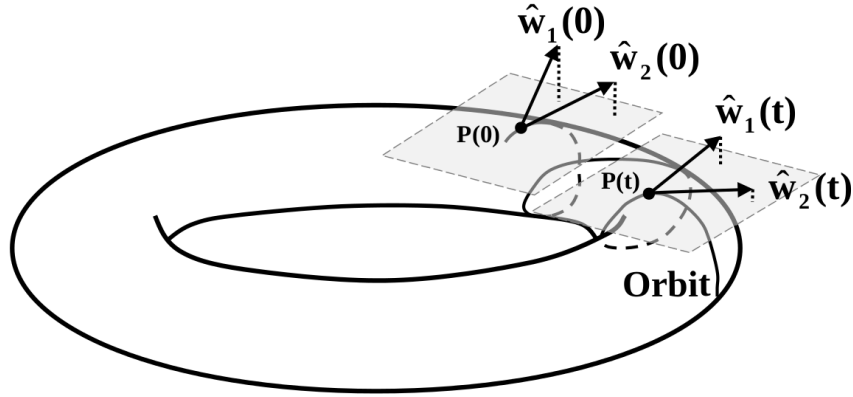}
\caption{Schematic representation of the evolution of two deviation
  vectors for a regular orbit. The motion takes place on a torus. We
  consider two initially distinct unit deviation vectors
  $\hat{\vec{w}}_1(0)$, $\hat{\vec{w}}_2(0)$ from point $P(0)$, which
  are not necessarily on the tangent space of the torus (this space is
  depicted as a shaded parallelogram passing through $P(0)$). As time
  evolves the deviation vectors tend to fall on the torus' tangent
  space and the corresponding unit vectors $\hat{\vec{w}}_1(t)$,
  $\hat{\vec{w}}_2(t)$ at time $t>0$ are `closer' to the current
  tangent space (i.e.~the grey-shaded parallelogram passing through
  $P(t)$), as the shortening of the perpendicular to the tangent
  spaces dotted lines from the edges of the deviation vectors
  indicate. Since there is no reason for the alignment of the two
  deviation vectors, the SALI will not become zero}
\label{fig:SM_SALI_vects_reg}
\end{figure}

Naturally, in order for the SALI to be efficiently used as a chaos
indicator it should exhibit distinct behaviors for chaotic and regular
orbits. As explained before the SALI becomes zero for chaotic
orbits. On the other hand, in the case of regular orbits deviation
vectors fall on the tangent space of the torus on which motion occurs,
having in general different directions as there is no reason for them
to be aligned \cite{VCE_98,SABV_03}. This behavior is shown
schematically in Fig.~\ref{fig:SM_SALI_vects_reg}. Thus, in this case
the index should be always different from zero. In practice, the
values of the SALI exhibit bounded fluctuations around some constant, positive number.

Thus, in order to compute the SALI we follow the evolution of two initially
distinct, random, unit deviation vectors $\hat{\vec{w}}_1(0)$,
$\hat{\vec{w}}_2(0)$. Choosing these vectors to be also orthogonal sets the
initial SALI to its highest possible value (SALI$(0)=\sqrt{2}$) and ensures
that they are considerably different from each other, which has proved
to be a very good computational practice. Then, every $t=\tau$ time units we normalize the evolved vectors $\vec{w}_1(i \tau)$, $\vec{w}_2(i \tau)$, $i =1,2,\ldots$, to $\hat{\vec{w}}_1(i \tau)$, $\hat{\vec{w}}_2(i \tau)$ and evaluate the SALI$(i \tau)$ from (\ref{eq:SALI}).  This algorithm is described in pseudo-code in Table~\ref{tab:SALI} of the Appendix.  A MAPLE code for this
algorithm, developed specifically for the H\'{e}non-Heiles system
(\ref{eq:2DHam}) can be found in Chap.~5 of \cite{BS_12}.

\begin{figure}
\centerline{
\includegraphics[scale=0.805]{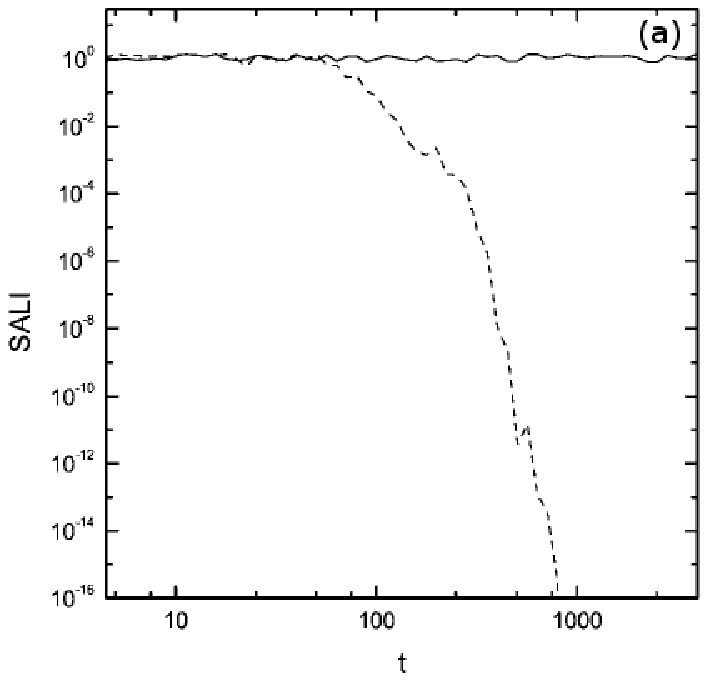}
\includegraphics[scale=0.324]{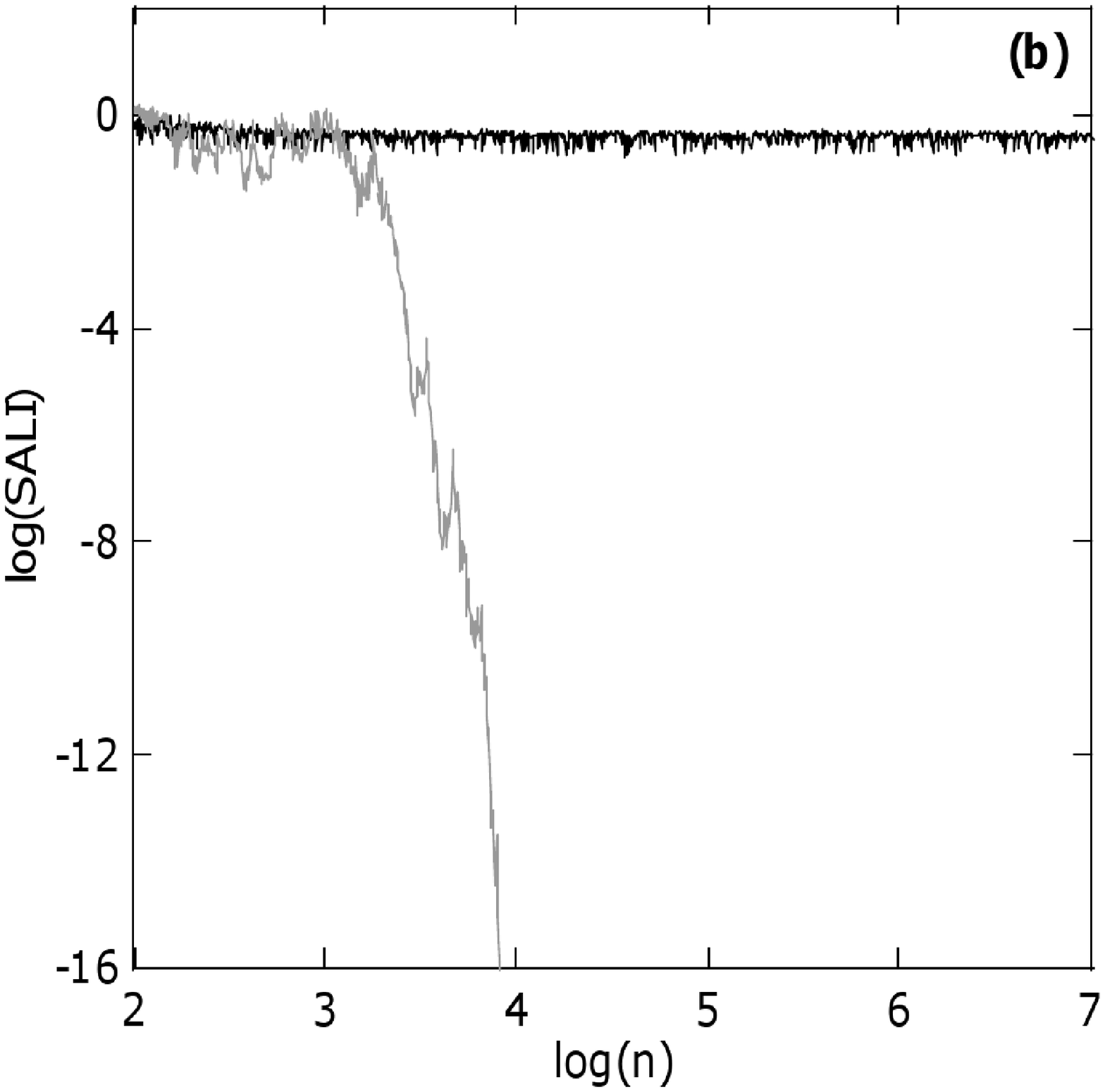} }
\caption{The time evolution of the SALI for a regular and a chaotic
  orbit of (a) the 2D Hamiltonian system (\ref{eq:2DHam}) for
  $H_2=0.125$ (after \cite{SABV_04}) and (b) the 6d map
  (\ref{eq:6d_map}) for $K=3$ and $\gamma=0.1$ (after \cite{S_01}).
  In (a) the time $t$ is continuous, while in (b) it is discrete and
  counts the map's iterations $n$.  The initial conditions of the
  orbits are: (a) $q_1=0$, $q_2=0.1$, $p_1=0.49058$, $p_2=0$ (regular
  orbit; solid curve) and $q_1=0$, $q_2=-0.25$, $p_1=0.42081$, $p_2=0$
  (chaotic orbit; dashed curve), and (b) $x_1=0.55$, $y_1=0.05$,
  $x_2=0.55$, $y_2=0.01$, $x_3=0.55$, $y_3=0$ (regular orbit; black
  curve) and $x_1=0.55$, $y_1=0.05$, $x_2=0.55$, $y_2=0.21$,
  $x_3=0.55$, $y_3=0$ (chaotic orbit; grey curve)}
\label{fig:SM_SALI_orbs}
\end{figure}

\begin{figure}
\sidecaption[t]
\includegraphics[scale=0.26]{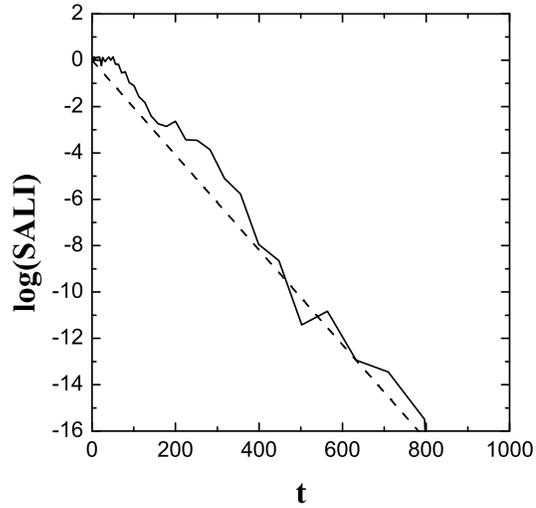}
\caption{The evolution of the SALI (solid curve) for the chaotic orbit
  of Fig.~\ref{fig:SM_SALI_orbs}(a) as a function of time $t$. The
  dashed line corresponds to a function proportional to $\exp
  {\left(-\lambda_1 t\right)}$ for $\lambda_1 = 0.047$. Note that the
  $t$-axis is linear (after \cite{SABV_04})}
\label{fig:SALI_exponential}
\end{figure}

The completely different behaviors of the SALI for regular and chaotic
orbits are clearly seen in Fig.~\ref{fig:SM_SALI_orbs}\footnote{We note
that throughout this chapter the logarithm to base 10 is denoted by $\log$.}, where some representative results are shown for the 2D Hamiltonian system
(\ref{eq:2DHam}) and the 6d symplectic map
\begin{equation}\label{eq:6d_map}
\begin{array}{ccl}
   \displaystyle x'_1 & \displaystyle = & \displaystyle x_1 +y'_1
   \\ \displaystyle y'_1 & \displaystyle = & \displaystyle y_1
   +\frac{K}{2 \pi} \sin \left( 2 \pi x_1 \right) - \frac{\gamma}{2
     \pi} \left\{ \sin \left[ 2 \pi \left( x_2 -x_1 \right) \right] +
   \sin \left[ 2 \pi \left( x_3 -x_1 \right) \right] \right\}
   \\ \displaystyle x'_2 & \displaystyle = & \displaystyle x_2 +y'_2
   \\ \displaystyle y'_2 & \displaystyle = & \displaystyle y_2
   +\frac{K}{2 \pi} \sin \left( 2 \pi x_2 \right) - \frac{\gamma}{2
     \pi} \left\{ \sin \left[ 2 \pi \left( x_3 -x_2 \right) \right] +
   \sin \left[ 2 \pi \left( x_1 -x_2 \right) \right] \right\}
   \\ \displaystyle x'_3 & \displaystyle = & \displaystyle x_3 +y'_3
   \\ \displaystyle y'_3 & \displaystyle = & \displaystyle y_3
   +\frac{K}{2 \pi} \sin \left( 2 \pi x_3 \right) - \frac{\gamma}{2
     \pi} \left\{ \sin \left[ 2 \pi \left( x_1 -x_3 \right) \right] +
   \sin \left[ 2 \pi \left( x_2 -x_3 \right) \right] \right\},
\end{array}
\end{equation}
obtained by considering $M=3$ coupled standard maps with
$K_1=K_2=K_3=K$ in (\ref{eq:Md_map}). From the results of
Fig.~\ref{fig:SM_SALI_orbs} we see that for both systems the SALI of
regular orbits (black, solid curves) remains practically constant and
positive, i.e.
\begin{equation}
\mbox{SALI} \propto \mbox{constant}.
\label{eq:SALI_reg}
\end{equation}
On the other hand, the SALI of chaotic orbits (black, dashed curve in
Fig.~\ref{fig:SM_SALI_orbs}(a) and grey, solid curve in
Fig.~\ref{fig:SM_SALI_orbs}(b)) exhibits a fast decrease to zero after
an initial transient time interval, reaching very small values around
the computer's accuracy ($10^{-16}$). Actually, it was shown in
\cite{SABV_04} that the SALI tends to zero exponentially fast in such
cases, following the law
\begin{equation}
\mbox{SALI}(t) \propto \exp{\left[-(\lambda_1-\lambda_2)t\right]},
\label{eq:SALI_ch}
\end{equation}
where $\lambda_1$, $\lambda_2$ ($\lambda_1 \geq \lambda_2$) are the
first (i.e.~the mLE) and the second largest LEs respectively. As an
example demonstrating the validity of this exponential-decay law we plot
in Fig.~\ref{fig:SALI_exponential} the evolution of the SALI (solid
curve) of the chaotic orbit of Fig.~\ref{fig:SM_SALI_orbs}(a) using a
linear horizontal axis for time $t$. Since for 2D Hamiltonian systems
$\lambda_2=0$, (\ref{eq:SALI_ch}) becomes
\begin{equation}
\mbox{SALI}(t) \propto \exp{\left(-\lambda_1 t\right)},
\label{eq:SALI_ch_2D}
\end{equation}
For this particular orbit the mLE was found to be $\lambda_1 \approx
0.047$ in \cite{SABV_04}. From Fig.~\ref{fig:SALI_exponential} we see
that (\ref{eq:SALI_ch_2D}) with $\lambda_1 = 0.047$ (dashed line)
reproduces correctly the evolution of the ALI\footnote{We note that
  here, as well as in several, forthcoming figures in this chapter,
  the evaluation of the LEs is done only for confirming the
  theoretical predictions for the time evolution of the SALI (equation
  (\ref{eq:SALI_ch_2D}) in the current case) and later on of the
  GALIs, and it is not needed for the computation of the SALI and the
  GALIs.}.

Thus, the completely different behavior of the SALI for regular
(\ref{eq:SALI_reg}) and chaotic (\ref{eq:SALI_ch}) orbits permits the
clear and efficient distinction between the two cases.  In
\cite{S_01,SABV_04} a comparison of the SALI's performance with
respect to other chaos detection techniques was presented and the
efficiency of the index was discussed.  A main advantage of the SALI
method is its ability to detect chaotic motion faster than other
techniques which depend on the whole time evolution of deviation
vectors, like the mLE and the spectral distance, because the SALI is
determined by the \emph{current} state of these vectors and is not
influenced by their evolution history. Hence, the moment the two
vectors are close enough to each other the SALI becomes practically
zero and guarantees the chaotic nature of the orbit beyond any
doubt. In addition, the evaluation of the SALI is simpler and more
straightforward with respect to other methods that require more
complicated computations. Such aspects were discussed in
\cite{SABV_04} where a comparison of the index with the Relative
Lyapunov Indicator (RLI) \cite{SESF_04} and the so-called `0--1' test
\cite{GM_04} was presented. Another crucial characteristic of the SALI
is that it attains values in a given interval, namely SALI$(t) \in [0,
\sqrt{2}]$, which does not change in time as is for example the case
for the Fast Lyapunov Indicator (FLI) \cite{FGL_97}.  Thus, setting a
realistic threshold value below which the SALI is considered to be
practically zero (and the corresponding orbit is characterized as
chaotic), allows the fast and accurate discrimination between regular
and chaotic motion. Due to all these features the SALI became a
reliable and widely used chaos indicator as its numerous applications
to a variety of dynamical systems over the years prove. Some of these
applications are discussed in Sect.~\ref{sect:appl}.

\section{The Generalized Alignment Index (GALI)}
\label{sect:GALI}

A fundamental difference between the SALI and other, commonly applied
chaos indicators, is that it uses information from the evolution of
two deviation vectors instead of just one. A consequence of this
feature is the appearance of the two largest LEs in
(\ref{eq:SALI_ch}). After performing this first leap from using only
one deviation vector, the question of going even further arises
naturally. To formulate this in other words: why should we stop in
using only two deviation vectors?  Can we extend the definition of the
SALI to include more deviation vectors? Assuming that this extension
is possible, what will we gain from it? Will the use of more than two
deviation vectors lead to the introduction of a new chaoticity index
which will permit the acquisition of a deeper understanding of the
system's dynamics, exhibiting at the same time a better numerical
performance than the SALI?  For instance, from (\ref{eq:SALI_ch}) we
realize that in the case of a chaotic orbit with $\lambda_1 \approx
\lambda_2$ the convergence of the SALI to zero will be extremely
slow. As a result long integrations would be required in order for the
index to distinguish this orbit from a regular one for which the SALI
remains practically constant. Although the existence of such chaotic
orbits is not very probable the drawback of the SALI remains. An
alternative way to state this problem is the following: can we
construct a new index whose behavior in the case of chaotic orbits
will depend on more LEs than the two largest ones so that it can
overcome the discrimination problem for $\lambda_1 \approx \lambda_2$?

Indeed, such an index can be constructed. The key point to its
development is the observation that the SALI is closely related to the
area of the parallelogram defined by the two deviation
vectors\footnote{Note that this parallelogram is not the usual 2d parallelogram on the plane because its sides (the deviation vectors) are not 2d vectors.}. From the schematic representation of the deviation vectors' evolution in Fig.~\ref{fig:SM_SALI_vects_ch} we see that when the SALI vanishes one of the diagonals of the parallelogram also vanishes, and consequently its area becomes zero. The area $A_2$ of a usual 2d parallelogram is equal to the norm of the exterior product of its two sides $\vec{v}_1$, $\vec{v}_2$, and also equal to the half of the product of its diagonals' lengths
\begin{equation}
A_2=\| \vec{v}_1 \times \vec{v}_2 \|= \frac{\| \vec{v}_1 +\vec{v}_2\|
  \cdot \| \vec{v}_1 - \vec{v}_2\|}{2}.
\label{eq:A2}
\end{equation}
In a similar way, the area $A$ of the parallelogram of Fig.~\ref{fig:SM_SALI_vects_ch} is given by the generalization of the
exterior product of vectors to higher dimensions, i.e.~the so-called
wedge product denoted by $(\wedge)$\footnote{For a brief introduction
  to the notion of the wedge product the reader is referred to the
  Appendix A of \cite{SBA_07} and the Appendix of \cite{S_10}.}, so
that
\begin{equation}
A=\| \hat{\vec{w}}_1 \wedge \hat{\vec{w}}_2 \|=\frac{\|
  \hat{\vec{w}}_1 +\hat{\vec{w}}_2\| \cdot \| \hat{\vec{w}}_1 -
  \hat{\vec{w}}_2\|}{2}.
\label{eq:area}
\end{equation}
Note the analogy of this equation to (\ref{eq:A2})\footnote{A proof of
  the second equality of (\ref{eq:area}) can be found in the Appendix
  B of \cite{SBA_07}.}.

Based on the fact that the SALI is related to the area of the
parallelogram defined by two unit deviation vectors, the
extension of the index to include more vectors is straightforward: the
new quantity is defined as the volume of the
parallelepiped formed by more than two deviation vectors. This volume
is computed as the norm of the wedge product of these vectors.  These
arguments led to the introduction in \cite{SBA_07} of the Generalized
Alignment Index of order $k$ (GALI$_k$) as
\begin{equation}
\mbox{GALI}_k(t)=\| \hat{\vec{w}}_1(t)\wedge \hat{\vec{w}}_2(t)\wedge
\ldots \wedge\hat{\vec{w}}_k(t) \|,
\label{eq:GALI}
\end{equation}
where $\hat{\vec{w}}_i$ are unit vectors as in (\ref{eq:SALI}). In
this definition the number of used deviation vectors should not exceed
the dimension of the system's phase space, because in this case the
$k$ vectors will become linearly dependent and the corresponding
volume will be by definition zero, as is for example the area defined
by two vectors having the same direction. Thus, for an $N$D
Hamiltonian system with $N\geq 2$ or a $2N$d symplectic map with
$N\geq 1$, we consider only GALIs with $2 \leq k \leq 2N$.

By its definition the GALI$_k$ is a quantity clearly indicating the
linear dependence (GALI$_k=0$) or independence (GALI$_k>0$) of $k$
deviation vectors. The SALI has the same discriminating ability as
SALI$=0$ indicates that the two vectors are aligned, i.e.~they are
linearly dependent, while $\mbox{SALI}>0$ implies that the vectors are
not aligned, which means that they are linearly independent. Actually,
the connection between the two indices can be quantified explicitly.
Indeed, it was proved in the Appendix B of \cite{SBA_07} that
\begin{equation}
\mbox{GALI}_2=\mbox{SALI} \cdot \frac{\max \left\{\|
  \hat{\vec{w}}_1(t) + \hat{\vec{w}}_2(t) \|, \| \hat{\vec{w}}_1(t) -
  \hat{\vec{w}}_2(t) \| \right\}}{2}.
\label{eq:SALI-GALI2}
\end{equation}
Since the $\max \left\{\| \hat{\vec{w}}_1(t) + \hat{\vec{w}}_2(t) \|,
  \| \hat{\vec{w}}_1(t) - \hat{\vec{w}}_2(t) \| \right\}$ is a number
in the interval $[\sqrt{2},2]$ we conclude that
\begin{equation}
\mbox{GALI}_2\propto\mbox{SALI},
\label{eq:SALI-GALI2_equi}
\end{equation}
which means that the GALI$_2$ is practically equivalent to the
SALI. This is another evidence that the GALI definition
(\ref{eq:GALI}) is a natural extension of the SALI for more than two
deviation vectors.

\subsection{Computation of the GALI}
\label{sect:GALI_compute}

Let us discuss now how one can actually calculate the value of the
GALI$_k$ for an $N$D Hamiltonian system ($N\geq 2$) or a $2N$d symplectic map
($N\geq 1$). For this purpose we consider the $k \times 2N$ matrix
\begin{equation}
\textbf{A}(t)= \left[
\begin{array}{cccc}
w_{11}(t) & w_{12}(t) & \cdots & w_{1\, 2N}(t) \\ w_{21}(t) &
w_{22}(t) & \cdots & w_{2\, 2N}(t) \\ \vdots & \vdots & & \vdots
\\ w_{k1}(t) & w_{k2}(t) & \cdots & w_{k\, 2N}(t) \end{array} \right]
\label{eq:GALI_A}
\end{equation}
having as rows the $2N$ coordinates of the $k$ unit deviation vectors
$\hat{\vec{w}}_i(t)$ with respect to the usual orthonormal basis
$\hat{\vec{e}}_1= (1,0,0,\ldots,0)$, $\hat{\vec{e}}_2=
(0,1,0,\ldots,0)$, ..., $\hat{\vec{e}}_{2N}= (0,0,0,\ldots,1)$. We
note that the elements of $\textbf{A}(t)$ satisfy the condition
$\sum_{j=1}^{2N} w_{ij}^2(t) =1$ for $i=1,2,\ldots,k$ as each
deviation vector has unit norm.

We can now follow two routes for evaluating the GALI$_k(t)$. According
to the first one we compute the GALI$_k$ by evaluating the norm of the
wedge product of $k$ vectors as
\begin{equation}
\mbox{GALI}_k(t) = \left\{\sum_{1 \leq i_1 < i_2 < \cdots < i_k \leq
  2N} \left( \det \left[
\begin{array}{cccc}
w_{1 i_1}(t) & w_{1 i_2}(t) & \cdots & w_{1 i_k}(t) \\ w_{2 i_1}(t) &
w_{2 i_2}(t) & \cdots & w_{2 i_k}(t) \\ \vdots & \vdots & & \vdots
\\ w_{k i_1}(t) & w_{k i_2}(t) & \cdots & w_{k i_k}(t) \end{array}
\right] \right)^2 \right\}^{1/2},
\label{eq:GALI_deriv}
\end{equation}
where the sum is performed over all the possible combinations of $k$
indices out of $2N$ (a proof of this equation can be found in
\cite{SBA_07}). In practice this means that in our calculation we
consider all the $k\times k$ determinants of $\textbf{A}(t)$.
Equation (\ref{eq:GALI_deriv}) is particularly useful for the
theoretical description of the GALI's behavior (actually expressions
(\ref{eq:GALI_ch}) and (\ref{eq:GALI_reg}) below were obtained by
using this equation), but not very efficient from a practical point of
view. The reason is that the number of determinants appearing in
(\ref{eq:GALI_deriv}) can increase enormously when $N$ grows, leading
to unfeasible numerical computations.

A simpler, straightforward and computationally more efficient approach
to evaluate the GALI$_k$ was developed in \cite{SBA_08}, where it was
proved that the index is equal to the product of the singular values
$z_i$, $i=1,2,\ldots, k$ of $\textbf{A}^T(t)$ (the transpose of matrix
$\textbf{A}(t)$), i.e.
\begin{equation}
\mbox{GALI}_k (t)=\prod_{i=1}^k z_i(t).
\label{eq:GALI_SVD}
\end{equation}
We note that the singular values of $\textbf{A}^T(t)$ are obtained by
performing the Singular Value Decomposition (SVD) procedure to
$\textbf{A}^T(t)$.  According to the SVD method (see for instance
Sect.~2.6 of \cite{NumRec}) the $2N \times k$ matrix $\textbf{A}^T$ is
written as the product of a $2N \times k$ column-orthogonal matrix
$\textbf{U}$ ($\textbf{U}^T \cdot \textbf{U} =\textbf{I}_k$, with
$\textbf{I}_k$ being the $k \times k$ unit matrix), a $k \times k$
diagonal matrix $\textbf{Z}$ having as elements the positive or zero
singular values $z_i$, $i=1,\ldots, k$, and the transpose of a $k
\times k$ orthogonal matrix $\textbf{V}$ ($\textbf{V}^T \cdot
\textbf{V} =\textbf{I}_k$), i.e.
\begin{equation}
\textbf{A}^{\mathrm{T}}=\textbf{U}\cdot \textbf{Z} \cdot
\textbf{V}^{\mathrm{T}}.
\label{eq:SVD}
\end{equation}

In practice, in order to compute the GALI of order $k$ we follow the evolution
of $k$ initially distinct, random, orthonormal deviation vectors
$\hat{\vec{w}}_1(0)$, $\hat{\vec{w}}_2(0)$, $\ldots$, $\hat{\vec{w}}_k(0)$.
Similarly to the computation of the SALI, choosing orthonormal vectors ensures
that all of them are sufficiently far from linear dependence and gives
to the GALI$_k$ its largest possible initial value GALI$_k=1$.  Afterwards,
every $t=\tau$ time units we normalize the evolved vectors $\vec{w}_1(i \tau)$,
$\vec{w}_2(i \tau)$, $\ldots$, $\vec{w}_k(i \tau)$, $i =1,2,\ldots$, to
$\hat{\vec{w}}_1(i \tau)$, $\hat{\vec{w}}_2(i \tau)$, $\ldots$,
$\hat{\vec{w}}_k(i \tau)$ and set them as rows of a matrix $\textbf{A}(i \tau)$
(\ref{eq:GALI_A}). Then, according to (\ref{eq:GALI_SVD}) the GALI$_k(i \tau)$
is computed as the product of the singular values of matrix $\textbf{A}^T(i
\tau)$. This algorithm is described in pseudo-code in Table \ref{tab:GALI} of
the Appendix. A MAPLE code computing all the possible GALIs (i.e.~GALI$_2$,
GALI$_3$ and GALI$_4$) for the 2D Hamiltonian (\ref{eq:2DHam}) can be found in
Chap.~5 of \cite{BS_12}.

\subsection{Behavior of the GALI for Chaotic and Regular Orbits}
\label{sect:GALI_ch_reg}

After defining the new index and explaining a practical way to evaluate it, let us discuss its ability to discriminate between chaotic and regular
motion. As we have already mentioned, in the case of a chaotic orbit all
deviation vectors eventually become aligned to the direction defined by the
largest LE. Thus, they become linearly dependent and consequently the volume
they define vanishes, meaning that the GALI$_k$, $2 \leq k \leq 2N$, will
become zero. Actually, in \cite{SBA_07} it was shown analytically that in this
case the the GALI$_k(t)$ decreases to zero exponentially fast with an exponent
which depends on the $k$ largest LEs as
\begin{equation}
\label{eq:GALI_ch}
\mbox{GALI}_k(t) \propto \exp \left\{-\left[ (\lambda_1-\lambda_2) +
  (\lambda_1-\lambda_3)+ \cdots+ (\lambda_1-\lambda_k)\right]t
\right\}.
\end{equation}
Note that for $k=2$ we get the exponential law (\ref{eq:SALI_ch}) in
agreement with the equivalence between the GALI$_2$ and the SALI
(\ref{eq:SALI-GALI2_equi}).

Let us now consider the case of regular motion in a $N$D Hamiltonian
system or a 2$N$d symplectic map with $N\geq 2$. In general, this
motion occurs on an $N$d torus in the system's 2$N$d phase space. As
we discussed in Sect.~\ref{sect:SALI}, in this case any deviation
vector eventually falls on the $N$d tangent space of the torus
(Fig.~\ref{fig:SM_SALI_vects_reg}).  Consequently, the $k$ initially
distinct, linearly independent deviation vectors we follow in order
to compute the evolution of the GALI$_k$ eventually falls on the
$N$d tangent space of the torus, without necessarily having the same
directions. Thus, if we do not consider more deviation vectors than
the dimension of the tangent space ($k \leq N$) we end up with $k$
linearly independent vectors on the torus' tangent space and
consequently the volume of the parallelepiped they define
(i.e.~the GALI$_k$) will be different from zero. As we  see later
on, numerical simulations show that the GALI$_k$ exhibits small fluctuations around some positive value. If, on the other hand, we consider more deviation vectors than the dimension of the tangent space ($N < k \leq 2N$) the deviation vectors eventually become linearly dependent, as we end
up with more vectors in the torus' tangent space than the space's
dimension. Thus, the volume that these vectors define will vanish and
the GALI$_k$ will become zero. Specifically, in \cite{SBA_07} it was
shown analytically that in this case the GALI$_k$ tends to zero
following a power law whose exponent depends on the torus dimension
and on the number $k$ of deviation vectors considered,
i.e.~GALI$_k\propto t^{-2(k-N)}$.  In summary the behavior of the
GALI$_k$ for regular orbits is
\begin{equation}
\label{eq:GALI_reg}
\mbox{GALI}_k (t) \propto \left\{ \begin{array}{ll} \mbox{constant} &
 \mbox{if $ 2 \leq k \leq N$ } \\   \frac{1}{t^{2(k-N)}} & \mbox{if $N< k \leq
    2N .$} \\
\end{array} \right.
\end{equation}
From this equation we see that $\mbox{SALI} \propto \mbox{GALI}_2
\propto \mbox{constant}$, in accordance to (\ref{eq:SALI_reg}).

\subsubsection{Some Illustrative Paradigms}
\label{sect:GALI_paradigms}

In what follows we illustrate the different behaviors of the GALI$_k$
by computing its evolution for some representative chaotic and regular
orbits of various $N$D autonomous Hamiltonians and 2$N$d symplectic
maps. Before doing so let us note that for these systems the LEs comes
in pairs of values having opposite signs
\begin{equation}
\label{eq:LEs_pairs}
\lambda_i=-\lambda_{2N-i+1}, \,\,\,\, i=1,2,\ldots,N,
\end{equation}
while, moreover
\begin{equation}
\label{eq:LEs_zero}
\lambda_N=\lambda_{N+1}=0
\end{equation}
for Hamiltonian systems \cite{BGGS_80a,H_83,S_10}.

\paragraph{Hamiltonian systems}
\label{sect:GALI_Ham paradigms}

\begin{figure}
\centerline{\hspace{0.5cm}
  \includegraphics[scale=0.245]{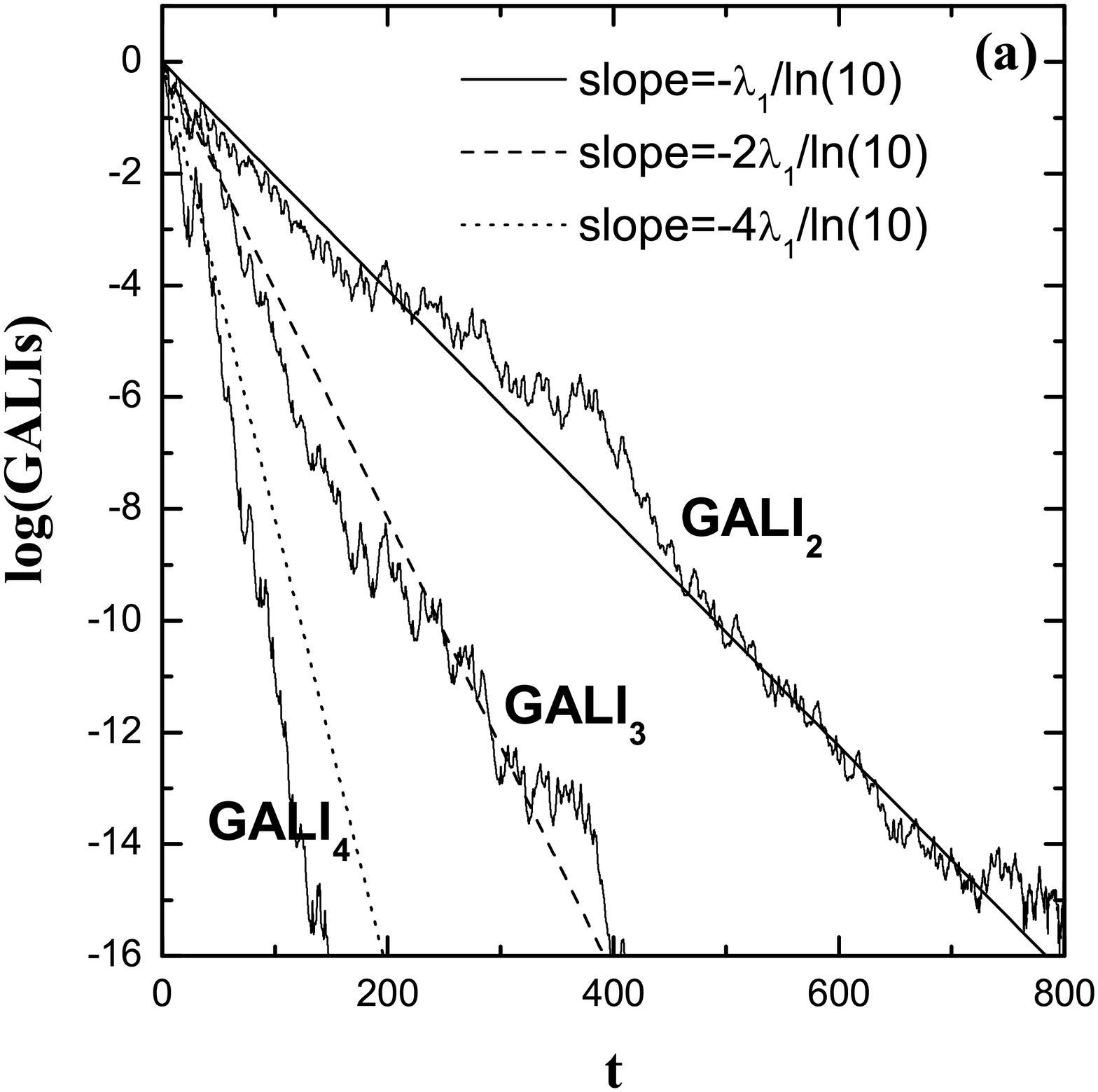}
\hspace{-1.2cm}
\includegraphics[scale=0.245]{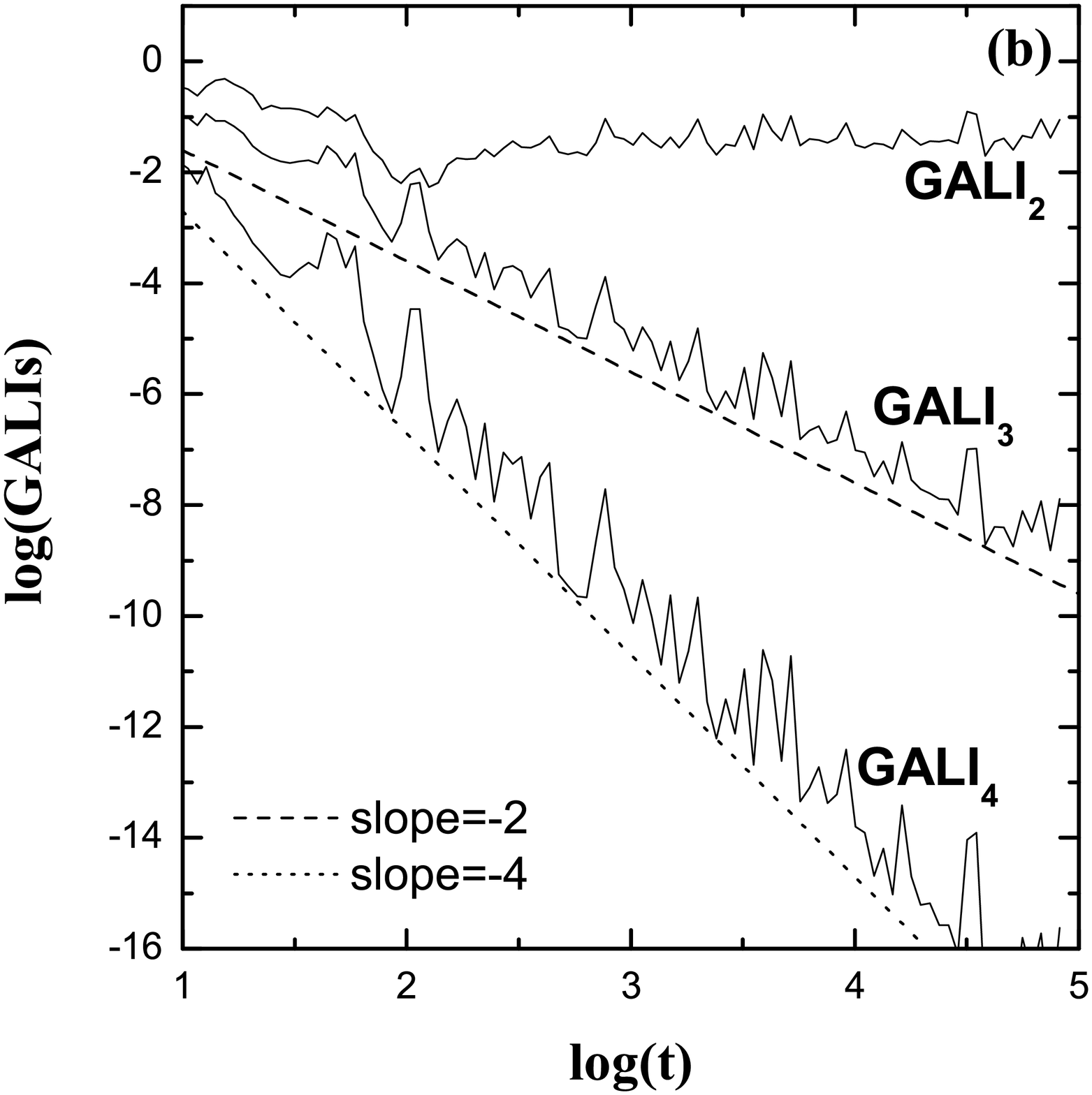} }
\caption{The time evolution of the GALI$_2$, the GALI$_3$ and the
  GALI$_4$ for (a) a chaotic and (b) a regular orbit of the 2D
  Hamiltonian (\ref{eq:2DHam}) for $H_2=0.125$. The chaotic orbit is
  the one considered in Fig.~\ref{fig:SM_SALI_orbs}a, while the
  initial conditions of the regular orbit are $q_1=0$, $q_2=0$,
  $p_1=0.5$, $p_2=0$. The straight lines correspond in (a) to
  functions proportional to $\exp(-\lambda_1t)$, $\exp(-2\lambda_1t)$
  and $\exp(-4\lambda_1t)$, for $\lambda_1=0.047$ and in (b) to
  functions proportional to $t^{-2}$ and $t^{-4}$. The slope of each
  line is mentioned in the legend. Note that the horizontal, time axis
  in (a) is linear, while in (b) is logarithmic (after \cite{SBA_07})}
\label{fig:GALI_behavior_2D}
\end{figure}

Initially, we consider the 2D Hamiltonian (\ref{eq:2DHam}) which has a
4d phase space. For this system we can define the GALI$_k$ for $k=2$,
3 and 4. Then, according to (\ref{eq:LEs_pairs}) and
(\ref{eq:LEs_zero}), the LEs satisfy the conditions
$\lambda_1=-\lambda_4$, $\lambda_2=\lambda_3=0$. Thus, according to
(\ref{eq:GALI_ch}) the evolution of the GALIs for a chaotic orbit is
given by
\begin{equation}
\mbox{GALI}_2(t) \propto e^{-\lambda_1 t},\,\,\, \mbox{GALI}_3(t)
\propto e^{-2\lambda_1 t},\,\,\,\mbox{GALI}_4(t) \propto
e^{-4\lambda_1 t}. \label{eq:2DHam_ch_GALIs}
\end{equation}
On the other hand, for a regular orbit (\ref{eq:GALI_reg}) indicates
that
\begin{equation}
\mbox{GALI}_2(t) \propto \mbox{constant},\,\,\, \mbox{GALI}_3(t)
\propto \frac{1}{t^2},\,\,\,\mbox{GALI}_4(t) \propto
\frac{1}{t^4}. \label{eq:2DHam_reg_GALIs}
\end{equation}
From the results of Fig.~\ref{fig:GALI_behavior_2D}, where the time evolution
of the GALI$_2$, the GALI$_3$ and the GALI$_4$ for a chaotic orbit (actually the one considered in Figs.~\ref{fig:SM_SALI_orbs}(a) and
\ref{fig:SALI_exponential}) and a regular orbit are plotted, we see that the
laws (\ref{eq:2DHam_ch_GALIs}) and (\ref{eq:2DHam_reg_GALIs}) describe quite
accurately the obtained numerical data.

\begin{figure}
\centerline{\hspace{0.5cm}
\includegraphics[scale=0.245]{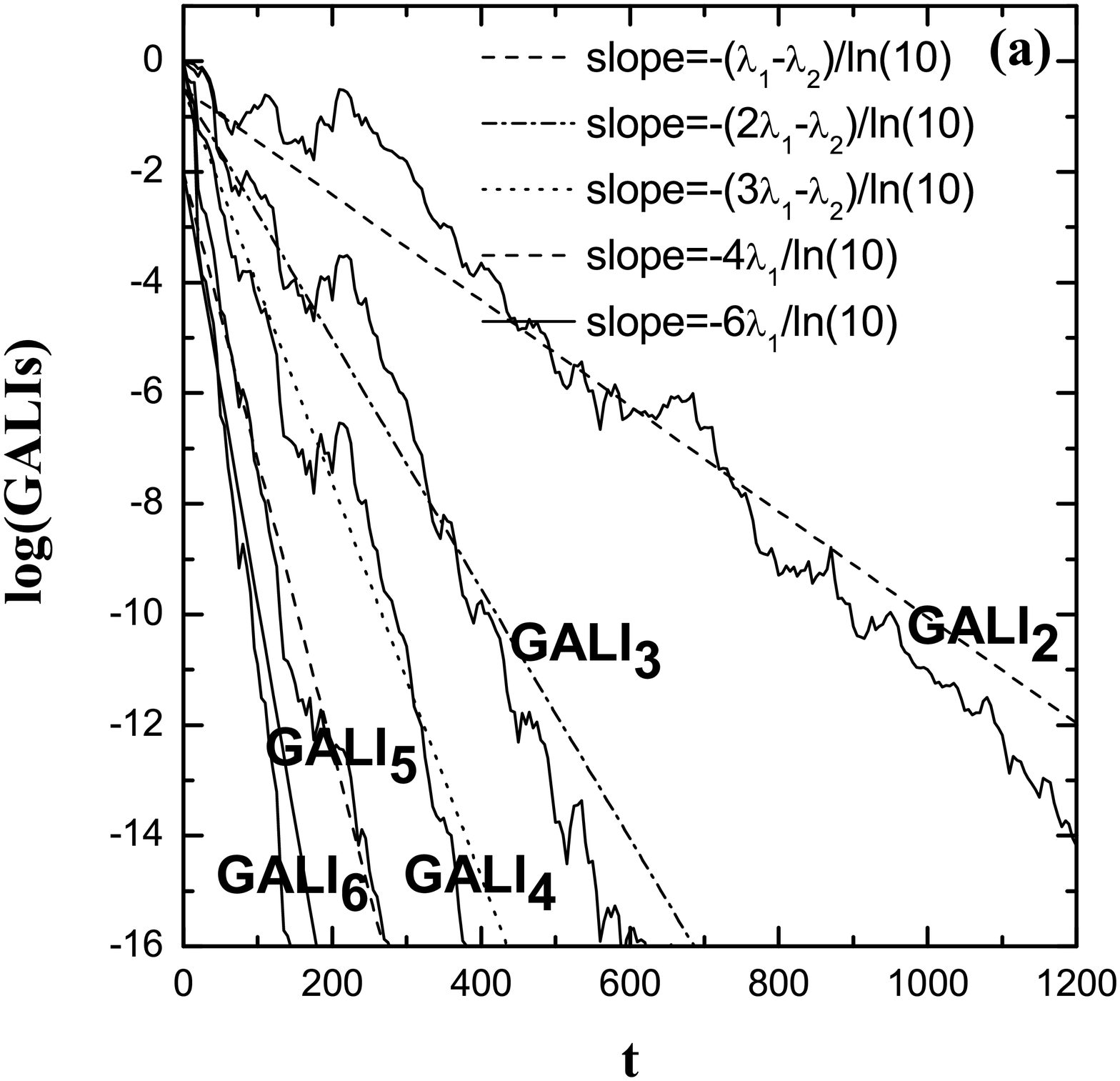}
\hspace{-1.2cm}
\includegraphics[scale=0.245]{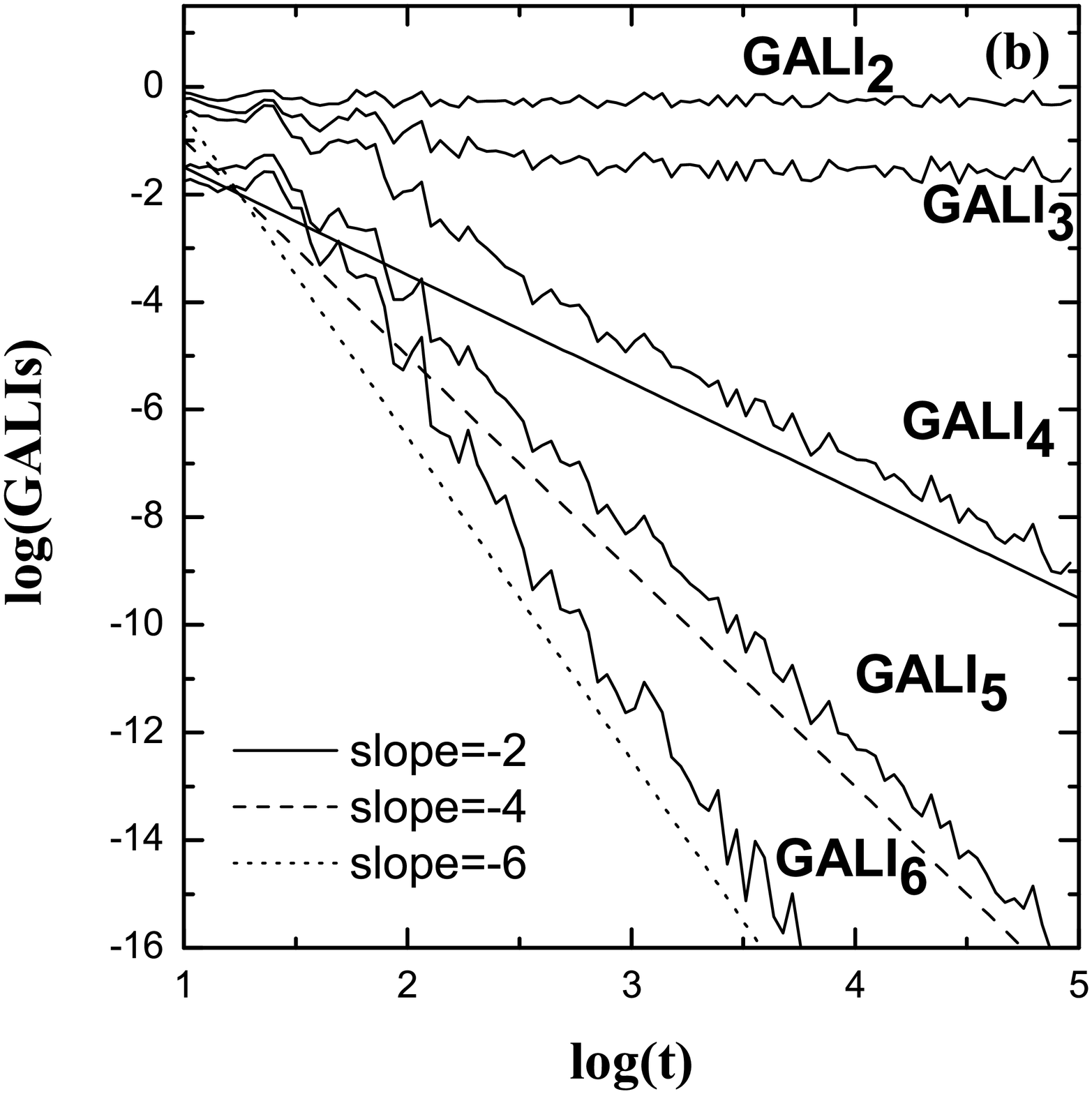} }
\caption{The time evolution of the GALI$_k$, $k=2,3,\ldots,6$ for (a)
  a chaotic and (b) a regular orbit of the 3D Hamiltonian
  (\ref{eq:3DHam}) with $H_3=0.09$, $\omega_1=1$, $\omega_2=\sqrt{2}$
  and $\omega_3=\sqrt{3}$. The initial conditions of the orbits are:
  (a) $q_1=0$, $q_2=0$, $q_3=0$, $E_1=0.03$, $E_2=0.03$, $E_3=0.03$,
  and (b) $q_1=0$, $q_2=0$, $q_3=0$, $E_1=0.005$, $E_2=0.085$,
  $E_3=0$, where the quantities $E_1$, $E_2$, $E_3$ (usually referred
  as the `harmonic energies') are related to the momenta $p_1$, $p_2$,
  $p_3$ through $p_i=\sqrt{2E_i/\omega_i}$, $i=1,2,3$. The straight
  lines in (a) correspond to functions proportional to
  $\exp[-(\lambda_1-\lambda_2)t]$, $\exp[-(2\lambda_1-\lambda_2)t]$,
  $\exp[-(3\lambda_1-\lambda_2)t]$, $\exp(-4\lambda_1t)$ and
  $\exp(-6\lambda_1t)$ for $\lambda_1=0.03$, $\lambda_2=0.008$, which
  are accurate numerical estimations of the orbit's two largest LEs
  (see \cite{SBA_07} for more details). The straight lines in (b)
  correspond to functions proportional to $t^{-2}$, $t^{-4}$ and
  $t^{-6}$. The slope of each line is mentioned in the legend.  The
  horizontal, time axis is linear in (a) and logarithmic in (b) (after
  \cite{SBA_07})}
\label{fig:GALI_behavior_3D}
\end{figure}

For a 3D Hamiltonian like (\ref{eq:3DHam}) the theoretical prediction
(\ref{eq:GALI_ch}) gives
\begin{equation}
\begin{array}{c}
\mbox{GALI}_2(t) \propto e^{-(\lambda_1 - \lambda_2)t},\,\,\,
\mbox{GALI}_3(t) \propto e^{-(2 \lambda_1 - \lambda_2)t}
,\,\,\,\mbox{GALI}_4(t) \propto e^{-(3 \lambda_1 -
  \lambda_2)t}, \vspace{0.2cm} \\ \mbox{GALI}_5(t) \propto
e^{-4\lambda_1 t},\,\,\, \mbox{GALI}_6(t) \propto e^{-6\lambda_1 t},
\end{array}
\label{eq:3DHam_ch_GALIs}
\end{equation}
for a chaotic orbit, because, according to (\ref{eq:LEs_pairs}) and
(\ref{eq:LEs_zero}), $\lambda_1=-\lambda_6$, $\lambda_2=-\lambda_5$
and $\lambda_3=\lambda_4=0$.  On the other hand, a regular orbit lies
on a 3d torus and according to (\ref{eq:GALI_reg}) the GALIs should
behave as
\begin{equation}
\begin{array}{c}
\displaystyle \mbox{GALI}_2(t) \propto \mbox{constant},\,\,\,
\mbox{GALI}_3(t) \propto \mbox{constant} ,\,\,\,\mbox{GALI}_4(t)
\propto \frac{1}{t^2}, \\ \displaystyle \mbox{GALI}_5(t) \propto
\frac{1}{t^4},\,\,\, \mbox{GALI}_6(t) \propto \frac{1}{t^6}.
\end{array}
\label{eq:3DHam_reg_GALIs}
\end{equation}
In Fig.~\ref{fig:GALI_behavior_3D} we plot the time evolution of the
various GALIs for a chaotic (Fig.~\ref{fig:GALI_behavior_3D}(a)) and a
regular (Fig.~\ref{fig:GALI_behavior_3D}(b)) orbit of the 3D
Hamiltonian (\ref{eq:3DHam}).  From the plotted results we see that
the behaviors of the GALIs are very well approximated by
(\ref{eq:3DHam_ch_GALIs}) and (\ref{eq:3DHam_reg_GALIs}). We note here
that the constant values that the GALI$_2$ and the GALI$_3$ eventually
attain in Fig.~\ref{fig:GALI_behavior_3D}(b) are not the same.
Actually, the limiting value of GALI$_3$ is
smaller than the one of GALI$_2$.

\begin{figure}
\centerline{
\begin{tabular}{cc}
\hspace{0.2cm}
\includegraphics[scale=0.245]{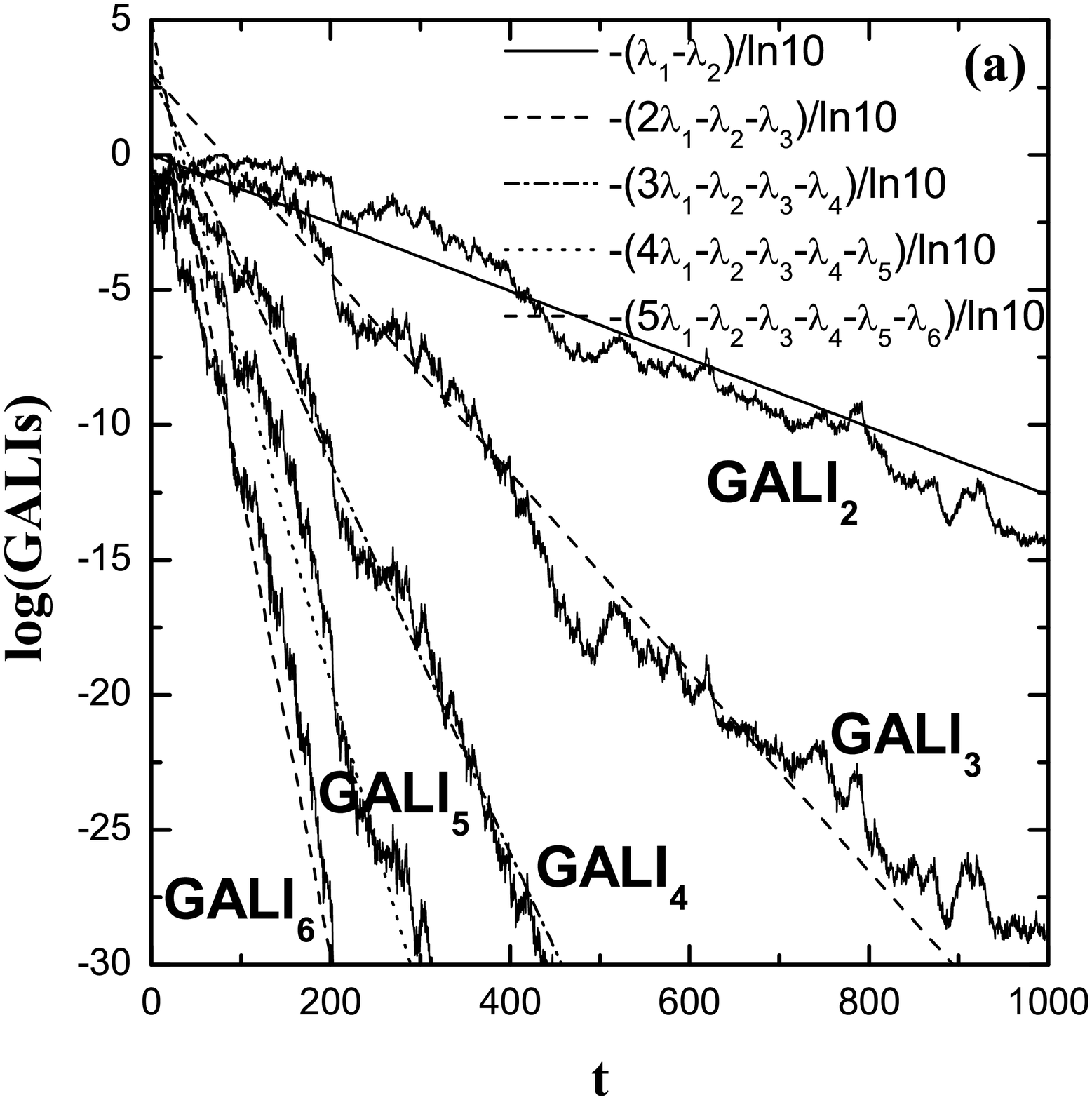} &
\hspace{-1.2cm}
\includegraphics[scale=0.245]{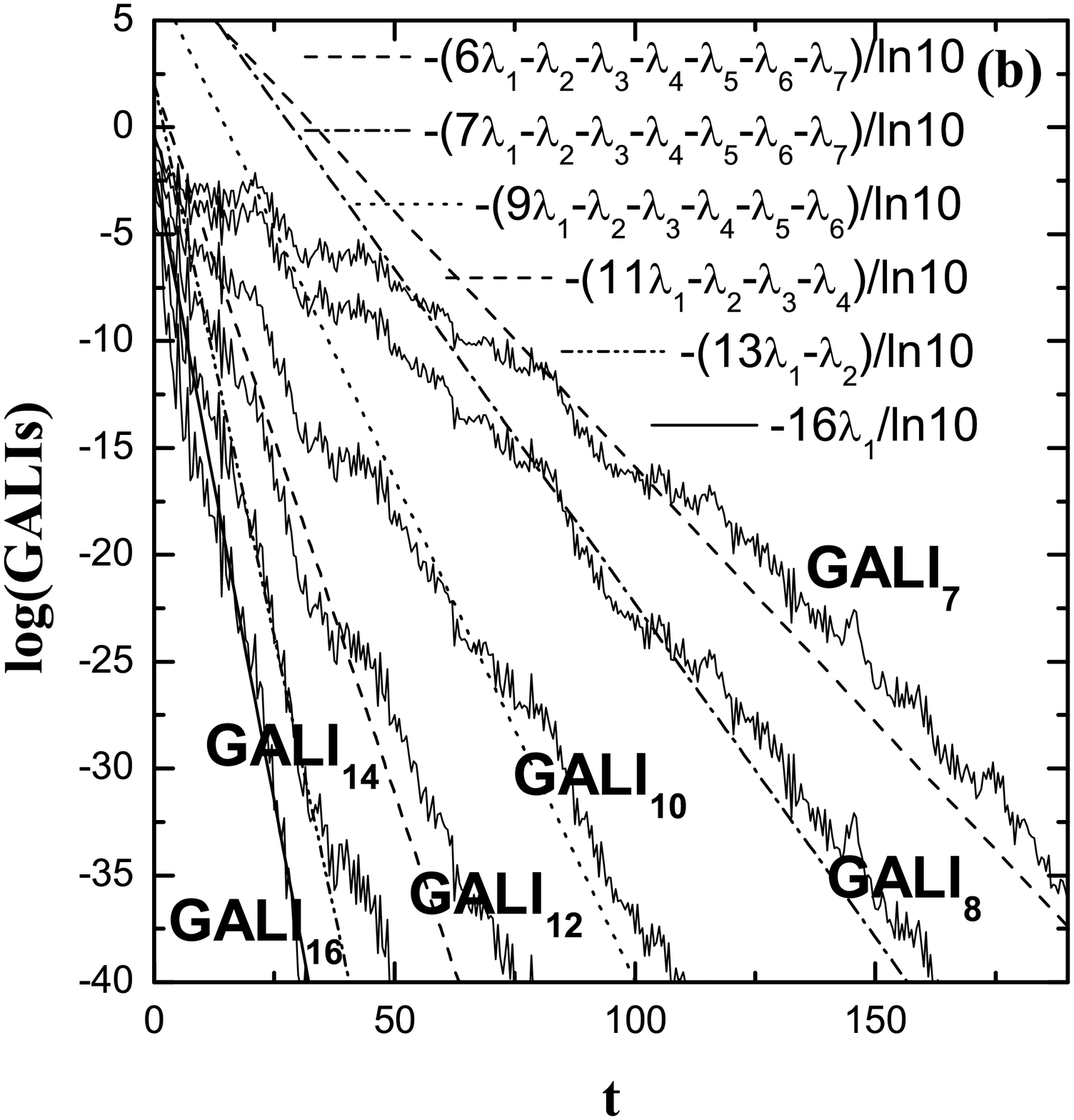}
\vspace{-1.0cm} \\
\hspace{0.2cm}
\includegraphics[scale=0.245]{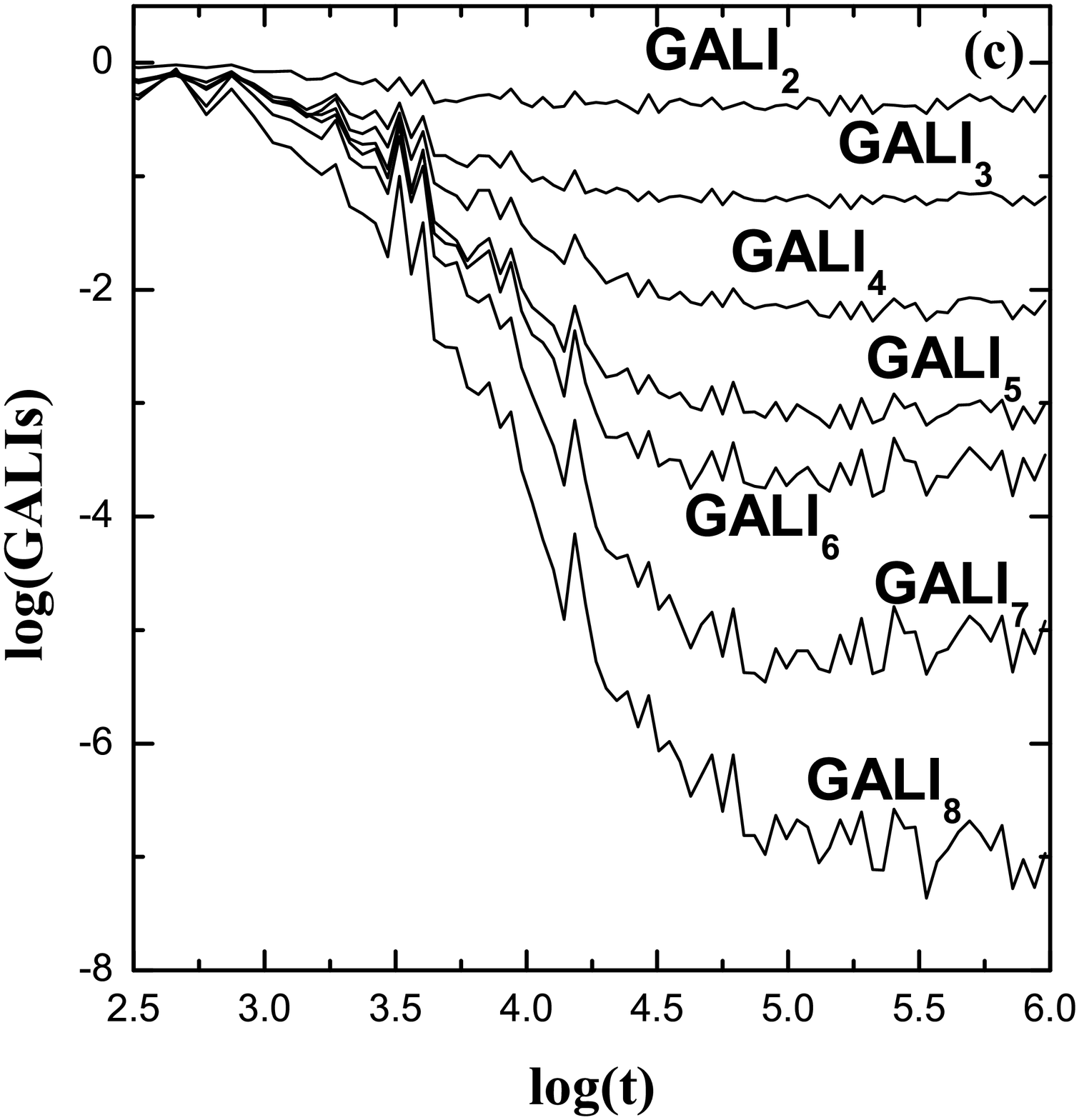} &
\hspace{-1.2cm}
\includegraphics[scale=0.245]{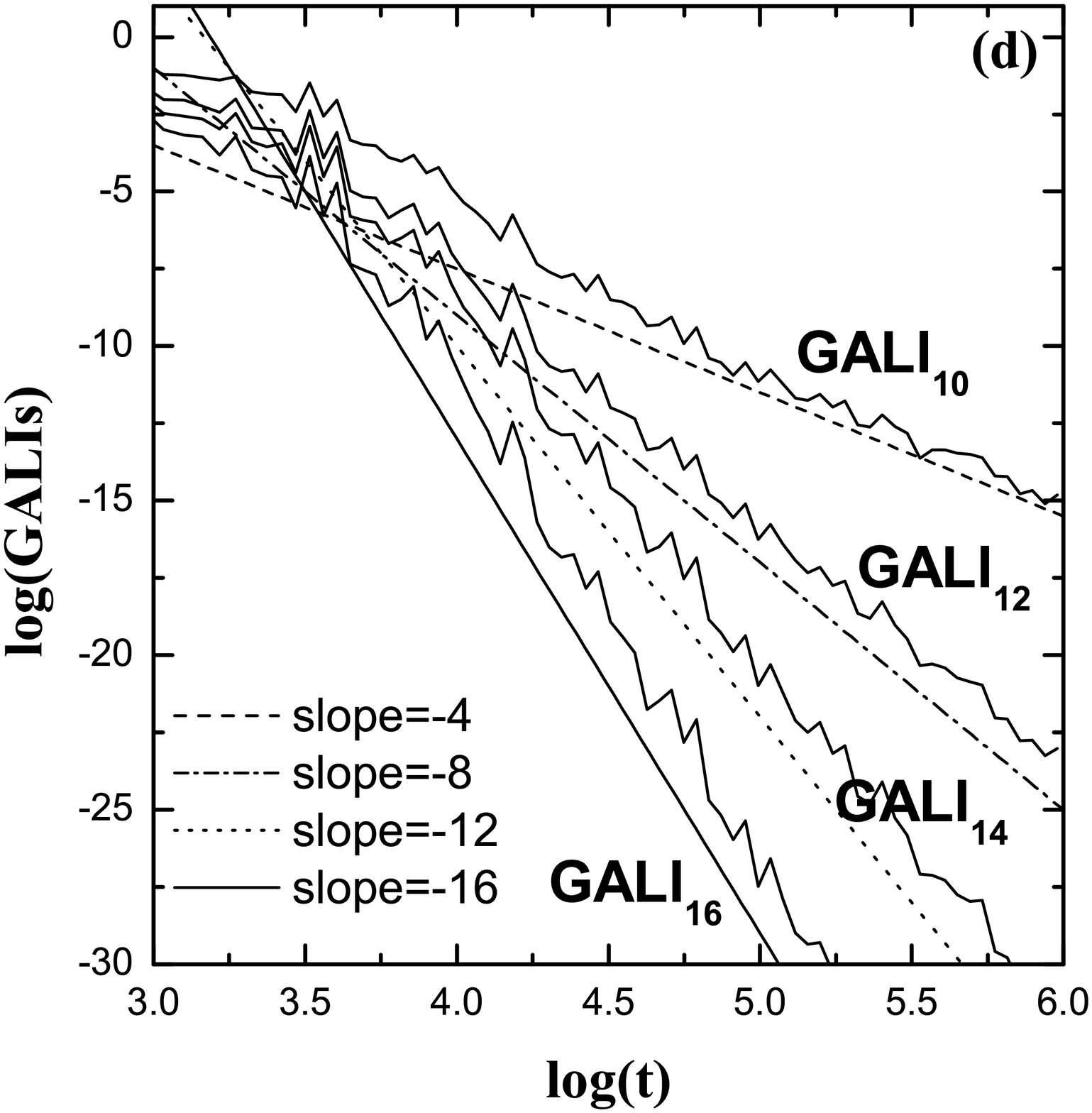}
\end{tabular}}
\caption{The time evolution of the GALI$_k$,
  $k=2,\ldots,8,10,12,14,16$ for a chaotic (panels (a) and (b)) and a
  regular orbit (panels (c) and (d)) of the $N$D Hamiltonian
  (\ref{eq:NDHam}) with $N=8$ and $\beta=1.5$. The initial conditions
  of the chaotic orbit are $Q_1=Q_4=2$, $Q_2=Q_5=1$, $Q_3=Q_6=0.5$,
  $Q_7=Q_8=0.1$, $P_i=0$ where $\displaystyle
  Q_i=\frac{2}{3}\sum_{j=1}^8 q_j \sin \left(\frac{i j
      \pi}{9}\right)$, $\displaystyle P_i=\frac{2}{3}\sum_{j=1}^8 p_j
  \sin \left(\frac{i j \pi}{9}\right)$, $i=1,\ldots,8$ (see
  \cite{SBA_08} for more details). The initial conditions of the
  regular orbit are $q_1=q_2=q_3=q_8=0.05$, $q_4=q_5=q_6=q_7=0.1$,
  $p_i=0$, $i=1,\ldots,8$. The straight lines in (a) and (b)
  correspond to exponential functions of the form (\ref{eq:GALI_ch})
  for $\lambda_1=0.170$, $\lambda_2=0.141$, $\lambda_3=0.114$,
  $\lambda_4=0.089$, $\lambda_5=0.064$, $\lambda_6=0.042$,
  $\lambda_7=0.020$ , which are estimations (obtained in
  \cite{SBA_08}) of the orbit's seven largest LEs. The straight lines
  in (d) correspond to functions proportional to $t^{-4}$, $t^{-8}$,
  $t^{-12}$ and $t^{-16}$. The slope of each line is mentioned in the
  legend.  Note the huge range differences in the horizontal, time
  axes between panels (a) and (b), where the axes are linear, and
  panels (c) and (d) where the axes are logarithmic (after
  \cite{SBA_08})}
\label{fig:GALI_behavior_8D}
\end{figure}

As an example of evaluating the GALIs for multidimensional
Hamiltonians we consider model (\ref{eq:NDHam}) for $N=8$
particles. This corresponds to an 8D Hamiltonian system $H_8$, having
a 16d phase space, which allows the definition of several GALIs:
starting from GALI$_2$ up to GALI$_{16}$. In
Fig.~\ref{fig:GALI_behavior_8D} the time evolution of several of these
indices are shown for a chaotic (Figs.~\ref{fig:GALI_behavior_8D}(a)
and (b)) and a regular (Figs.~\ref{fig:GALI_behavior_8D}(c) and (d))
orbit.  From these results we again conclude that the laws
(\ref{eq:GALI_ch}) and (\ref{eq:GALI_reg}) are quite accurate in
describing the time evolution of the GALIs.

The first seven indices, GALI$_2$ up to GALI$_8$, exhibit completely
different behaviors for chaotic and regular motion: they tend
exponentially fast to zero for a chaotic orbit
(Figs.~\ref{fig:GALI_behavior_8D}(a) and (b)), while they attain
constant, positive values for a regular one
(Fig.~\ref{fig:GALI_behavior_8D}(c)).  This characteristic makes them
ideal numerical tools for discriminating between the two cases, as we
 see in Sect.~\ref{sect:global_2-N} where some specific numerical
examples are discussed in detail.

Although the constancy of the GALI$_k$, $k=1,\ldots,8$ for regular
orbits is predicted from (\ref{eq:GALI_reg}), nothing is yet said
about the actual values of these constants. It is evident from
Fig.~\ref{fig:GALI_behavior_8D}(c) that these values decrease as the
order $k$ of the GALI$_k$ increases, something which was also observed
in Fig.~\ref{fig:GALI_behavior_3D}(b) for the 3D Hamiltonian
(\ref{eq:3DHam}). For the regular orbit of
Fig.~\ref{fig:GALI_behavior_8D}(c) we see that GALI$_8 \approx
10^{-7}$.  One might argue that this very small value could be
considered to be practically zero and that the orbit might be
(wrongly) classified as chaotic. The flaw in this argumentation is
that the possible smallness of GALI$_8 \approx 10^{-7}$ is of relative
nature as this value should be compared with the values that the index
reaches for actual chaotic orbits. For instance, the chaotic orbit of
Fig.~\ref{fig:GALI_behavior_8D}(b) has GALI$_8 \approx 10^{-40}$,
after only $t\approx 160$ time units! At the same time we get GALI$_8
\approx 10^{-1}$ for the regular orbit
(Fig.~\ref{fig:GALI_behavior_8D}(c)). In addition, extrapolating the
results of GALI$_8$ for the chaotic orbit in
Fig.~\ref{fig:GALI_behavior_8D}(b) to e.g.~$t\approx 10^5$ we would
obtain values extremely smaller than the value GALI$_8 \approx
10^{-7}$ archived for the regular orbit in
Fig.~\ref{fig:GALI_behavior_8D}(c).

The necessity to determine an appropriate threshold value for the
GALI$_k$, $2 \leq k \leq N$, below which orbits will be securely
classified as chaotic, becomes evident from the above analysis. Since
a theoretical, or even an empirical (numerical) relation between the
order $k$ of the GALI$_k$ and the constant value it reaches for
regular orbits is still lacking, one efficient way to determine this
threshold value is by computing the GALI$_k$ for some representative
chaotic and regular orbits of each studied system.  Then, a safe
policy is to define this threshold to be a few orders of magnitude
smaller than the minimum value obtained by the GALI$_k$ for the tested
regular orbits. For example, based on the results of
Fig.~\ref{fig:GALI_behavior_3D} for the 3D Hamiltonian (\ref{eq:3DHam}) this threshold value for the GALI$_3$ could be set to be $\leq 10^{-8}$, while for the system of Fig.~\ref{fig:GALI_behavior_8D} a reliable threshold value for the GALI$_8$ could be $\leq 10^{-16}$.

The results of Fig.~\ref{fig:GALI_behavior_8D} verify the predictions of
(\ref{eq:GALI_ch}) and (\ref{eq:GALI_reg}) that the GALIs of order $8< k \leq
16$ tend to zero both for chaotic and regular orbits. Nevertheless, the
completely different way they do so, i.e.~they decay exponentially fast for
chaotic orbits, while they follow a power law decay for regular ones, allows us again to develop a well-tailored strategy to discriminate between the two cases. The different decay laws result in enormous differences in the time the
indices need to reach any predefined low value. Thus, the measurement of this time can be used to characterize the nature of the orbits, as we see in Sect.~\ref{sect:global_N-2N}. For example, for the chaotic orbit of Fig.~\ref{fig:GALI_behavior_8D}(b) GALI$_{16} \approx 10^{-30}$ after
about $t\approx 25$ time units, while it reaches the same small value after
about $t\approx 10^{5}$ time units for the regular orbit of
Fig.~\ref{fig:GALI_behavior_8D}(d); a time interval which is larger by a factor $\approx 4,000$ with respect to the chaotic orbit!

\paragraph{Symplectic Maps}
\label{sect:GALI_maps paradigms}

\begin{figure}
\centerline{
\hspace{0.5cm}
\includegraphics[scale=0.28]{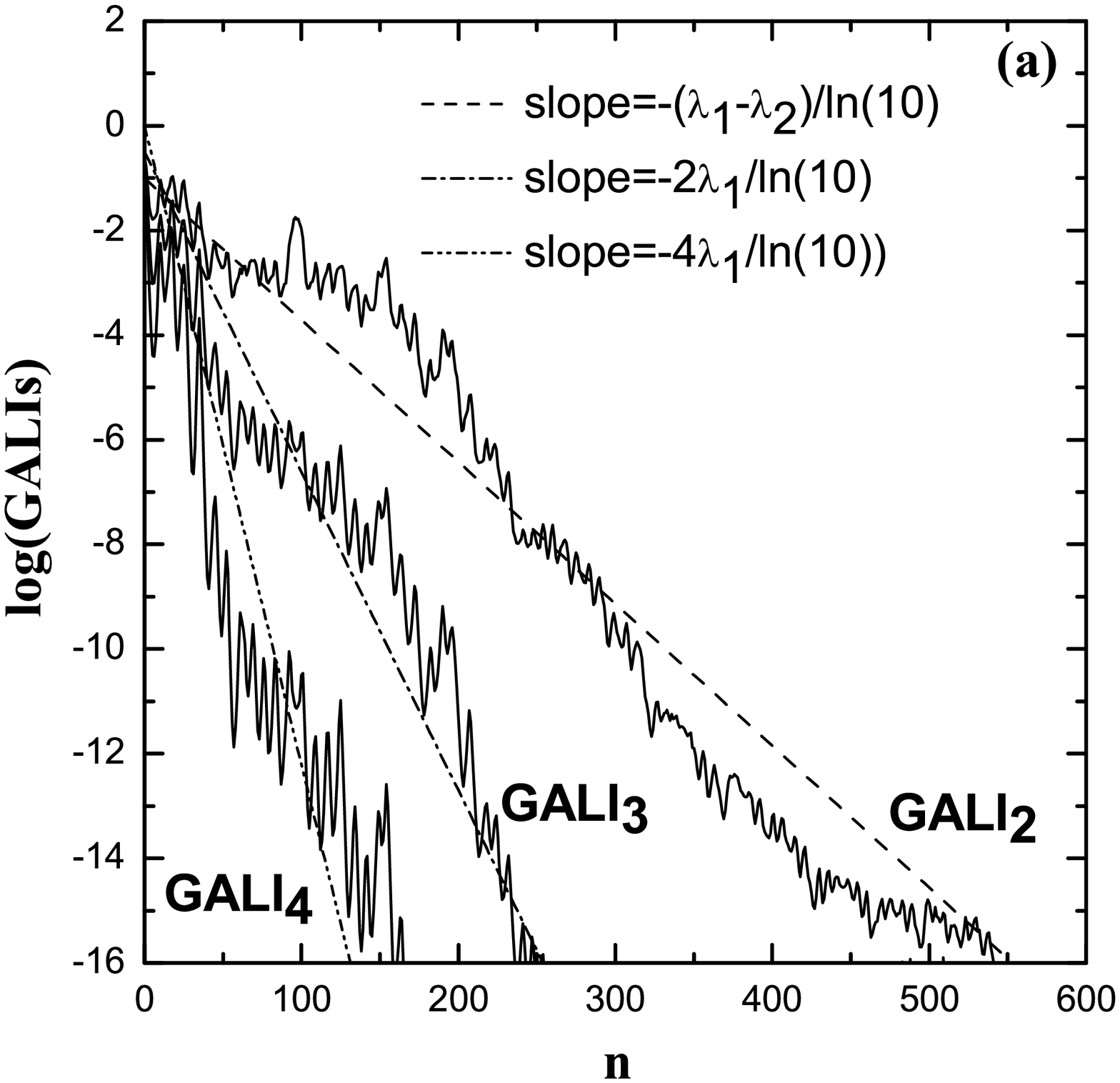}
\hspace{-2.2cm}
\includegraphics[scale=0.28]{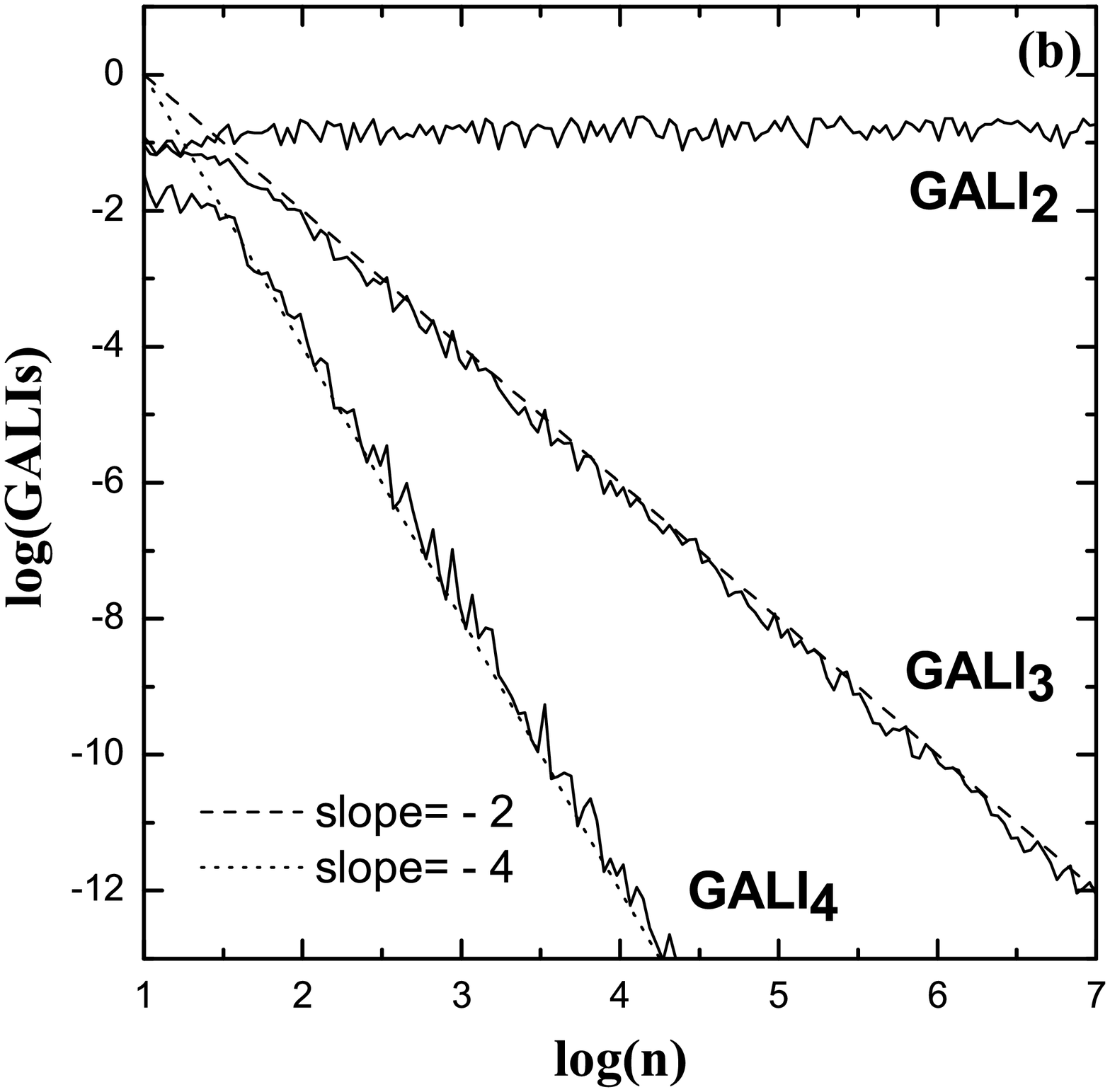}
}
\caption{The evolution of the GALI$_2$, the GALI$_3$ and the GALI$_4$
  with respect to the number of iterations $n$ for (a) a chaotic and
  (b) a regular orbit of the 4d map (\ref{eq:4d_map}) with $K=0.5$ and
  $\gamma=0.05$. The initial conditions of the orbits are: (a)
  $x_1=0.55$, $y_1=0.1$, $x_2=0.005$, $y_2=0.01$, and (b) $x_1=0.55$,
  $y_1=0.1$, $x_2=0.54$, $y_2=0.01$. The straight lines in (a)
  correspond to functions proportional to
  $\exp[-(\lambda_1-\lambda_2)n]$, $\exp(-2\lambda_1 n)$ and
  $\exp(-4\lambda_1 n)$ for $\lambda_1=0.07$, $\lambda_2=0.008$, which
  are the orbit's LEs obtained in \cite{MSB_09}. The straight lines in
  (b) represent functions proportional to $n^{-2}$ and $n^{-4}$. The
  slope of each line is mentioned in the legend. Note that the
  horizontal axis is linear in (a) and logarithmic in (b) (after
  \cite{MSB_09})}
\label{fig:GALI_behavior_4d}
\end{figure}

\begin{figure}
\centerline{
\hspace{0.5cm}
\includegraphics[scale=0.28]{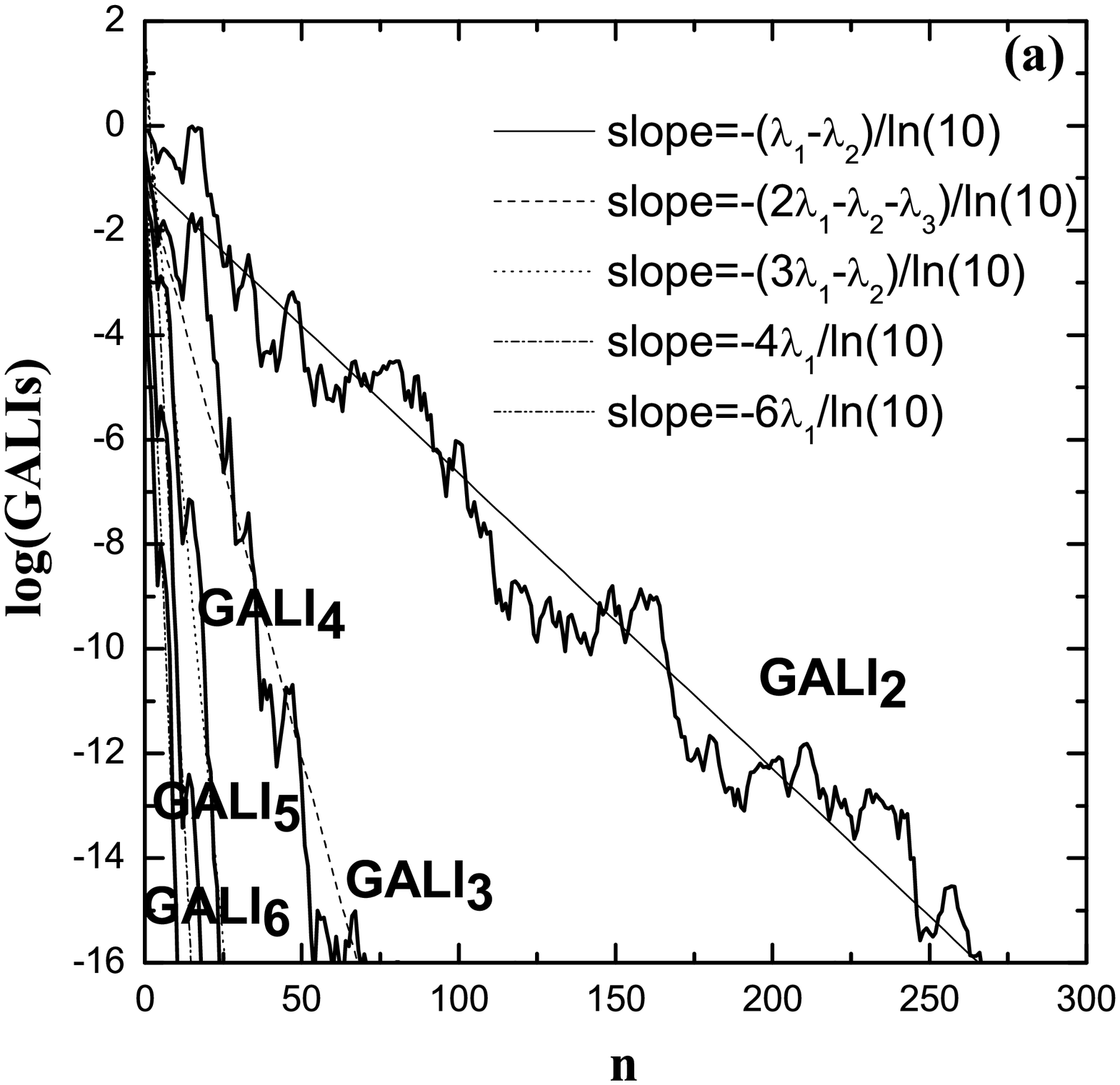}
\hspace{-2.2cm}
\includegraphics[scale=0.28]{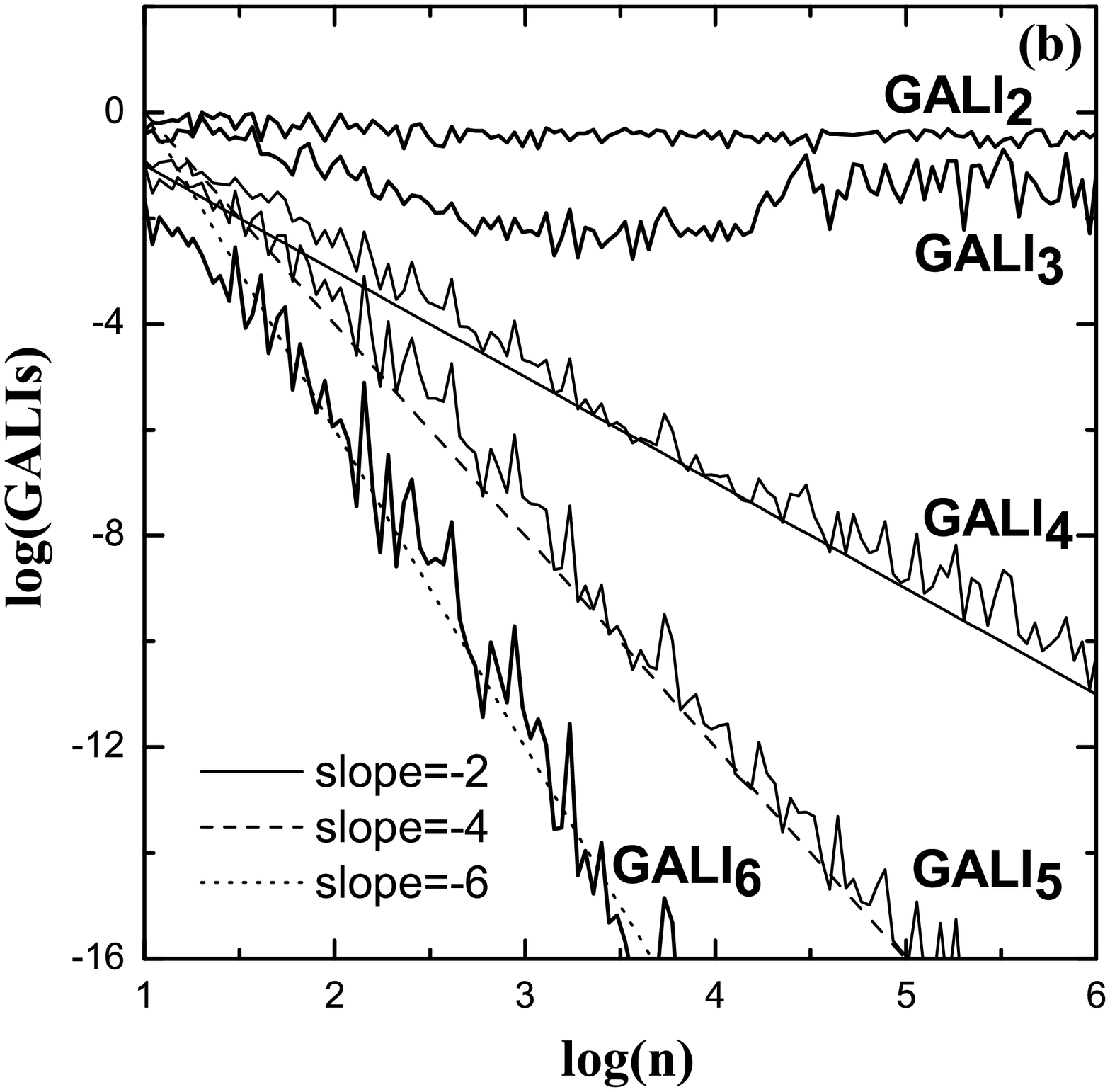}
  }
  \caption{The evolution of the GALI$_k$, $k=2,3,\ldots,6$ with
    respect to the number of iterations $n$ for (a) a chaotic (after
    \cite{MSB_08}) and (b) a regular orbit of the 6d map
    (\ref{eq:6d_map}) with $K=3$ and $\gamma=0.1$.  The initial
    conditions of the orbits are: (a) $x_1=x_2=x_3=0.8$, $y_1=0.05$,
    $y_2=0.21$, $y_3=0.01$, and (b) $x_1=x_2=x_3=0.55$, $y_1=0.05$,
    $y_2=0.21$, $y_3=0$. The straight lines in (a) correspond to
    functions proportional to $\exp[-(\lambda_1-\lambda_2)n]$,
    $\exp[-(2\lambda_1-\lambda_2-\lambda_3)n]$,
    $\exp[-(3\lambda_1-\lambda_2)n]$, $\exp(-4\lambda_1 n)$ and
    $\exp(-6\lambda_1 n)$ for $\lambda_1=0.70$, $\lambda_2=0.57$,
    $\lambda_3=0.32$, which are the orbit's LEs obtained in
    \cite{MSB_08}.  The straight lines in (b) represent functions
    proportional to $n^{-2}$, $n^{-4}$ and $n^{-6}$. The slope of each
    line is mentioned in the legend. Note that the horizontal axis is
    linear in (a) and logarithmic in (b) }
\label{fig:GALI_behavior_6d}
\end{figure}

Although up to now our discussion concerned the implementation of the
GALIs to Hamiltonian systems, the indices follow laws
(\ref{eq:GALI_ch}) and (\ref{eq:GALI_reg}) also for symplectic maps
(with the obvious substitution of the continuous time $t$ by a
discrete one which counts the map's iterations $n$) as the
representative results of Figs.~\ref{fig:GALI_behavior_4d} and
\ref{fig:GALI_behavior_6d} clearly verify. In particular, in
Fig.~\ref{fig:GALI_behavior_4d} we see the behavior of the GALIs for a
chaotic (Fig.~\ref{fig:GALI_behavior_4d}(a)) and a regular
(Fig.~\ref{fig:GALI_behavior_4d}(b)) orbit of the 4d map
\begin{equation}
\label{eq:4d_map}
\begin{array}{ccl}
   \displaystyle x'_1 & \displaystyle = & \displaystyle x_1 +y'_1
   \\ \displaystyle y'_1 & \displaystyle = & \displaystyle y_1
   +\frac{K}{2 \pi} \sin \left( 2 \pi x_1 \right) - \frac{\gamma}{2
     \pi} \sin \left[ 2 \pi \left( x_2 -x_1 \right) \right]
   \\ \displaystyle x'_2 & \displaystyle = & \displaystyle x_2 +y'_2
   \\ \displaystyle y'_2 & \displaystyle = & \displaystyle y_2
   +\frac{K}{2 \pi} \sin \left( 2 \pi x_2 \right) - \frac{\gamma}{2
     \pi} \sin \left[ 2 \pi \left( x_1 -x_2 \right) \right],
\end{array}
\end{equation}
obtained from (\ref{eq:Md_map}) for $M=2$ and $K_1=K_2=K$, while in
Fig.~\ref{fig:GALI_behavior_6d} a chaotic
(Fig.~\ref{fig:GALI_behavior_6d}(a)) and a regular
(Fig.~\ref{fig:GALI_behavior_6d}(b)) orbit of the 6d map
(\ref{eq:6d_map}) are considered.

These results illustrate the fact that the GALI$_k$ has the same
behavior for Hamiltonian flows and symplectic maps. For instance, even
by simple inspection we conclude that the GALIs behave similarly in
Figs.~\ref{fig:GALI_behavior_2D} and \ref{fig:GALI_behavior_4d}, which
refer to a 2D Hamiltonian and a 4d map respectively, as well as in
Figs.~\ref{fig:GALI_behavior_3D} and \ref{fig:GALI_behavior_6d}, which
refer to a 3D Hamiltonian and a 4d map respectively.

\subsubsection{The Case of 2d Maps}
\label{sect:2d_maps}

Equations (\ref{eq:GALI_ch}) and (\ref{eq:GALI_reg}) describe the
behavior of the GALIs for $N$D Hamiltonian systems and $2N$d
symplectic maps with $N\geq 2$. What happens if $N=1$? The case of an
1D, time independent Hamiltonian is not very interesting because such
systems are integrable and chaos does not appear. But, this is not the
case for 2d maps, which can exhibit chaotic behavior.

\begin{figure}
\centerline{
\hspace{2.2cm}
\includegraphics[scale=0.3]{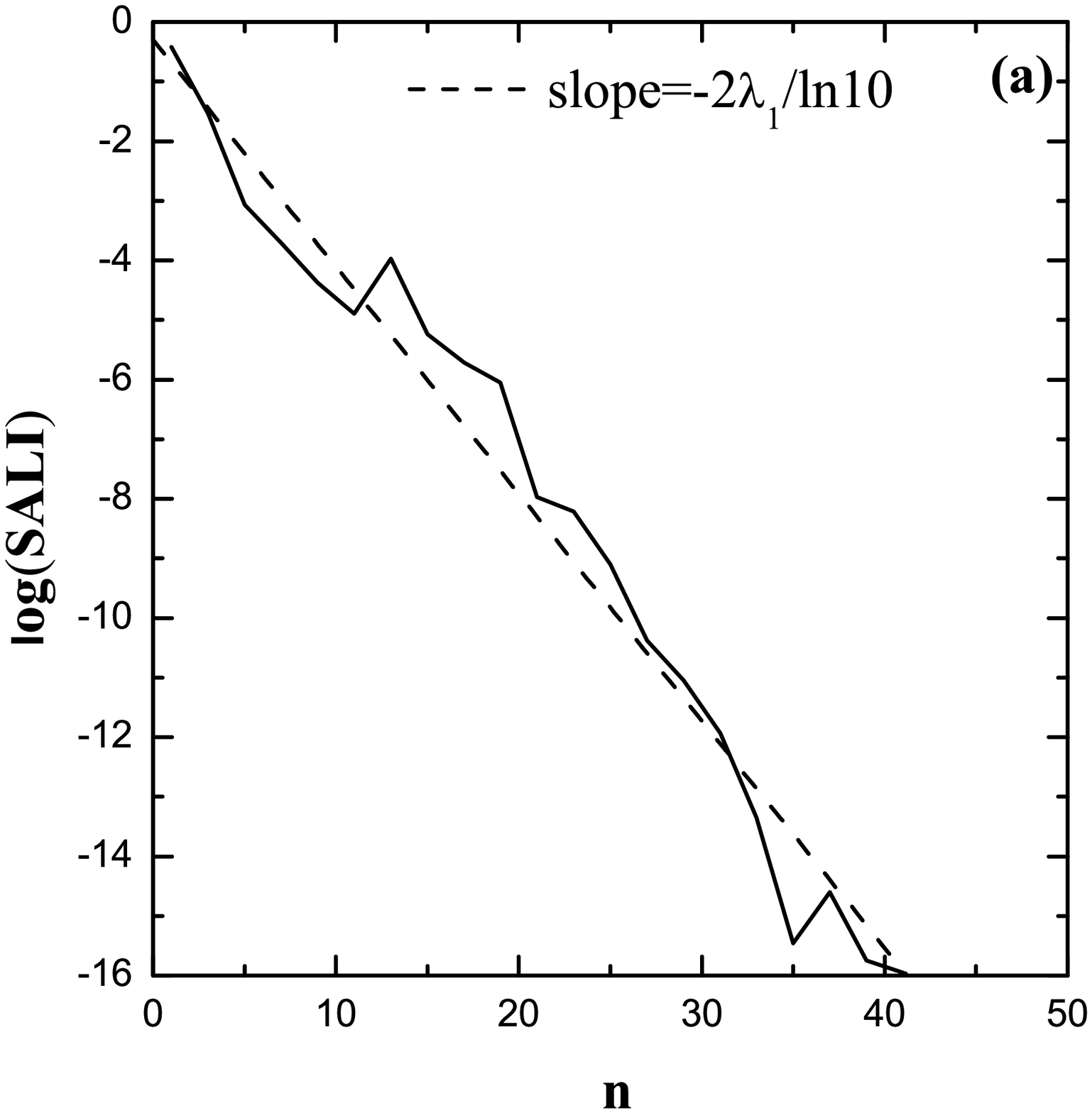}
\hspace{-2.2cm}
\includegraphics[scale=0.3]{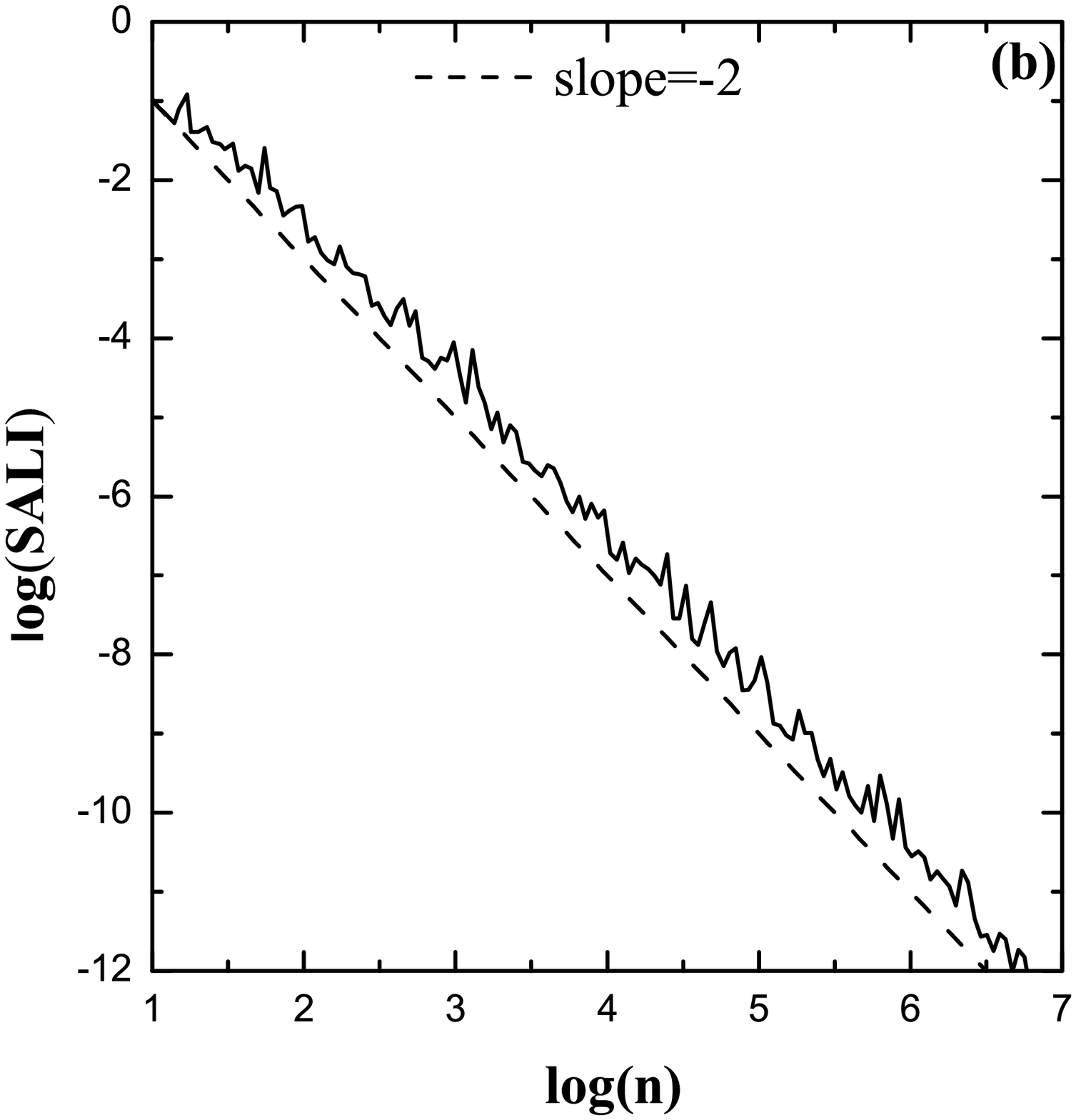}
\vspace{-0.5cm}}
\caption{The evolution of the SALI (which in practice is the GALI$_2$)
  with respect to the number of iterations $n$ for (a) a chaotic and
  (b) a regular orbit of the 2d map (\ref{eq:2d_map}) with $K=2$. The
  initial conditions of the orbits are: (a) $x_1=y_1=0.2$, and (b)
  $x_1=0.4$, $y_1=0.8$. The straight line in (a) corresponds to a
  function proportional to $\exp(-2\lambda_1 n)$ for
  $\lambda_1=0.438$, which is the orbit's mLE obtained in
  \cite{MSAB2008}, while the line in (b) represents a function
  proportional to $n^{-2}$. The slope of each line is mentioned in the
  legend. Note that the horizontal axis is linear in (a) and
  logarithmic in (b) (after \cite{MSAB2008})}
\label{fig:GALI_behavior_2d}
\end{figure}

In 2d maps only the GALI$_2$ (which, according to
(\ref{eq:SALI-GALI2_equi}) is equivalent to the SALI) is defined. For
chaotic orbits the GALI$_2$ decreases exponentially to zero according
to (\ref{eq:GALI_ch}), which becomes
\begin{equation}
\label{eq:GALI_ch_2dmap}
\mbox{GALI}_2(n) \propto \mbox{SALI}(n) \propto \exp \left(
-2\lambda_1 n \right),
\end{equation}
in this particular case, since, according to (\ref{eq:LEs_pairs})
$\lambda_1=-\lambda_2>0$.  Note that in (\ref{eq:GALI_ch_2dmap}) we
have substituted the continuous time $t$ of (\ref{eq:GALI_ch}) by the
number $n$ of map's iterations.  The agreement between the prediction
(\ref{eq:GALI_ch_2dmap}) and actual, numerical data can be seen for
example in Fig.~\ref{fig:GALI_behavior_2d}(a) where the evolution of
the SALI ($\propto \mbox{GALI}_2$) is plotted for a chaotic orbit of
the 2d standard map
\begin{equation}
\label{eq:2d_map}
\begin{array}{ccl}
   \displaystyle x'_1 & \displaystyle = & \displaystyle x_1 +y'_1
   \\ \displaystyle y'_1 & \displaystyle = & \displaystyle y_1
   +\frac{K}{2 \pi} \sin \left( 2 \pi x_1 \right),
\end{array}
\end{equation}
obtained from (\ref{eq:Md_map}) for $M=1$. Thus, we conclude that
(\ref{eq:GALI_ch}) is also valid for 2d maps.

But what happens in the case of regular orbits? Is (\ref{eq:GALI_reg})
still valid for $k=2$ and $N=1$? First of all let us note that for
these particular values of $k$ and $N$ only the second branch of
(\ref{eq:GALI_reg}) is meaningful, and it provides the prediction that
the GALI$_2$ tends to zero as $n^{-2}$. This result is interesting, as
this is the first case of regular motion for which no GALI remains
constant. But actually the vanishing of the GALI$_2$ in this case is
not surprising. Regular motion in 2d maps occurs on 1d invariant
curves. So, any deviation vector from a regular orbit eventually
falls on the tangent space of this curve, which of course has dimension
1. Thus, the two deviation vectors needed for the computation of the
GALI$_2$ eventually becomes collinear and consequently
GALI$_2\rightarrow 0$. Actually the prediction obtained by
(\ref{eq:GALI_reg}), that for regular orbits of 2d maps
\begin{equation}
\label{eq:GALI_reg_2dmap}
\mbox{GALI}_2(n) \propto \mbox{SALI}(n) \propto \frac{1}{n^2},
\end{equation}
is correct, as for example the results of
Fig.~\ref{fig:GALI_behavior_2d}(b) show.

In conclusion we note that the behavior of the SALI/GALI$_2$ for
chaotic and regular orbits in 2d maps is respectively given by
(\ref{eq:GALI_ch_2dmap}) and (\ref{eq:GALI_reg_2dmap}), which are
obtained from (\ref{eq:GALI_ch}) and (\ref{eq:GALI_reg}) for $k=2$ and
$N=1$. The different behaviors of the index for chaotic (exponential
decay) and regular motion (power law decay) were initially observed in
\cite{S_01}, although the exact functional laws
(\ref{eq:GALI_ch_2dmap}) and (\ref{eq:GALI_reg_2dmap}) were derived
later \cite{SABV_04,SBA_07}. As was pointed out even from the first
paper on the SALI \cite{S_01}, these differences allow us to use the
SALI/GALI$_2$ to distinguish between chaotic and regular motion also
in 2d maps (see for instance \cite{S_01,MSAB2008}).

\subsection{Regular Motion on Low Dimensional Tori}
\label{sect:low}

An important feature of the GALIs is their ability to identify regular
motion on low dimensional tori. In order to explain this capability
let us assume that a regular orbit lies on an $s$d torus, $2\leq s
\leq N$, in the $2N$d phase space on an $N$D Hamiltonian system or a
$2N$d map with $N\geq 2$. Then, following similar arguments to the
ones made in Sect.~\ref{sect:GALI_ch_reg} for regular motion on an
$N$d torus, we conclude that the GALI$_k$ eventually remains
constant for $2\leq k \leq s$, because in this case the $k$ deviation
vectors will remain linearly independent when they eventually
fall on the $s$d tangent space of the torus. On the other hand, any
$s< k \leq 2N$ deviation vectors eventually become linearly
dependent as there will be more vectors on the torus' tangent space
than the space's dimension, and consequently the GALI$_k$ will
vanish. In this case, the way the GALI$_k$ tends to zero depends not
only on $k$ and $N$, as in (\ref{eq:GALI_reg}), but also on the
dimension $s$ of the torus. Actually, it was shown analytically in
\cite{CB2006,SBA_08} that for regular orbits on an $s$d torus the
GALI$_k$ behaves as
\begin{equation} \mbox{GALI}_k (t) \propto
  \left\{ \begin{array}{ll} \mbox{constant} & \mbox{if $2\leq k \leq
      s$} \\ \frac{1}{t^{k-s}} & \mbox{if $s< k \leq 2N-s$}
    \\ \frac{1}{t^{2(k-N)}} & \mbox{if $2N-s< k \leq 2N .$} \\
\end{array}\right.
\label{eq:GALI_reg_low_tor}
\end{equation}
It is worth noting that for $s=N$ we retrieve (\ref{eq:GALI_reg}) as
the second branch of (\ref{eq:GALI_reg_low_tor}) becomes meaningless,
while by setting $k=2$, $s=1$ and $N=1$ we get
(\ref{eq:GALI_reg_2dmap}).

\begin{figure}
\centerline{
\begin{tabular}{cc}
\hspace{0.2cm}
\includegraphics[scale=0.245]{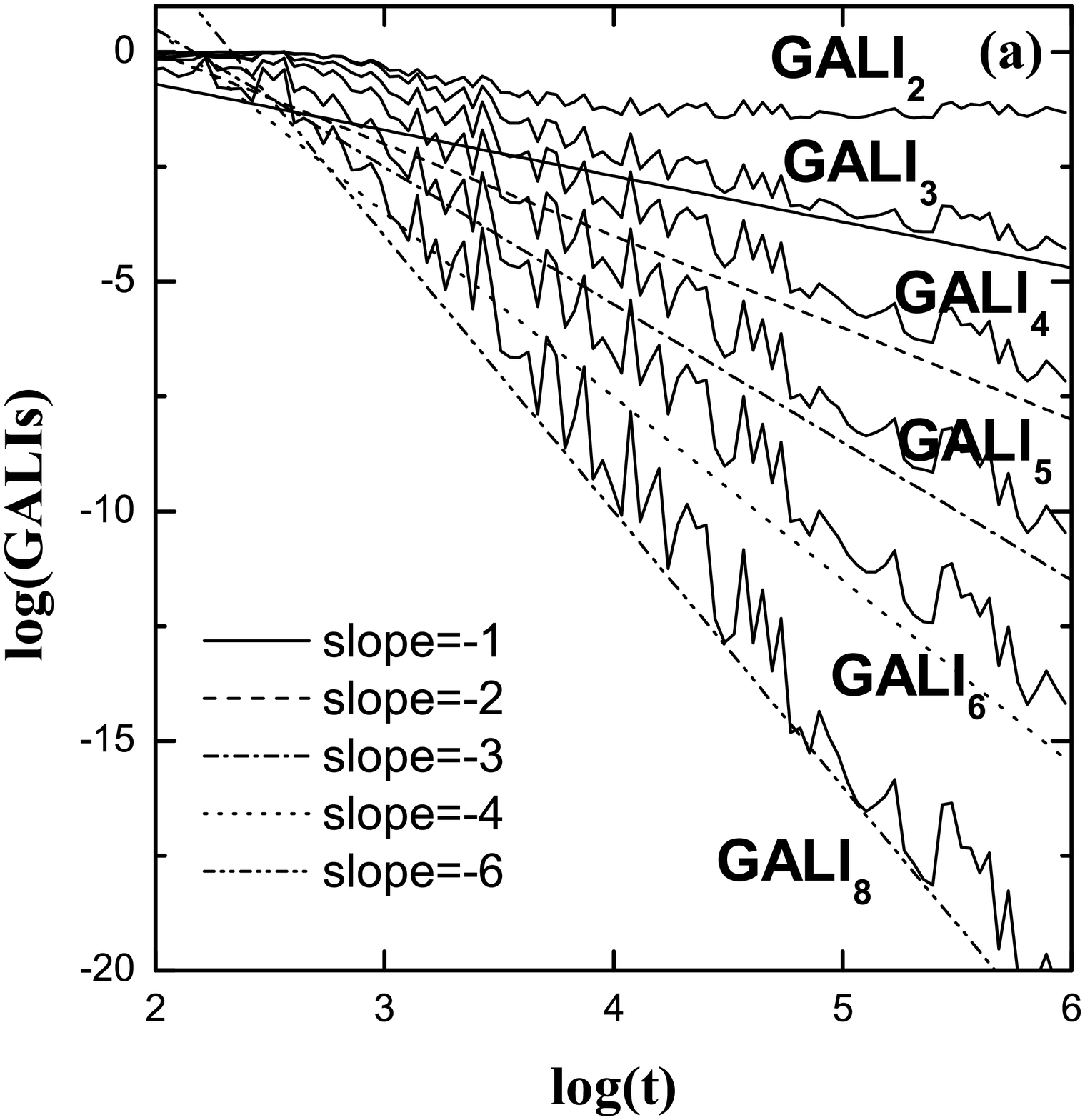} &
\hspace{-1.2cm}
\includegraphics[scale=0.245]{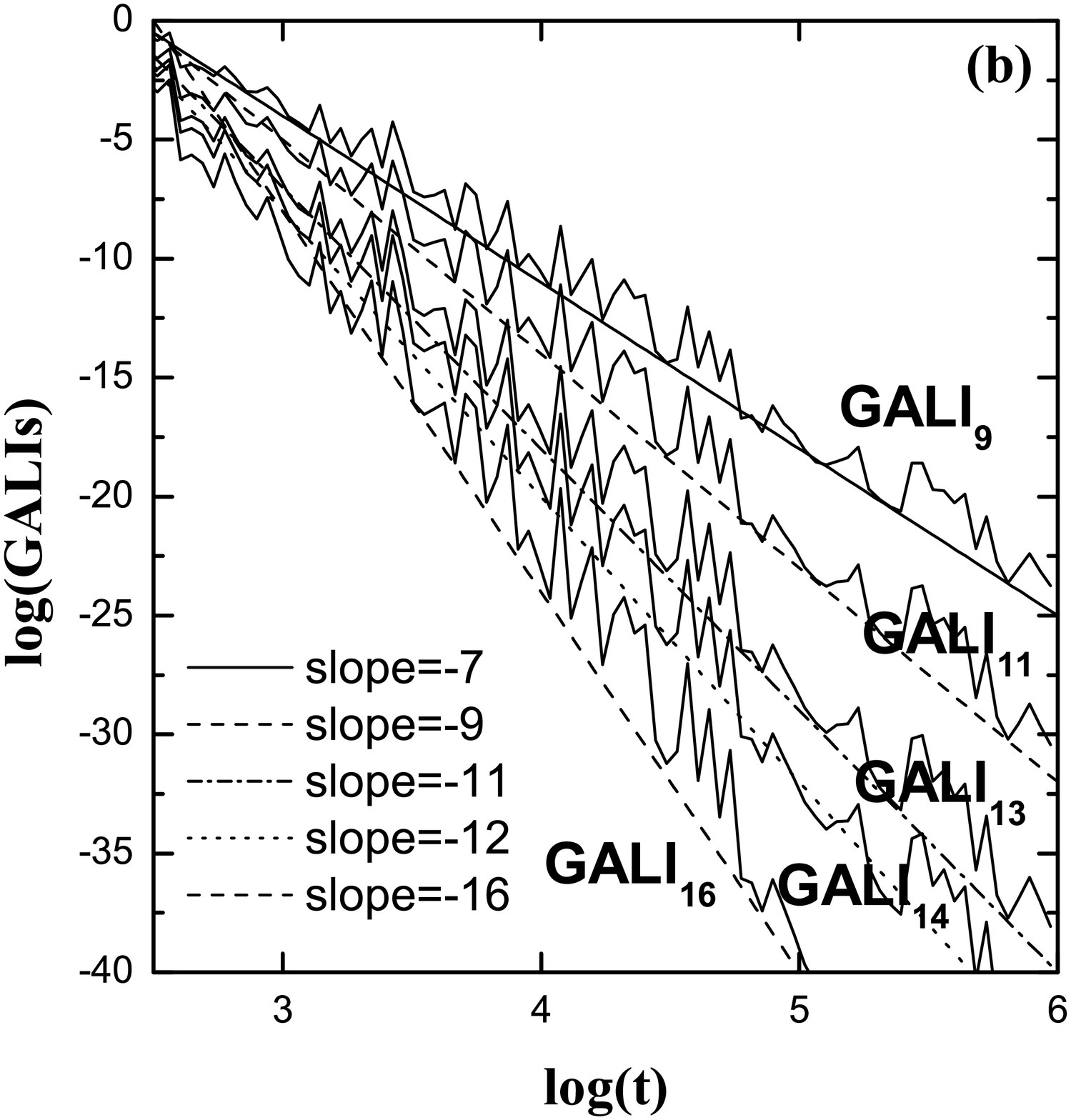}
\vspace{-1.0cm} \\
\hspace{0.2cm}
\includegraphics[scale=0.245]{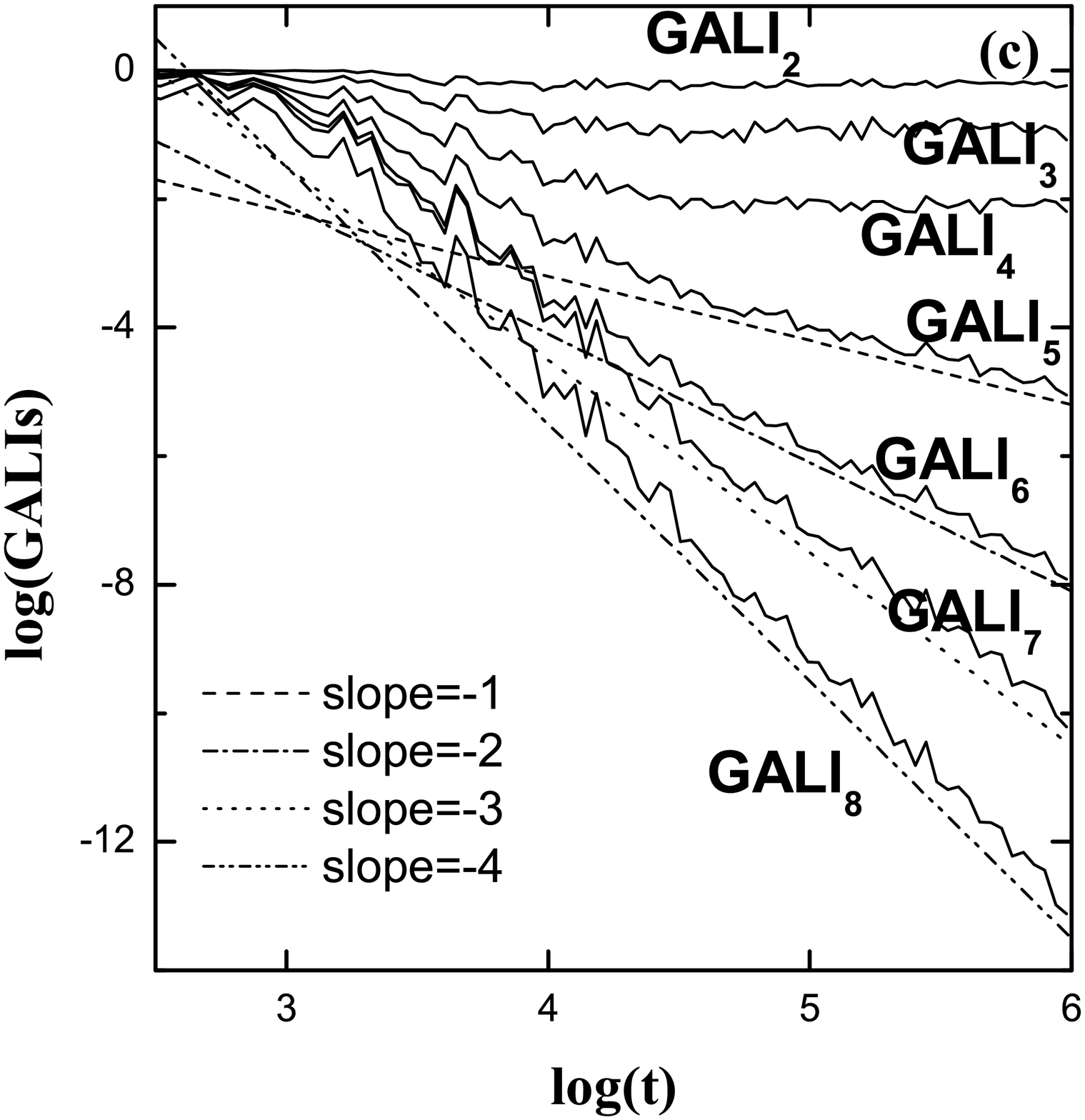} &
\hspace{-1.2cm}
\includegraphics[scale=0.245]{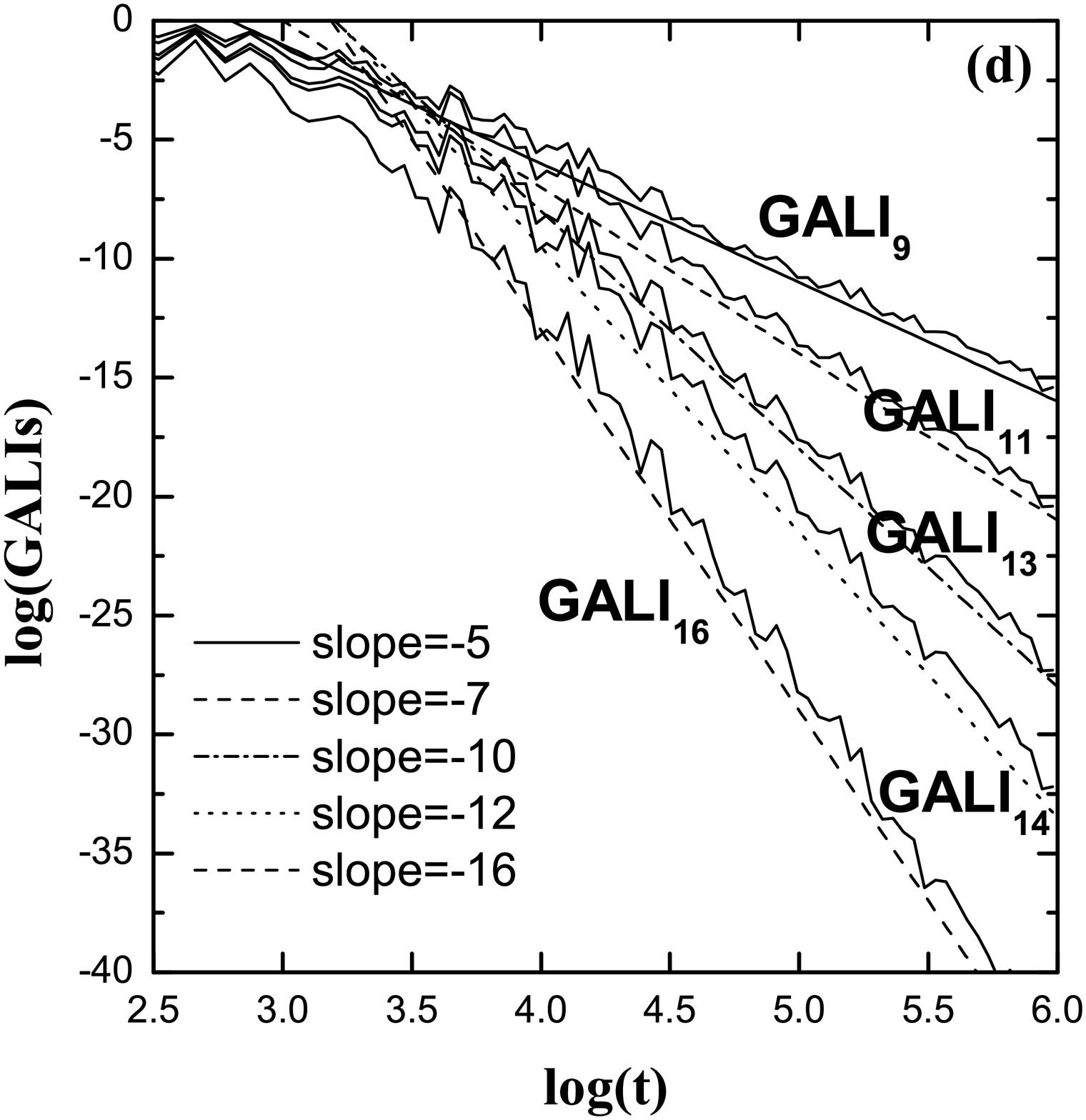}
\end{tabular}}
\caption{The time evolution of the GALI$_k$,
  $k=2,\ldots,9,11,13,14,16$ for a regular orbit lying on a 2d torus
  (panels (a) and (b)) and for another one lying on a 4d torus (panels
  (c) and (d)) of the 8D Hamiltonian $H_8$ considered in
  Fig.~\ref{fig:GALI_behavior_8D}.  The initial conditions of the
  first orbit are $Q_1=2$, $P_1=0$, $Q_i=P_i=0$, $i=2,\ldots,8$ (the
  definition of these variables is given in the caption of
  Fig.~\ref{fig:GALI_behavior_8D}). The initial conditions of the
  second orbit are $q_i=0.1$, $p_i=0$, $i=1,\ldots,8$.  The plotted
  straight lines correspond to the power law predictions
  (\ref{eq:GALI_reg_low_tor}) for $N=8$, $s=2$ (panels (a) and (b))
  and for $N=8$, $s=4$ (panels (c) and (d)). The slope of each line is
  mentioned in the legend (after \cite{SBA_08}) }
\label{fig:GALI_behavior_FPU_low}
\end{figure}

The validity of (\ref{eq:GALI_reg_low_tor}) is supported by the
results of Fig.~\ref{fig:GALI_behavior_FPU_low} where two
representative regular orbits of the $H_8$ Hamiltonian, obtained by
setting $N=8$ in (\ref{eq:NDHam}), are considered (we note that
Fig.~\ref{fig:GALI_behavior_8D} refers to the same model). The first
orbit (Figs.~\ref{fig:GALI_behavior_FPU_low}(a) and (b)) lies on a 2d
torus as the constancy of only GALI$_2$ indicates. The decay of the
remaining GALIs is well reproduced by the power laws
(\ref{eq:GALI_reg_low_tor}) for $N=8$ and $s=2$. The second orbit
(Figs.~\ref{fig:GALI_behavior_FPU_low}(c) and (d)) lies on a 4d torus
and consequently the GALI$_2$, the GALI$_3$ and the GALI$_4$ remain
constant, while all other indices follow power law decays according to
(\ref{eq:GALI_reg_low_tor}) for $N=8$ and $s=4$.

\begin{figure}
\centerline{
\hspace{1.5cm}
\includegraphics[scale=0.31]{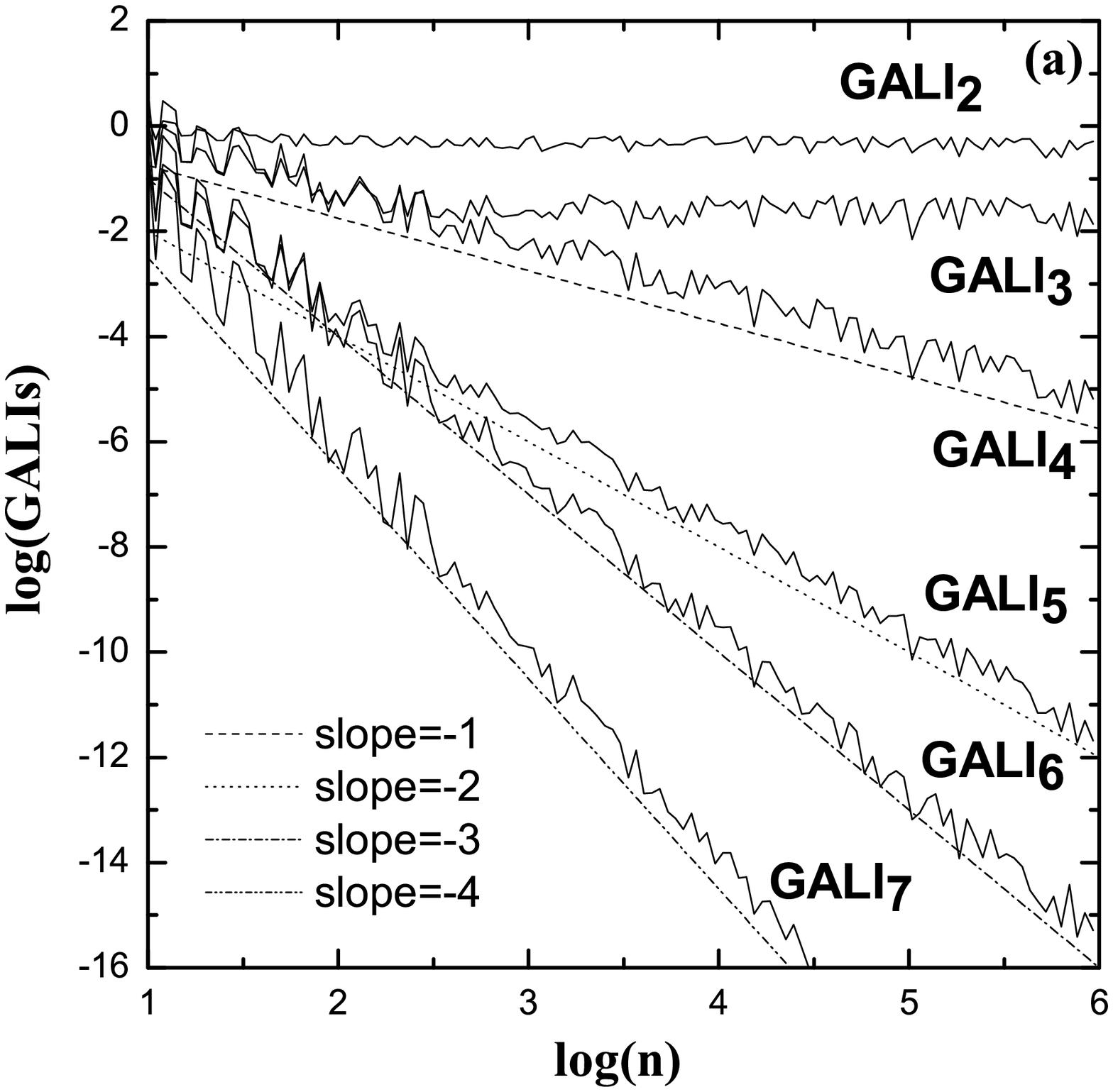}
\hspace{-2.9cm}
\includegraphics[scale=0.31]{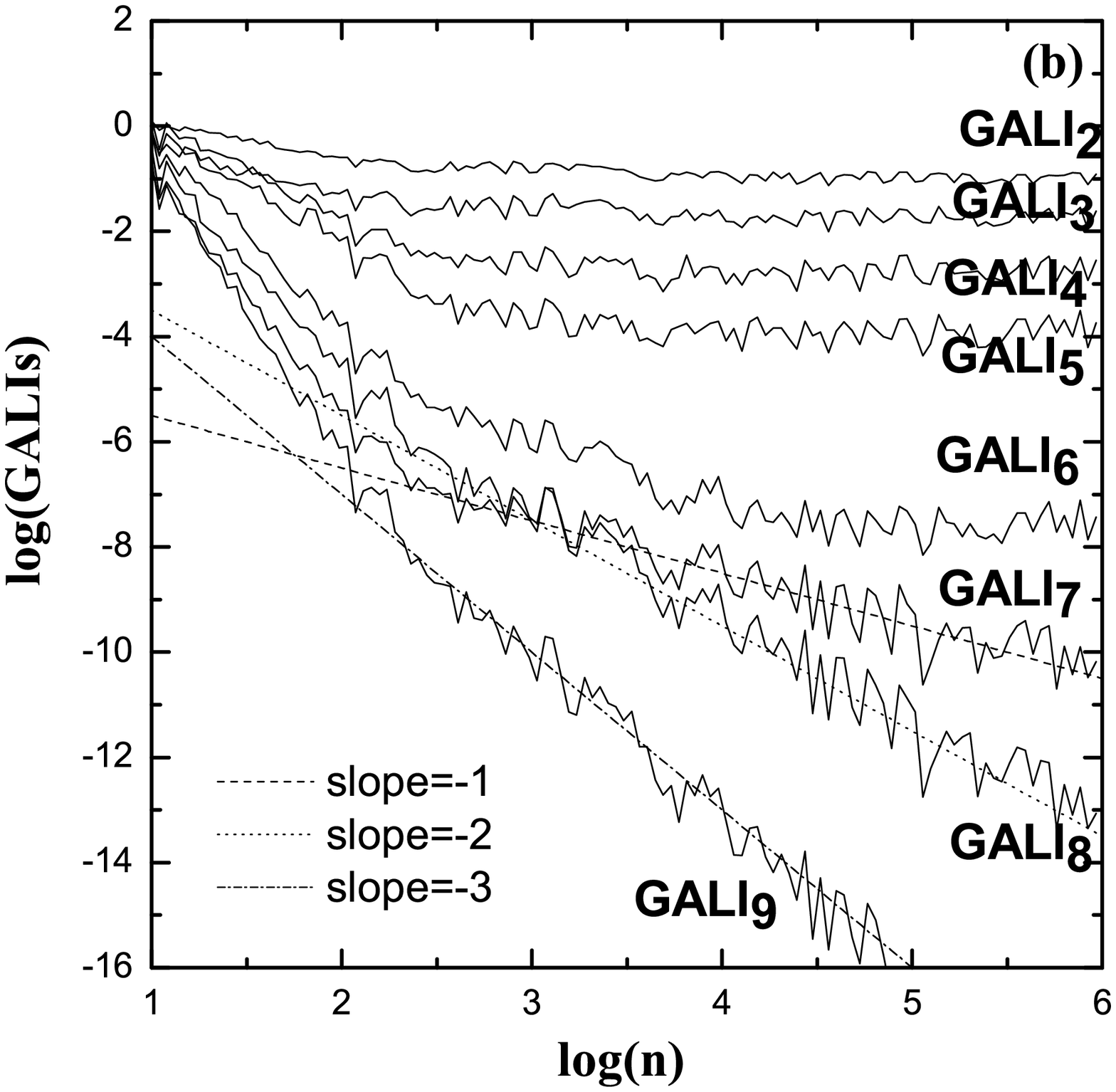}
}
\caption{The evolution of several GALIs for a regular orbit lying (a)
  on a 3d torus and (b) on a 6d torus of the 40d map obtained by
  setting $M=20$ in (\ref{eq:Md_map}). In (a) the initial conditions
  of the orbit are $x_{11}=0.65$, $x_{12}=0.55$, $x_i=0.5$ $\forall i
  \neq 11,12$, and $y_i=0$, $i=1,\ldots,20$, while the parameters of
  the map are set to $\gamma=0.001$ and $K_i=K=2$, $i=1,\ldots,
  20$. In (b) $\gamma=0.00001$ and $K_i$ are set in triplets of -1.35,
  -1.45, -1.55 (i.e.~$K_1=-1.35$, $K_2=-1.45$, $K_3=-1.55$,
  $K_4=-1.35$, $\ldots$, $K_{20}=-1.45$), while the orbit's exact
  initial conditions can be found in \cite{BMC2009}. The plotted
  straight lines correspond to the power law predictions
  (\ref{eq:GALI_reg_low_tor}) for (a) $N=20$, $s=3$ and (b) $N=20$,
  $s=6$. The slope of each line is mentioned in the legend (after
  \cite{BMC2009})}
\label{fig:GALI_behavior_maps_low}
\end{figure}

In Fig.~\ref{fig:GALI_behavior_maps_low} we see the evolution of some
GALIs for regular motion on low dimensional tori of the 40d map
obtained by (\ref{eq:Md_map}) for $M=20$. The results of
Fig.~\ref{fig:GALI_behavior_maps_low}(a) denote that the orbit lies on
a 3d torus in the 40d phase space of the map, while in the case of
Fig.~\ref{fig:GALI_behavior_maps_low}(b) the motion takes place on a
6d torus. The plotted straight lines help us verify that for both
orbits the behaviors of the decaying GALIs are accurately reproduced
by (\ref{eq:GALI_reg_low_tor}) for $N=20$, $s=3$
(Fig.~\ref{fig:GALI_behavior_maps_low}(a)) and $N=20$, $s=6$
(Fig.~\ref{fig:GALI_behavior_maps_low}(b)).

\subsubsection{Searching for Regular Motion on Low Dimensional Tori}
\label{sect:low_search}

Equation (\ref{eq:GALI_reg_low_tor}), as well as the results of
Figs.~\ref{fig:GALI_behavior_FPU_low} and
\ref{fig:GALI_behavior_maps_low} imply that the GALIs can be also used
for identifying regular motion on low dimensional tori. From
(\ref{eq:GALI_reg_low_tor}) we deduce that the dimension of the torus
on which the regular motion occurs coincides with the largest order
$k$ of the GALIs for which the GALI$_k$ remains constant. Based on
this remark we can develop a strategy for locating low dimensional
tori in the phase space of a dynamical system. The GALI$_k$ of initial
conditions resulting in motion on an $s$d torus eventually will remain
constant for $2\leq k \leq s$, while it will decay to zero following
the power law (\ref{eq:GALI_reg_low_tor}) for $k>s$. So, after some
relatively long time interval, all the GALIs of order $k>s$ will have
much smaller values than the ones of order $k \leq s$.  Thus, in order
to identify the location of $s$d tori, $2 \leq s \leq N$, in the $2N$d
phase space of a dynamical system we evaluate at first various GALIs
for several initial conditions and then find the initial conditions
which result in large GALI$_k$ values for $k \leq s$ and small values
for $k>s$.

As was mentioned in Sect.~\ref{sect:GALI_paradigms}, the constant, final
values of the GALIs for regular motion decrease with the order of the
GALI (see Figs.~\ref{fig:GALI_behavior_3D}(b),
\ref{fig:GALI_behavior_8D}(c), \ref{fig:GALI_behavior_6d}(b),
\ref{fig:GALI_behavior_FPU_low}(c) and
\ref{fig:GALI_behavior_maps_low}). Since this decrease has not been
quantified yet, a good computational approach in the quest for low
dimensional tori is to `normalize' the values of the GALIs for each
individual orbit by dividing them by the largest GALI$_k$ value, $\max
\left(\mbox{GALI}_k \right)$, obtained by all orbits in the studied
ensemble at the end time $t=t_e$ of the integration.  In this way we
define the `normalized GALI$_k$'
\begin{equation}\label{eq:g_k}
    g_k(t)=\frac{\mbox{GALI}_k (t)}{\max \left[\mbox{GALI}_k (t_e)\right]}.
\end{equation}
Then, by coloring each initial condition according to
its $g_k(t_e)$ value we can construct phase space charts where the
position of low dimensional tori is easily located.

\begin{figure}
\centerline{\hspace{-0.1cm}\includegraphics[width=\textwidth]{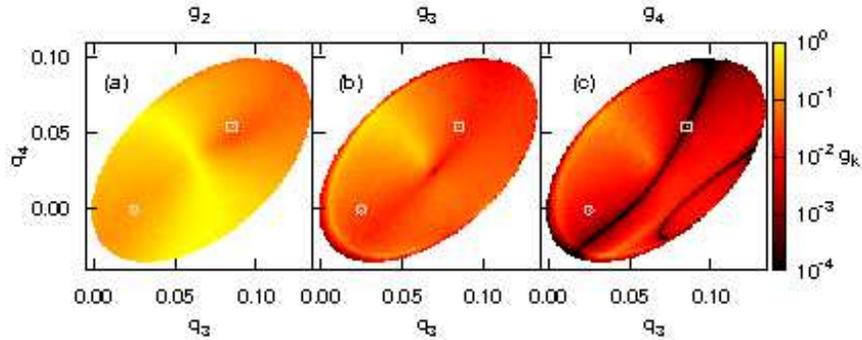}}
\caption{Regions of different $g_k$ (\ref{eq:g_k}) values for (a)
  $k=2$, (b) $k=3$, (c) $k=4$, in the subspace $(q_3, q_4)$ of the 4D
  Hamiltonian $H_4$ obtained from (\ref{eq:NDHam}) for $N=4$ and
  $\beta=1.5$. The remaining coordinates of the considered initial
  conditions are set to $q_1=q_2=0.1$, $p_1=p_2=p_3=0$, while $p_4>0$
  is evaluated so that $H_4=0.010075$. White regions correspond to
  forbidden initial conditions. The color scales shown at the right of
  the panels are used to color each point according to the orbit's
  $g_k$ value at $t=10^6$. The points with coordinates $q_3=0.106$,
  $q_4=0.0996$ (marked by a triangle), $q_3=0.085109$, $q_4=0.054$
  (marked by a square) and $q_3=0.025$, $q_4=0$ (marked by a circle)
  correspond to regular orbits on a 2d, a 3d and a 4d torus
  respectively (after \cite{GES_12}) }
\label{fig:GALI_search_low_1}
\end{figure}

To illustrate this method we present (following \cite{GES_12}) the
search for low dimensional tori in a subspace of the 8d phase space of
the 4D Hamiltonian system $H_4$ obtained by setting $N=4$ and
$\beta=1.5$ in (\ref{eq:NDHam}). In order to facilitate the
visualization of the whole procedure we restrict our search in the
subspace $(q_3, q_4)$ by setting the other initial conditions of the
studied orbits to $q_1=q_2=0.1$, $p_1=p_2=p_3=0$, while $p_4>0$ is
evaluated so that $H_4=0.010075$. In Fig.~\ref{fig:GALI_search_low_1}
we color each permitted initial condition in the $(q_3, q_4)$ plane
according to its $g_2$, $g_3$ and $g_4$ value at $t=t_e=10^6$ time
units (panels (a), (b) and (c) respectively).

\begin{figure}
\centerline{\hspace{-0.1cm}
\includegraphics[scale=1.29]{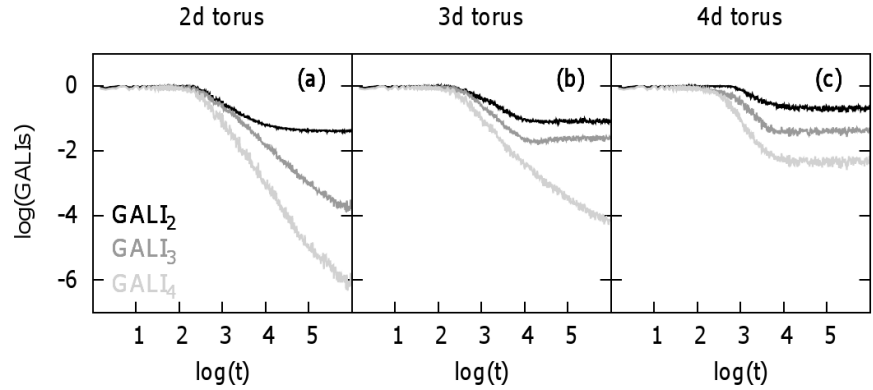}}
\caption{The time evolution of the GALI$_2$, the GALI$_3$ and the
  GALI$_4$ of regular orbits lying on a (a) 2d, (b) 3d, (c) 4d torus
  of the 4D Hamiltonian considered in
  Fig.~\ref{fig:GALI_search_low_1}. The initial conditions of these
  orbits are respectively marked by a triangle, a square and a circle
  in Fig.~\ref{fig:GALI_search_low_1} (after \cite{GES_12}) }
\label{fig:GALI_search_low_2}
\end{figure}

For this particular Hamiltonian we can have regular motion on 2d, 3d and 4d
tori. Let us see now how we can exploit the results of
Fig.~\ref{fig:GALI_search_low_1} to locate such tori.  Motion on 2d tori
results in large final $g_2$ values and to small $g_3$ and $g_4$. So,
such tori should be located in regions colored in yellow or light red in
Fig.~\ref{fig:GALI_search_low_1}(a) and in black in
Figs.~\ref{fig:GALI_search_low_1}(b) and (c).  A region which satisfies these
requirements is located at the upper border of the colored areas in
Fig.~\ref{fig:GALI_search_low_1}.  The evolution of the GALIs of an orbit with
initial conditions in that region (denoted by a triangle in
Fig.~\ref{fig:GALI_search_low_1}) is shown in
Fig.~\ref{fig:GALI_search_low_2}(a) and it verifies that the motion takes place
on a 2d torus, as only the GALI$_2$ remains constant.

Extending the same argumentation to higher dimensions we see that
motion on a 3d torus can occur in regions colored in yellow or light
red in both Figs.~\ref{fig:GALI_search_low_1}(a) and (b) and in black
in Fig.~\ref{fig:GALI_search_low_1}(c). The initial condition of an
orbit of this kind is marked by a small square in
Fig.~\ref{fig:GALI_search_low_1}. The evolution of this orbit's GALIs
(Fig.~\ref{fig:GALI_search_low_2}(b)) verifies that the orbit lies on
a 3d torus, because only the GALI$_2$ and the GALI$_3$ remain
constant.  Orbits on 4d tori is the most common situation of regular
motion for this 4D Hamiltonian system. This is evident from the
results of Fig.~\ref{fig:GALI_search_low_1} because most of the
permitted area of initial conditions correspond to high $g_2$, $g_3$
and $g_4$ values. A randomly chosen initial condition in this region
(marked by a circle in Fig.~\ref{fig:GALI_search_low_1}) results
indeed to regular motion on a 4d torus as the constancy of its
GALI$_k$, $k=2,3,4$ in Fig.~\ref{fig:GALI_search_low_2}(c) clearly
indicates.

We note that initial conditions leading to chaotic motion in this
system would correspond to very small $g_2$, $g_3$ and $g_4$ values
(due to the exponential decay of the associated GALIs) and
consequently would be colored in black in \emph{all} panels of
Fig.~\ref{fig:GALI_search_low_1}. The lack of such regions in
Fig.~\ref{fig:GALI_search_low_1} signifies that all considered initial
conditions lead to regular motion. This happens because regions
of chaotic motion occupy a tiny fraction of the system's phase space,
because its nonlinearity strength is very small. Therefore, chaotic
motion is not captured by the grid of initial conditions of
Fig.~\ref{fig:GALI_search_low_1}.

\subsection{Behavior of the GALI for Periodic Orbits}
\label{sect:GALI_po}

Let us now discuss the behavior of the GALIs for periodic orbits of period $T$; i.e.~orbits satisfying the condition $\vec{x}(t+T)= \vec{x}(t)$, with
$\vec{x}(t)$ being the coordinate vector in the system's phase space. In the
presentation of this topic we mainly follow the analysis performed in
\cite{MSA_12}. The linear stability of periodic orbits is defined by the eigenvalues of the so-called monodromy matrix, which is obtained by the solution of the variational equations (for Hamiltonian systems) or by the evolution of the tangent map (for symplectic maps) for one period $T$ (see for example \cite{B_69,S_01b} and Sect.~3.3 of \cite{LL_92}). When all eigenvalues lie on the unit circle in the complex plane the orbit is characterized as
elliptic, while otherwise it is called hyperbolic (unstable). For a detailed presentation of the various stability types of periodic orbits the reader is referred for example to \cite{B_69,H_75,HM_87,HD_98,S_01b}.

The presence of periodic orbits influence significantly the dynamics. In most systems we observe that the majority of non-periodic orbits in the vicinity of an elliptic one are regular. So, although initial conditions near an elliptic orbit can lead to chaos, regular orbits exhibiting a time evolution similar to the elliptic orbit itself prevail. If one assumes that the elliptic orbit is integrable and in its vicinity the
Kolmogorov-Arnold-Moser (KAM) theorem (see for example Sect.~3.2 of \cite{LL_92} and references therein) can be applied (for which one needs to check a non-degeneracy condition which is typically satisfied), then there is large measure of orbits on KAM tori nearby. In Hamiltonian systems of dimension larger than 2 the phenomenon of Arnold diffusion (see for example Chap.~6 of \cite{LL_92} and references therein) typically would lead to an escape of orbits from the neighborhood of the elliptic orbit. However, it is generally believed that Arnold diffusion occurs on a slow time scale, and we do not expect interference with the GALI method. Of course, regular behavior on nearby KAM tori does not imply that the elliptic orbit itself is stable (e.g.~Appendix of \cite{DMS00}). On the other hand, in chaotic Hamiltonian systems and symplectic maps orbits in the vicinity of an unstable periodic orbit typically behave chaotically and diverge from the periodic one exponentially fast.
This divergence is characterized by LEs (with at least one of them being positive) which are determined by the eigenvalues of the monodromy matrix (e.g.~\cite{BFS_79,SBA_07} and Sect.~5.2b of \cite{LL_92}). Thus, following arguments similar to
the ones developed in Sect.~\ref{sect:GALI_ch_reg} for chaotic orbits,
we easily see that the GALI$_k$ of unstable periodic orbits decreases
to zero following the exponential law (\ref{eq:GALI_ch}), i.~e.
\begin{equation}
\label{eq:GALI_upo}
\mbox{GALI}_k(t) \propto \exp \left\{-\left[ (\lambda_1-\lambda_2) +
  (\lambda_1-\lambda_3)+ \cdots+ (\lambda_1-\lambda_k)\right]t
\right\},
\end{equation}
where $\lambda_i$, $i=1,\ldots,k$ are the periodic orbit's $k$ largest
LEs.

\begin{figure}
\centerline{
\hspace{2.cm}
\includegraphics[scale=0.305]{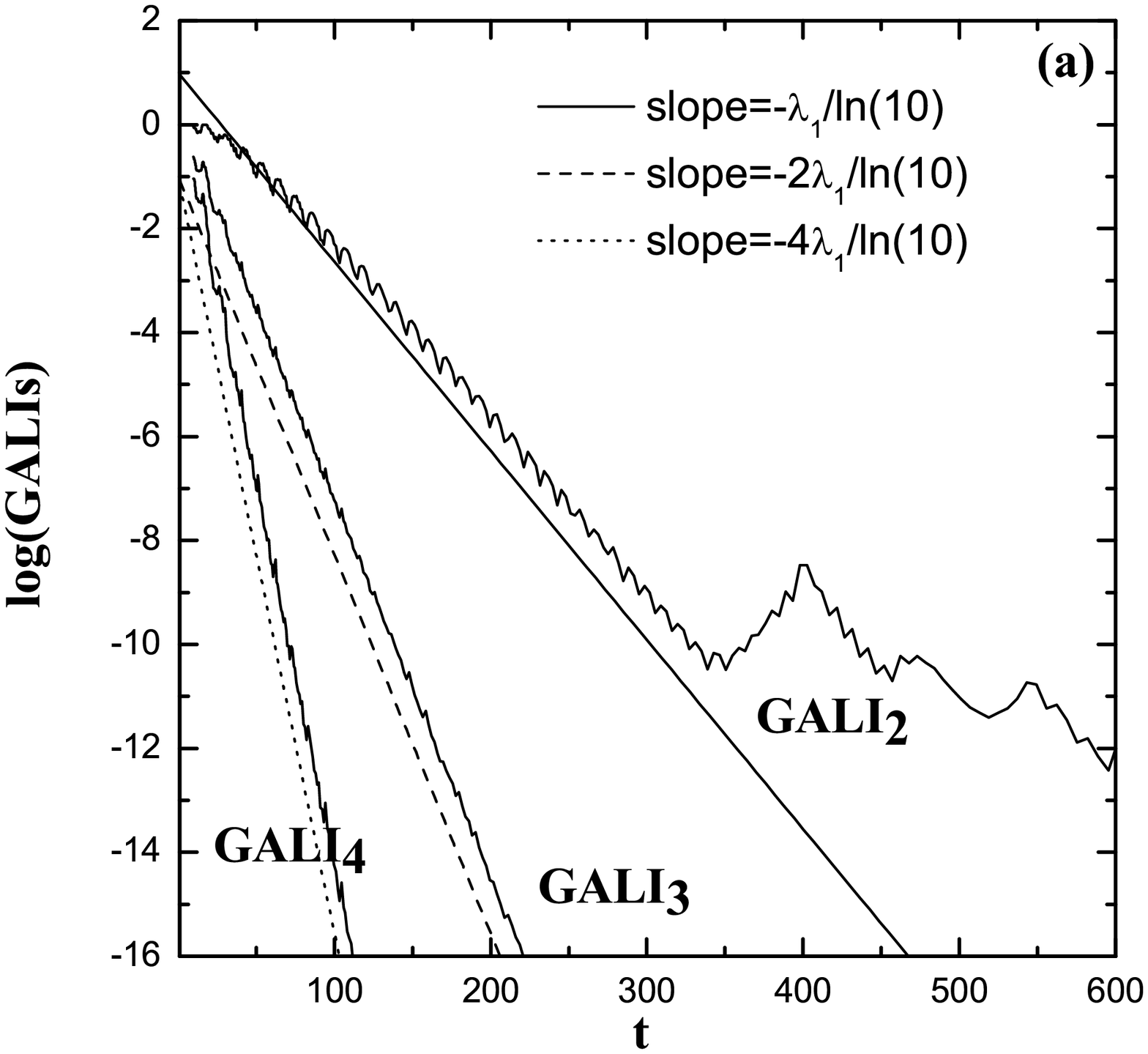}
\hspace{-2.9cm}
\includegraphics[scale=0.305]{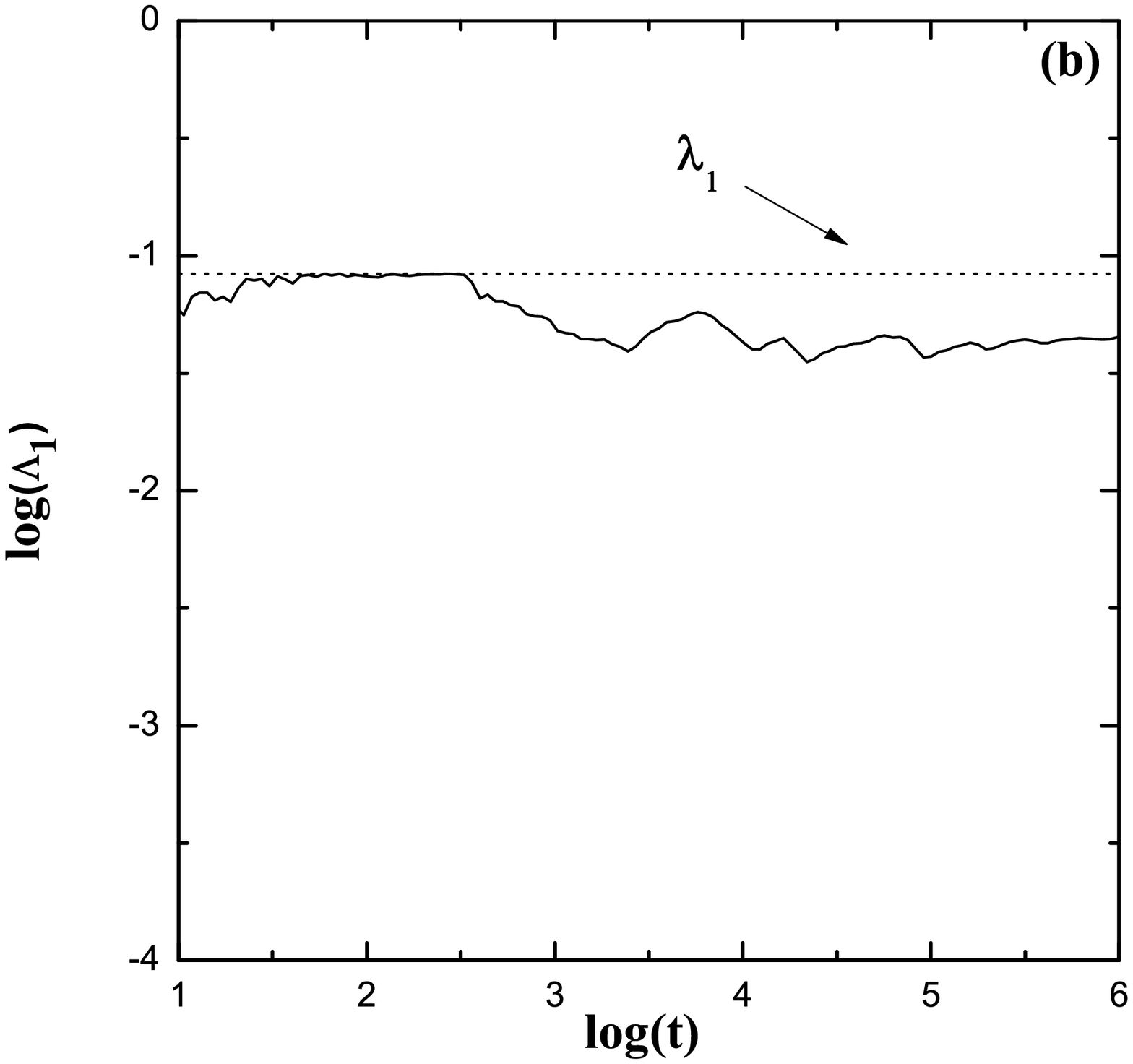}
}
\caption{The time evolution of (a) the GALI$_2$, the GALI$_3$, the
  GALI$_4$ and (b) the finite time mLE $\Lambda_1$ of an unstable
  periodic orbit of the 2D Hamiltonian (\ref{eq:2DHam}) for
  $H_2=0.125$. The initial conditions of the orbit are $q_1=0$,
  $q_2=0.2083772012$, $p_1=0.4453146996$, $p_2=0.1196065752$.  The
  straight lines in (a) correspond to functions proportional to
  $\exp(-\lambda_1t)$, $\exp(-2\lambda_1t)$ and $\exp(-4\lambda_1t)$,
  for $\lambda_1=0.084$, which is the mLE of the periodic orbit. The
  slope of each line is mentioned in the legend. The horizontal dotted
  line in (b) indicates the value $\lambda_1=0.084$ (after
  \cite{MSA_12}) }
\label{fig:GALI_unstable_po}
\end{figure}

In Fig.~\ref{fig:GALI_unstable_po}(a) we see that the evolution of the
GALIs for an unstable periodic orbit of the 2D Hamiltonian
(\ref{eq:2DHam}) is well approximated by (\ref{eq:GALI_upo}) for
$\lambda_1=0.084$. This value is the orbit's mLE determined by the
eigenvalues of the corresponding monodromy matrix (see \cite{MSA_12}
for more details). We also note that according to (\ref{eq:LEs_pairs})
and (\ref{eq:LEs_zero}) we set $\lambda_1=-\lambda_4$, and
$\lambda_2=\lambda_3=0$ in (\ref{eq:GALI_upo}). The agreement between
the numerical data and the theoretical prediction (\ref{eq:GALI_upo})
is lost after about $t\approx 350$ time units. This happens because
the numerically computed orbit eventually deviates from the unstable
periodic one due to unavoidable computational inaccuracies and enters
the chaotic region around the periodic orbit. In general, this region
is characterized by different LEs with respect to the ones of the
periodic orbit. The effect of this behavior on the orbit's finite time
mLE $\Lambda_1$ (\ref{eq:ftLE}) is seen in
Fig.~\ref{fig:GALI_unstable_po}(b). The computed $\Lambda_1$ deviates
from the value $\lambda_1=0.084$ (marked by a horizontal dotted line)
at about the same time the GALI$_2$ changes its decreasing rate in
Fig.~\ref{fig:GALI_unstable_po}(a).  Eventually, $\Lambda_1$
stabilizes at another positive value, which characterizes the
chaoticity of the region around the periodic orbit.

On the other hand, the case of stable periodic orbits is a bit more
complicated, because the GALIs behave differently for Hamiltonian
flows and symplectic maps. In \cite{MSA_12} it was shown analytically
that for stable periodic orbits of $N$D Hamiltonian systems, with
$N\geq 2$, the GALIs decay to zero following the following power laws
\begin{equation}
\mbox{GALI}_k (t) \propto
\left\{ \begin{array}{ll}
\frac{1}{t^{k-1}} & \mbox{if $2\leq  k \leq 2N-1$} \\
\frac{1}{t^{2N}} & \mbox{if $k = 2N .$} \\
\end{array}\right.
\label{eq:GALI_stable_po_Ham}
\end{equation}
We observe that this equation can be derived from
(\ref{eq:GALI_reg_low_tor}), which describes the behavior of the GALIs
for motion on an $s$d tori, by setting $s=1$. We note that the first
branch of (\ref{eq:GALI_reg_low_tor}) is meaningless for $s=1$, while
the other two branches take the forms appearing in
(\ref{eq:GALI_stable_po_Ham}). The connection between
(\ref{eq:GALI_reg_low_tor}) and (\ref{eq:GALI_stable_po_Ham}) is not
surprising if we notice that a periodic orbit is nothing more than an
1d closed curve in the system's phase space, having the some dimension
with an 1d torus.

Small, random perturbations from the stable periodic orbit generally
results in regular motion on an $N$d torus. So, the GALIs of the
perturbed orbit will follow (\ref{eq:GALI_reg}). Thus, in general, the GALIs of regular orbits in the vicinity of a stable periodic orbit behave differently
with respect to the indices of the periodic orbit itself (except from the
GALI$_{2N}$ and the GALI$_{2N-1}$, which respectively follow the laws $\propto
t^{-2N}$ and $\propto t^{-(2N-2)}$ in both cases). The most profound change
happens for the GALIs of order $2 \leq k \leq N$ because, according to
(\ref{eq:GALI_reg}), they remain constant in the neighborhood of the periodic
orbit, while they decay to zero following the power law
(\ref{eq:GALI_stable_po_Ham}) for the periodic orbit.

\begin{figure}
\centerline{
\hspace{2.1cm}
\includegraphics[scale=0.3]{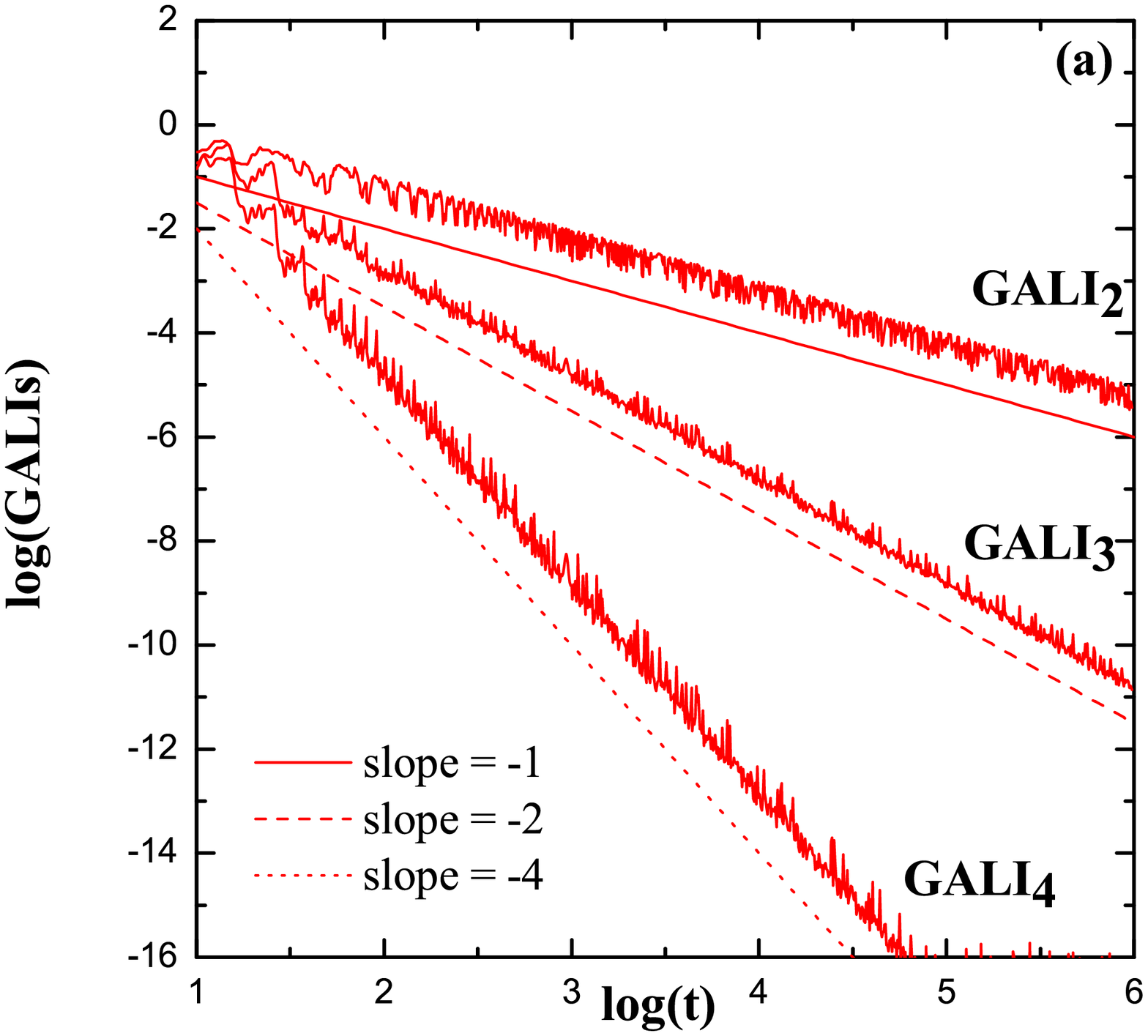}
\hspace{-2.7cm}
\includegraphics[scale=0.3]{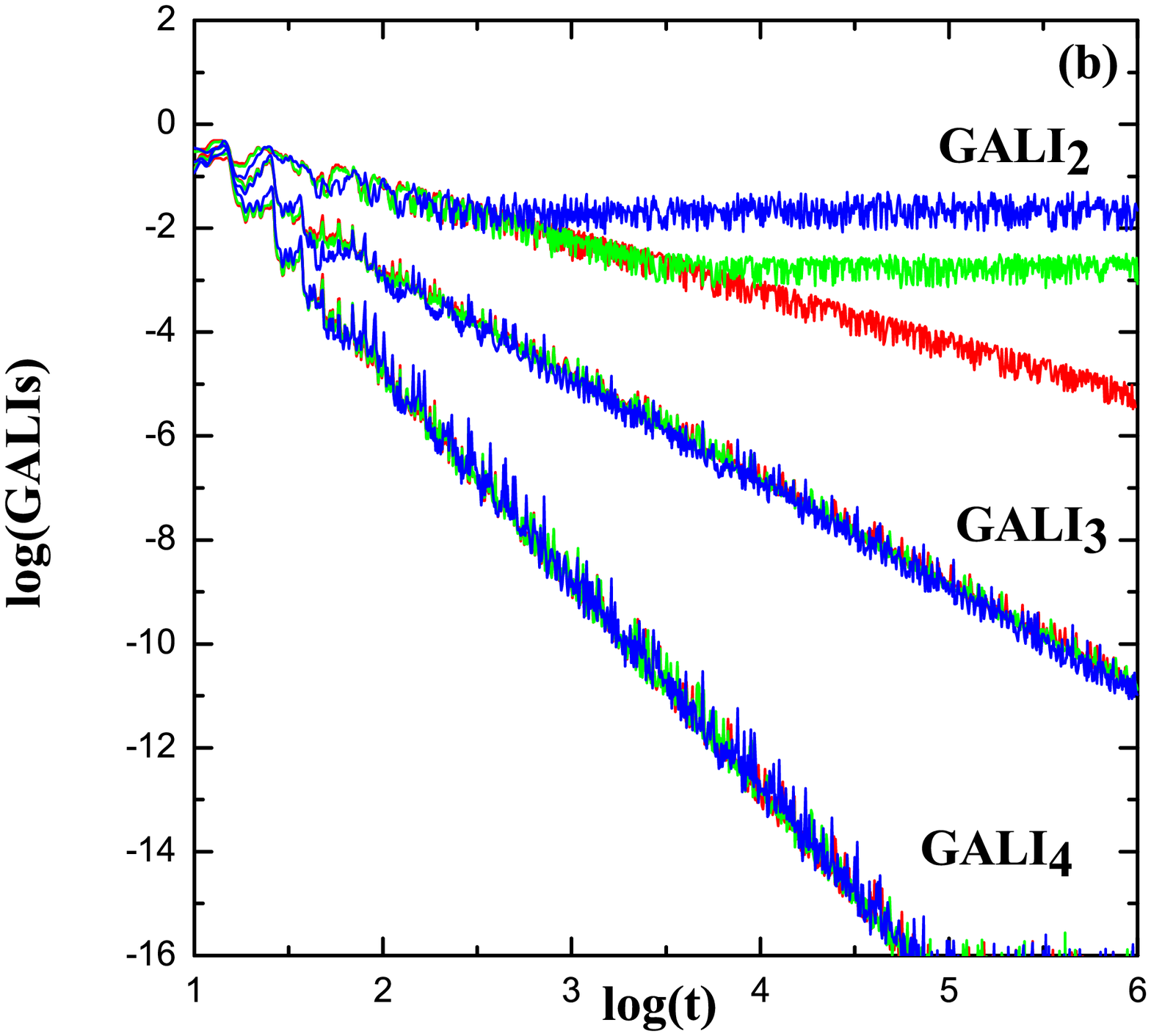}
}
\caption{(a) The time evolution of the GALI$_2$, the GALI$_3$ and the
  GALI$_4$ for a stable periodic orbit of the 2D Hamiltonian
  (\ref{eq:2DHam}) for $H_2=0.125$. The orbit's initial conditions are
  $q_1=0=q_{10}$, $q_2=0.35207=q_{20}$, $p_1=0.36427=p_{10}$,
  $p_2=0.14979=p_{20}$. The straight lines correspond to functions
  proportional to $t^{-1}$, $t^{-2}$ and $t^{-4}$. The slope of each
  line is mentioned in the legend. (b) The same plot as in (a) where
  apart from the GALIs of the stable periodic orbit (red curves) the
  indices of two neighboring, regular orbits are also plotted. Their
  initial conditions are $q_1=q_{10}$, $p_2=p_{20}$ for both of them,
  while $q_2=q_{20}+ 0.00793$ (green curves), and $q_2=q_{20}+
  0.02793$ (blue curves). In both cases the $p_1>0$ initial condition
  is set so that $H_2=0.125$. Note that the curves of the GALI$_3$ and
  the GALI$_4$ for all three orbits overlap each other (after
  \cite{MSA_12}) }
\label{fig:Ham_spo_neighbor}
\end{figure}

The correctness of (\ref{eq:GALI_stable_po_Ham}) becomes evident from
the results of Fig.~\ref{fig:Ham_spo_neighbor}(a), where the time
evolution of the GALIs of a stable periodic orbit of the 2D
Hamiltonian (\ref{eq:2DHam}) is shown. In particular, we see that the
indices decay to zero following the power laws GALI$_2 \propto
t^{-1}$, GALI$_3 \propto t^{-2}$, GALI$_4 \propto t^{-4}$ predicted
from (\ref{eq:GALI_stable_po_Ham}).  According to (\ref{eq:GALI_reg})
the GALIs of regular orbits in the neighborhood of the stable periodic
orbit should behave as GALI$_2 \propto \mbox{constant}$, GALI$_3
\propto t^{-2}$ and GALI$_4 \propto t^{-4}$. Thus, only the GALI$_2$
is expected to behave differently for regular orbits in the vicinity of the periodic orbit of Fig.~\ref{fig:Ham_spo_neighbor}(a). The results of
Fig.~\ref{fig:Ham_spo_neighbor}(b) show that this is actually true. The GALI$_2$ of the neighboring regular orbits initially follows the
same power law decay of the periodic orbit (GALI$_2 \propto t^{-1}$),
but later on it stabilizes to a constant positive value. We see that
the further the orbit is located from the periodic one the sooner the
GALI$_2$ deviates from the power law decay.

\begin{figure}
\sidecaption[t]
\includegraphics[scale=0.265]{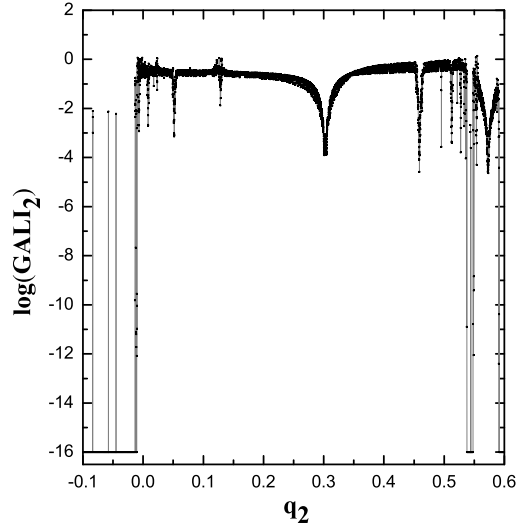}
\caption{The values of the GALI$_2$ at $t=10^5$ for several orbits of
  the 2D Hamiltonian (\ref{eq:2DHam}) as a function of the $q_2$
  coordinate of the orbits' initial conditions. The remaining
  coordinates are $q_1=p_2=0$, while $p_1>0$ is set so that
  $H_2=0.125$. Actually these initial conditions lie on the
 $p_2=0$ line of Figs.~\ref{fig:HH_global_GALI2} and
\ref{fig:HH_global_GALI4}. The numerical data (black points) are line connected   (grey line) in order to facilitate the visualization of the value
  changes (after \cite{MSA_12})}
\label{fig:Ham_spo_neighbor_scan}
\end{figure}

These differences of the GALI$_2$ values can be used to identify the
location of stable periodic orbits in the system's phase space, although the index was not developed for this particular
purpose\footnote{It is worth mentioning here that other chaos indicators,
like the Orthogonal Fast Lyapunov Indicator (OFLI) and its
variations \cite{B05,B06}, are quite successful in performing this
task as they were actually designed for this purpose.} This
becomes evident from the result of
Fig.~\ref{fig:Ham_spo_neighbor_scan} where the values of the GALI$_2$
at $t=10^5$ for several orbits of the H\'{e}non-Heiles system
(\ref{eq:2DHam}) are plotted as a function of the $q_2$ coordinate of
the orbits' initial conditions. The remaining coordinates are
$q_1=p_2=0$, while $p_1>0$ is set so that $H_2=0.125$. Actually
these initial conditions lie on the symmetry line of the
subspace defined by $q_1=0$, $p_1>0$, i.e.~the horizontal
line $p_2=0$ in Figs.~\ref{fig:HH_global_GALI2} and
\ref{fig:HH_global_GALI4} below. This line passes through the
initial condition of some periodic orbits of the system. For the
construction of Fig.~\ref{fig:Ham_spo_neighbor_scan} we considered an ensemble of $7\, 000$ orbits whose $q_2$ coordinates are equally distributed in the interval $-0.1 \leq q_2 \leq 0.6$. The data points are line connected, so that
the changes of the GALI$_2$ values become easily visible.

In Fig.~\ref{fig:Ham_spo_neighbor_scan} regions of relatively large
GALI$_2$ values ($\gtrsim 10^{-4}$) correspond to regular (periodic or
quasiperiodic) motion.  Chaotic orbits and unstable periodic orbits
have very small GALI$_2$ values ($\lesssim 10^{-12}$), while domains
with intermediate values ($10^{-12} \lesssim \mbox{GALI}_2 \lesssim
10^{-4}$) correspond to sticky chaotic orbits. An interesting feature
of Fig.~\ref{fig:Ham_spo_neighbor_scan} is the appearance of some
relatively narrow regions where the GALI$_2$ decreases abruptly
obtaining values $10^{-4} \lesssim \mbox{GALI}_2 \lesssim 10^{-1}$;
the most profound one being in the vicinity of $q_2 \approx
0.3$. These regions correspond to the immediate neighborhoods of
stable periodic orbits, with the periodic orbit itself been located at
the point with the smallest GALI$_2$ value.

The creation of these characteristic `pointy' shapes is due to the behavior
depicted in Fig.~\ref{fig:Ham_spo_neighbor}(b): the GALI$_2$ has relatively
small values on the stable periodic orbit, for which it decreases as $\propto
t^{-1}$, while it attains constant, positive values for regular orbits in the vicinity of the periodic orbit. These constant values increase as the orbit's initial conditions depart further away from the periodic orbit. So, more generally, the appearance of such `pointy' formations in GALI$_k$ plots ($2 \leq k \leq N$) provide good indications for the location of stable periodic
orbits.

\begin{figure}
\centerline{
\hspace{2.1cm}
\includegraphics[scale=0.3]{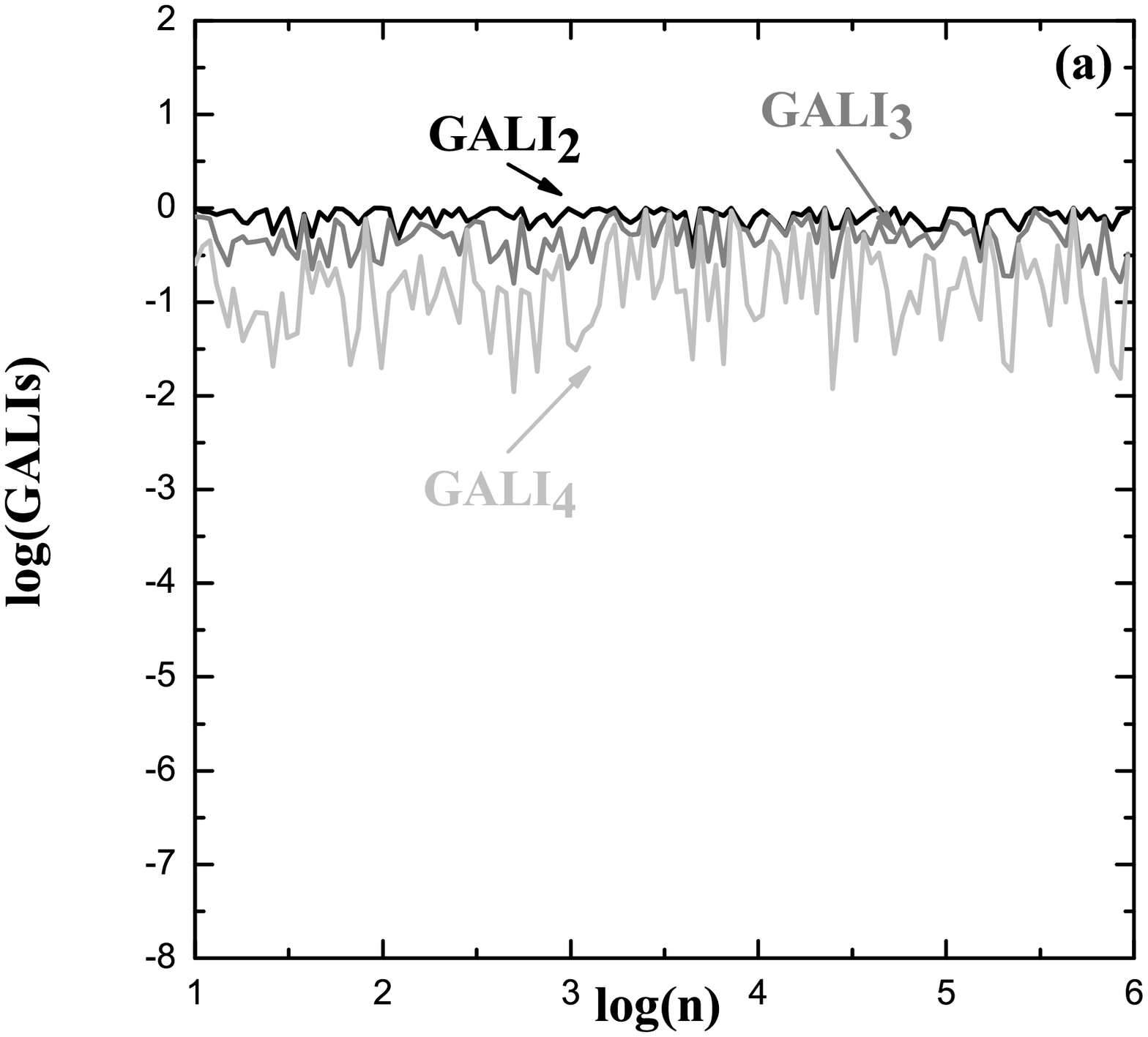}
\hspace{-2.7cm}
\includegraphics[scale=0.3]{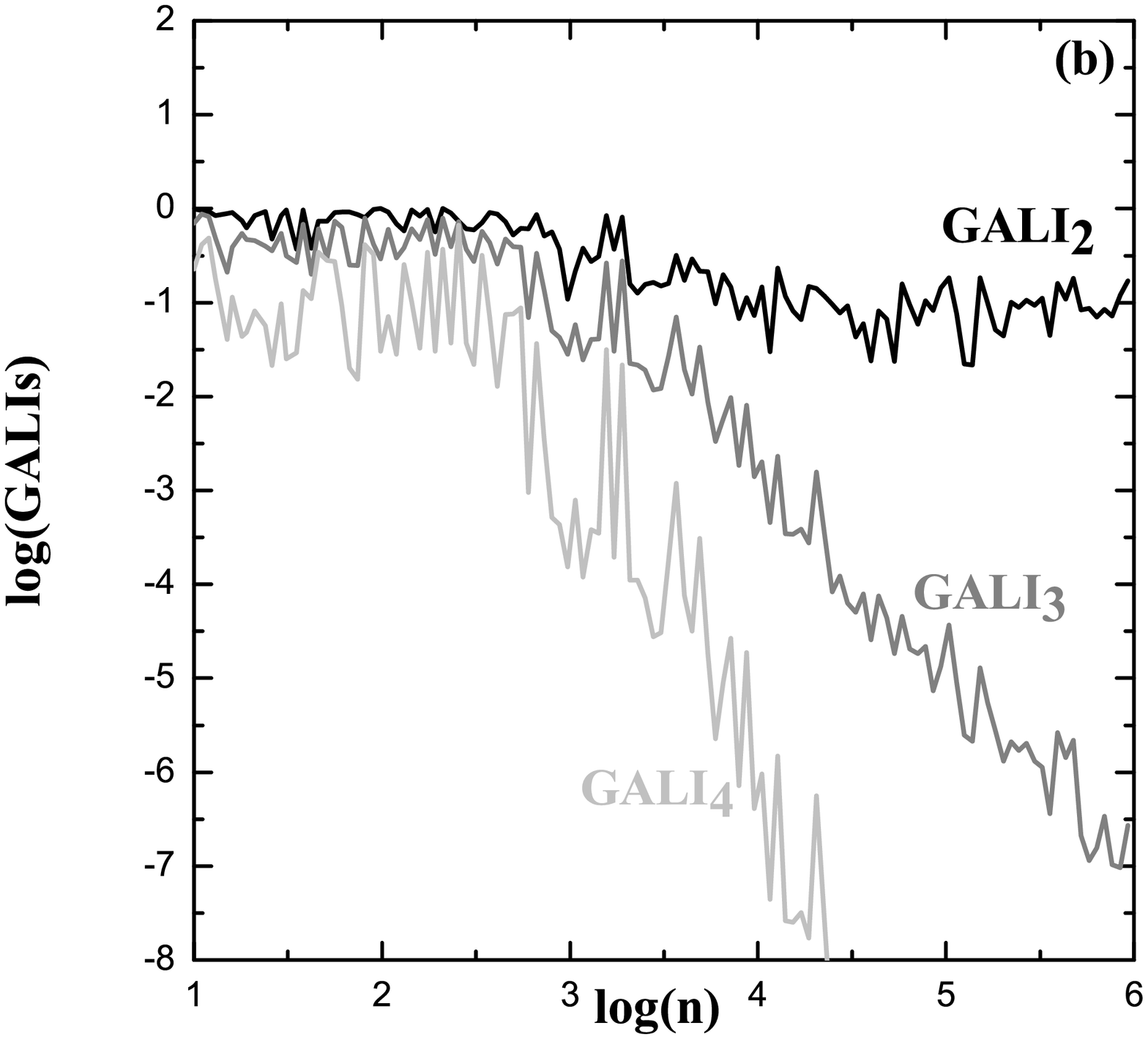}
}
\caption{The evolution of the GALI$_2$, the GALI$_3$ and the GALI$_4$
  with respect to the number of iterations $n$ for (a) a stable
  periodic orbit and (b) a nearby regular orbit, of the 4d map
  (\ref{eq:4d_map}) with $K=0.9$ and $\gamma=0.05$. The initial
  conditions of the orbits are: (a) $x_1=0.23666$, $y_1=0.0$,
  $x_2=0.23666$, $y_2=0.0$, and (b) $x_1=0.23$, $y_1=0.0$,
  $x_2=0.236$, $y_2=0.0$}
\label{fig:map_spo_neighbor}
\end{figure}

Let us now turn our attention to maps.  In $2N$d symplectic maps stable
periodic orbits of period $l$ correspond to $l$ distinct points (the so-called
stable fixed points of order $l$).  Any deviation vector from the periodic
orbit rotates around each fixed point. This behavior can be easily seen in the
case of 2d maps where the tori around a stable fixed point correspond to closed invariant curves which can be represented, through linearization, by ellipses
(see for example Sect.~3.3b of \cite{LL_92}). Thus, any $k$ initially
distinct deviation vectors needed for the computation of the GALI$_k$ will
rotate around the fixed point keeping on average the angles between them
constant. Consequently the volume of the parallelepiped they define, i.e.~the value of the GALI$_k$, will remain practically constant. Thus, in the case of stable periodic orbits of $2N$d maps, with $N\geq 1$ we have
\begin{equation}
\label{eq:GALI_stable_po_map}
\mbox{GALI}_k (t) \propto \mbox{constant}, \,\,\,\, \mbox{ for $2\leq k \leq
    2N$}.
\end{equation}
This behavior is clearly seen in Fig.~\ref{fig:map_spo_neighbor}(a)
where the evolution of the GALI$_2$, the GALI$_3$ and the GALI$_4$ for
a stable periodic orbit of period 7 of the 4d map (\ref{eq:4d_map}) is
plotted.

Again small perturbations of the periodic orbit's initial conditions
generally result in motion on an $N$d tori. Then, the evolution of the
corresponding GALIs is provided by (\ref{eq:GALI_reg}) for $N\geq 2$,
while the GALI$_2$ will decrease to zero according to
(\ref{eq:GALI_reg_2dmap}) for 2d maps. So, the most striking
difference between the behavior of the GALI$_k$ of a stable periodic
orbit and of a neighboring, regular orbit appears for $k>N$, because
in this case the GALI$_k$ remains constant for the periodic orbit,
while it decays to zero for the neighboring one. Differences of this
kind can be observed in Fig.\ref{fig:map_spo_neighbor}(b).

\section{Applications}
\label{sect:appl}

The ability of the SALI and the GALI methods to efficiently
discriminate between chaotic and regular motion was described in
detail in the previous sections, where some exemplary Hamiltonian
systems and symplectic maps were considered. In what follows we
present applications of this ability to various dynamical systems
originating from different research fields.

\subsection{Global Dynamics}
\label{sect:global}

In Sect.~\ref{sect:GALI_ch_reg} we discussed how one can use the various GALIs
to reveal the chaotic or regular nature of individual orbits in the $2N$d phase space of a dynamical system. Additionally, in Sect.~\ref{sect:GALI_po} we saw
how the measurement of the GALI$_2$ values for an ensemble of orbits
can facilitate the uncovering of some dynamical properties of the studied
system, in particular the pinpointing of stable periodic orbits
(Fig.~\ref{fig:Ham_spo_neighbor_scan}), while in Sect.~\ref{sect:low_search} we described how a more general search can help us locate motion on low
dimensional tori.

Now we  see how one can use the GALIs in order to study the global
dynamics of a system. For simplicity we  use in our analysis the
2D Hamiltonian system (\ref{eq:2DHam}), but the methods presented
below can be (and actually have already been) implemented to
higher-dimensional systems.

\subsubsection{Investigating Global Dynamics by the GALI$_k$ with $2 \leq k \leq N$}
\label{sect:global_2-N}

According to (\ref{eq:GALI_ch}) and (\ref{eq:GALI_reg}) the GALI$_k$,
with $2 \leq k \leq N$, behaves in a completely different way for
chaotic (exponential decay) and regular (remains practically constant)
orbits. Thus, by coloring each initial condition of an ensemble of
orbits according to its GALI$_k$ value at the end of a fixed
integration time we can produce color plots where regions of chaotic
and regular motion are easily seen. In addition, by choosing an
appropriate threshold value for the GALI$_k$, below which the orbit is
characterized as chaotic (see Sect.~\ref{sect:GALI_paradigms} on how
to set up this threshold), we can efficiently determine the `strength'
of chaos by calculating the percentage of chaotic orbits in the
studied ensemble.  Then, by performing the same analysis for different
parameter values of the system we can determine its physical
mechanisms that increase or suppress chaotic behavior.

A practical question arises though: which index should one use for
this kind of analysis? The obvious advantage of the GALI$_2$/SALI is
its easy computation according to (\ref{eq:SALI}), which requires the
evolution of only two deviation vectors. On the other hand, evaluating
the GALIs of order up to $k=N$ is more CPU-time consuming as the
computation of the index from (\ref{eq:GALI_SVD}) requires the
evolution of more deviation vectors, as well as the implementation of
the SVD algorithm. An advantage of these higher order indices is that
they tend to zero faster than the GALI$_2$/SALI for chaotic
orbits. So, reaching their threshold value which characterizes an
orbit as chaotic, requires in general, less computational effort. This
feature is particularly useful when we want to estimate the percentage
of chaotic orbits, as there is no need to continue integrating orbits
which have been characterized as chaotic (see Sect.~5.2 of
\cite{SBA_07} for an example of this kind).  Thus, we conclude that
the reasonable choices for such global studies are the GALI$_2$/SALI
and the GALI$_N$.

\begin{figure}
\centerline{
\begin{tabular}{c}
\includegraphics[scale=0.4]{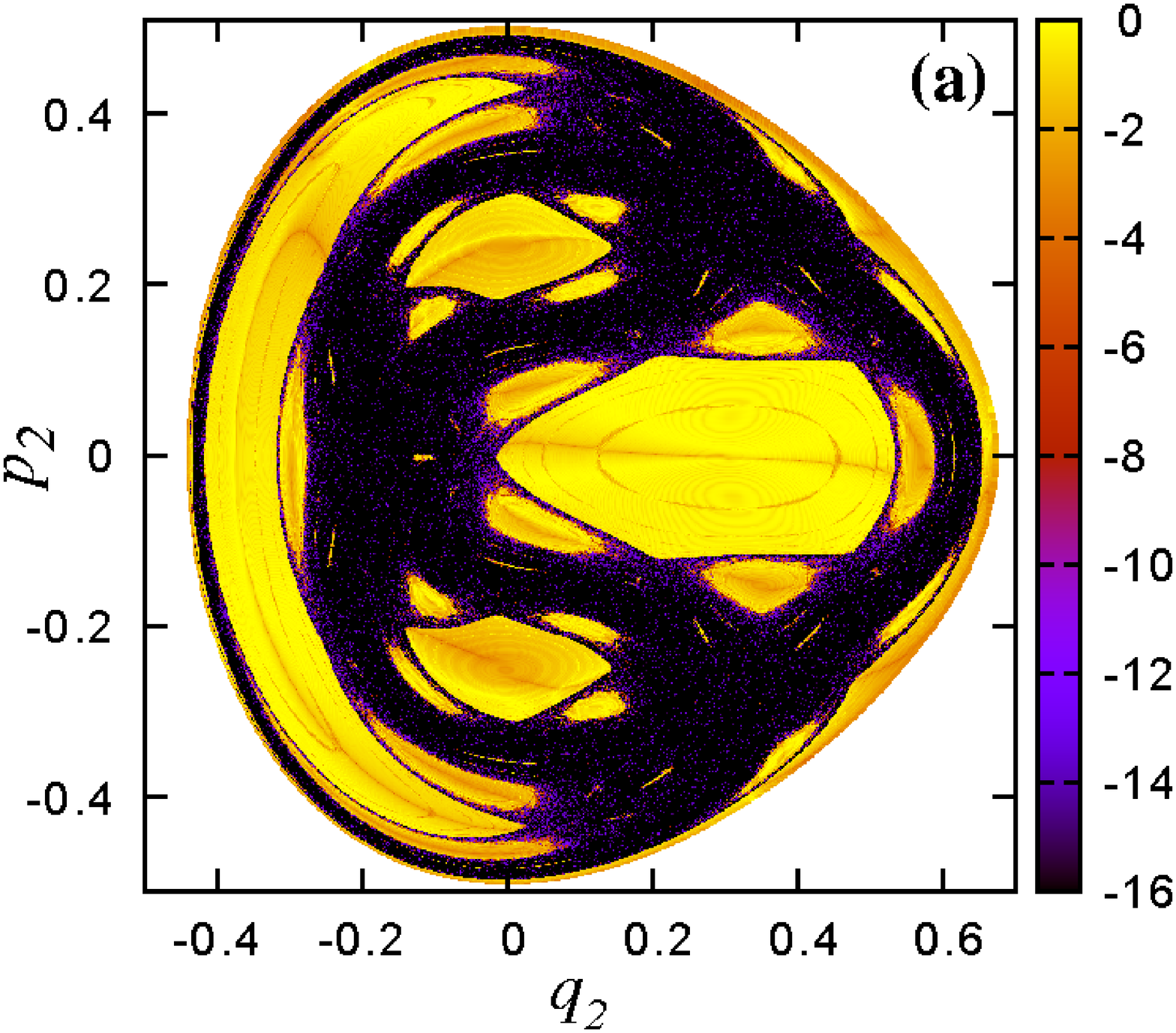}
\\
\includegraphics[scale=0.4]{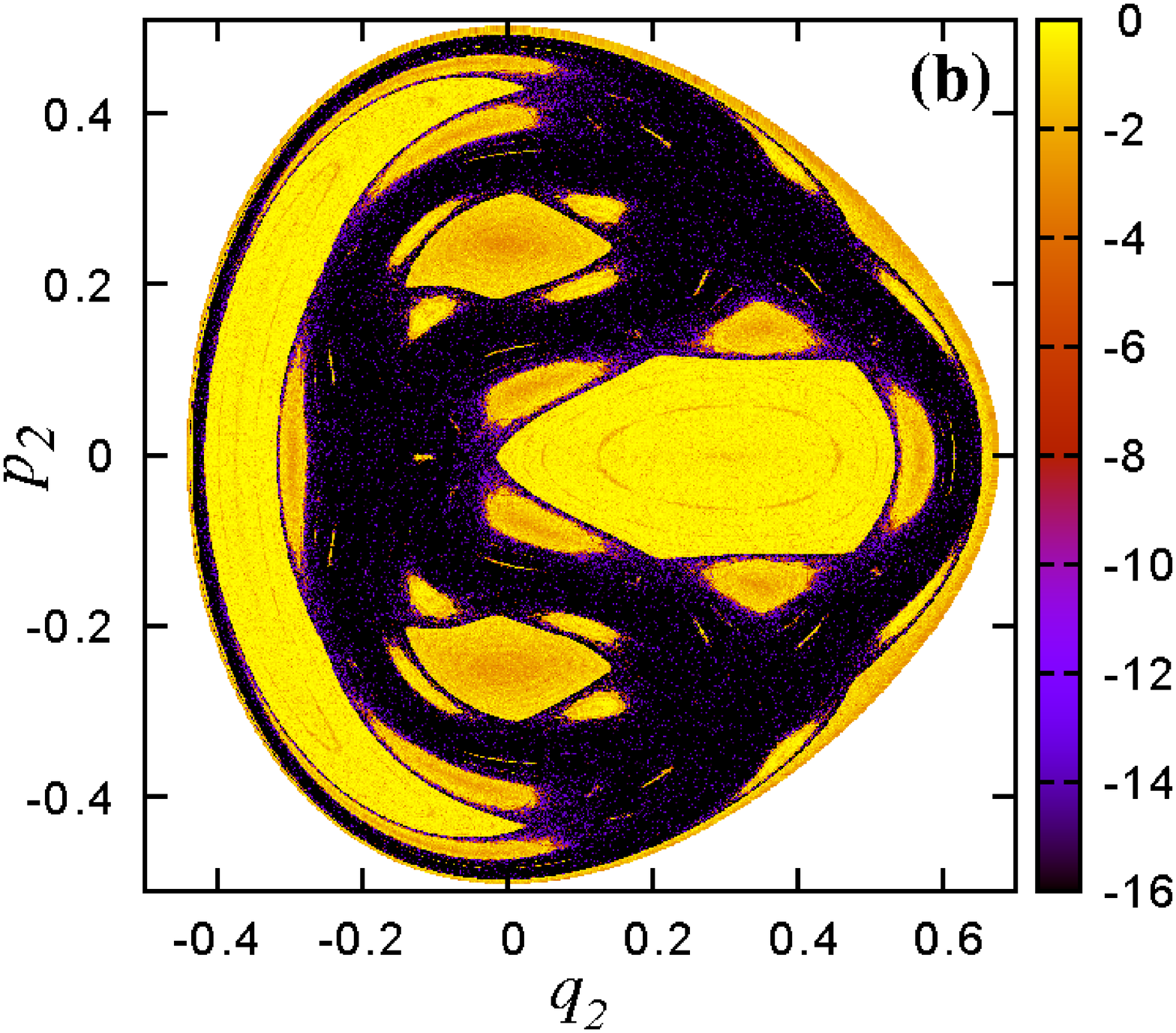}
\end{tabular}
}
\caption{Regions of different values of the GALI$_2$ on the
  Poincar\'{e} surface of section defined by $q_1=0$ of the 2D
  Hamiltonian (\ref{eq:2DHam}) for $H_2$=0.125. A set of approximately
  $350\,000$ equally spaced initial conditions on the grid $(q_2,p_2)
  \in [-0.5, 0.7 ]\times[-0.5, 0.5 ] $ is used.  White regions
  correspond to forbidden initial conditions. The color scales shown
  at the right of the panels are used to color each point according to
  the orbit's $\log (\mbox{GALI}_2)$ value at $t=2 \, 000$. In (a) the
  same set of initial orthonormal deviation vectors was used for the
  computation of the GALI$_2$ of each initial condition, while in (b)
  a different, randomly produced set of vectors was used for each
  orbit }
\label{fig:HH_global_GALI2}
\end{figure}

In order to illustrate this process, let us consider the 2D
H\'{e}non-Heiles system (\ref{eq:2DHam}), for which $\mbox{GALI}_N
\,\equiv \,\mbox{GALI}_2$, since $N=2$. In
Fig.~\ref{fig:HH_global_GALI2} we see color plots of its Poincar\'{e}
surface of section defined by $q_1=0$ (a concise description of the
construction of a surface of section can be found for instance in
Sect.~1.2b of \cite{LL_92}). The remaining initial conditions of each
orbit are its coordinates on the $(q_2,p_2)$ plane of
Fig.~\ref{fig:HH_global_GALI2}, while $p_1>0$ is set so that
$H_2=0.125$. For each panel of Fig.~\ref{fig:HH_global_GALI2} a 2d
grid of approximately $350\,000$ equally distributed initial
conditions is considered.  Each point on the $(q_2,p_2)$ plane is
colored according to its $\log (\mbox{GALI}_2)$ value at $t=2\, 000$,
while white regions denote not permitted initial conditions. Regions
colored in yellow or light red correspond to regular orbits, while
dark blue and black domains contain chaotic ones. Intermediate colors
at the borders between these two regions indicate sticky chaotic
orbits.

This kind of color plots can reveal fine details of the underlying dynamics,
like for example the small yellow `islands' of regular motion inside the large,
black chaotic `sea', as well as allow the accurate estimation of the percentage
of chaotic or regular orbits in the studied ensemble. Naturally the denser the
used grid is, the finer the uncovered details become, but unfortunately the
higher the needed computational effort gets.  In an attempt to speed up the
whole process the following procedure was followed in \cite{AMS_05} where the
dynamics of the H\'{e}non-Heiles system (\ref{eq:2DHam}) was studied.  The
final GALI$_2$/SALI value and the corresponding color was assigned not only to
the initial condition of the studied orbit, but also to all intersection points of the orbit with the surface of section. This assignment can be extended even
further by additionally taking into account the symmetry of Hamiltonian
(\ref{eq:2DHam}) with respect to the $q_2$ variable, which results in
structures symmetric with respect to the $p_2=0$ axis in Fig.~\ref{fig:HH_global_GALI2}. Consequently, points symmetric to this axis
should have the same GALI$_2$/SALI value. So, orbits with initial conditions on grid points to which a color has already been assigned, as they were
intersection points with the surface of section of previously computed orbits,
are not computed again and so the construction of color plots like the ones of
Fig.~\ref{fig:HH_global_GALI2} is speeded up significantly. In \cite{AMS_05} it was shown that this approach achieves very accurate estimations of the
percentages of chaotic orbits with respect to the ones obtaining by coloring
each and every initial condition according to the index's value at the end of
the integration time (this is actually how Fig.~\ref{fig:HH_global_GALI2} was
produced).

Let us now discuss the differences between panels (a) and (b) of
Fig.~\ref{fig:HH_global_GALI2}. In both figures the chaotic regions
are practically the same. Nevertheless, in the yellow and light red
colored domains, where regular motion occurs, some `spurious'
structures appear in Fig.~\ref{fig:HH_global_GALI2}(a), which are not
present in Fig.~\ref{fig:HH_global_GALI2}(b). For example, inside the
large stability island with $0 \lesssim q_2 \lesssim 0.5$ at the right
side of Fig.~\ref{fig:HH_global_GALI2}(a) we observe an almost
horizontal formation colored in light red, while similar colored
`arcs' appear inside many other islands of regular motion.  These
artificial features emerge when one uses exactly the same set of
orthonormal, initial deviation vectors for every studied orbit, as we
did in Fig.~\ref{fig:HH_global_GALI2}(a).  The appearance of such
features in color plots of other chaos detection methods has already
been reported in the literature \cite{BBB_09}.  A simple way to avoid
them is to use a different, random set of initial, orthonormal vectors
for the computation of the GALI$_2$, as we did in
Fig.~\ref{fig:HH_global_GALI2}(b). By doing so, these spurious
features disappear and only structures related to the actual dynamics
of the system remain, like for instance the cyclical `chain' of the
light red colored, elongated regions inside the big stability island
at the right side of Fig.~\ref{fig:HH_global_GALI2}(b). This structure
indicates the existence of some higher order stability island, which
are surrounded by an extremely thin chaotic layer. This layer is not
visible for the resolution used in Fig.~\ref{fig:HH_global_GALI2}(b).
A magnification, and a much finer grid would reveal this tiny chaotic
region.

\subsubsection{Investigating Global Dynamics by the GALI$_k$ with $N < k \leq 2N$}
\label{sect:global_N-2N}

As was clearly explained in Sect.~\ref{sect:GALI_paradigms} the GALIs
of order $N < k \leq 2N$ tend to zero both for chaotic and regular
orbits, but with very different time rates as (\ref{eq:GALI_ch}) and
(\ref{eq:GALI_reg}) state. This deference can be also used to
investigate global dynamics, but following an alternative approach to
the one developed in Sect.~\ref{sect:global_2-N}. Since these GALIs
decay to zero exponentially fast for chaotic orbits, but follow a much
slower power law decay for regular ones, the time $t_{th}$ they need
to reach an appropriately chosen, small threshold value will be
significantly different for the two kinds of orbits.  We note that
both the exponential and the power law decays become faster with
increasing order $k$ of GALI$_k$. Consequently, the creation of huge
differences in the GALI$_k$ values, which allow the discrimination
between chaotic and regular motion, will appear earlier for larger $k$
values. So, in general, the overall required computational time
decreases significantly by using a higher order GALI$_k$, despite the
integration of more deviation vectors, since this integration will be
terminated earlier.  Thus, the best choice in investigations of this
kind is to use the GALI$_{2N}$.

\begin{figure}
\sidecaption[t]
\includegraphics[scale=0.41]{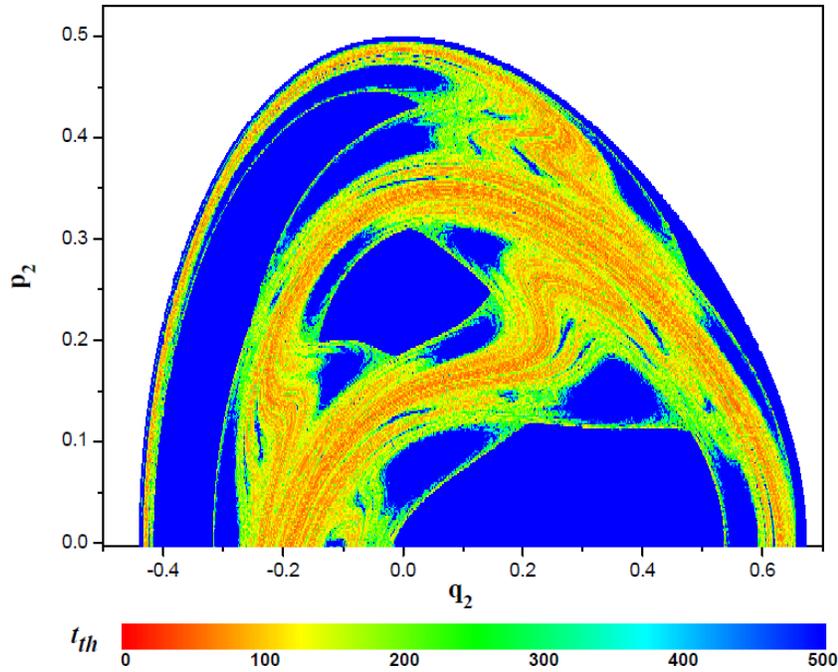}
\caption{Regions of different values of the time $t_{th}$ needed for
  the GALI$_4$ to become less than $10^{-12}$ on the $q_1=0$ surface
  of section of the 2D H\'enon-Heiles system (\ref{eq:2DHam}). Each
  orbit is integrated up to $t=500$ time units. White regions
  correspond to forbidden initial conditions. The color scales shown
  below the panel are used to color each point according to the
  orbit's $t_{th}$ value (after \cite{SBA_07}) }
\label{fig:HH_global_GALI4}
\end{figure}

Let us illustrate this approach by computing the GALI$_4$ for the 2D
H\'{e}non-Heiles system (\ref{eq:2DHam}), at a grid in its $q_1=0$
surface of section.  The outcome of this procedure is seen in
Fig.~\ref{fig:HH_global_GALI4}, where each initial condition is
colored according to the time $t_{th}$ needed for its GALI$_4$ to
become $\leq 10^{-12}$.  Each orbit is integrated up to $t=500$ time
units and if its GALI$_4$ value at the end of the integration is
larger than the threshold value $10^{-12}$ the corresponding $t_{th}$
value is set to $t_{th}=500$ and the initial condition is colored in
blue according to the color scales seen below the panel of
Fig.~\ref{fig:HH_global_GALI4}. Regions of regular motion correspond
to large $t_{th}$ values and are colored in blue, while all the
remaining colored domains contain chaotic orbits. Again, white regions
correspond to forbidden initial conditions.  This approach yields a
very detailed chart of the dynamics, analogous to the one seen in
Fig.~\ref{fig:HH_global_GALI2}.

An advantage of the current approach is its ability to clearly reveal
various `degrees' of chaotic behavior in regions not colored in blue.
Strongly chaotic orbits are colored in red and yellow as their
GALI$_4$ becomes $\leq 10^{-12}$ quite fast. Orbits with larger
$t_{th}$ values correspond to chaotic orbits which need more time in
order to show their chaotic nature, while the `sticky' chaotic regions
are characterized by even higher $t_{th}$ values and are colored in
light blue. We note that for every initial condition we used a
different, random set of orthonormal deviation vectors in order to
avoid the appearance of possible `spurious' structures, like the ones
seen in Fig.~\ref{fig:HH_global_GALI2}(a).

\subsection{Studies of Various Dynamical Systems}
\label{sect:appl_studies}

The SALI and the GALI methods have been used broadly for the study of
the phase space dynamics of several models originating from different
scientific fields. These studies include the characterization of
individual orbits as chaotic or regular, as well as the consideration
of large ensembles of initial conditions along the lines presented in
Sect.~\ref{sect:global}, whenever a more global understanding of the
underlying dynamics was needed.

In this section we present a brief, qualitative overview of such
investigations. For this purpose we focus mainly on the outcomes of
these studies avoiding a detailed presentation of mathematical
formulas and equations for each studied model.

\subsubsection{An Accelerator Map Model}
\label{sect:appl_studies_map_acc}

Initially, let us discuss two representative applications of the
SALI. The first one concerns the study of a 4d symplectic map which
describes the evolution of a charged particle in an accelerator ring
having a localized thin sextupole magnet.  The specific form of this
map can be found in \cite{BS_06} where the SALI method was used for
the construction of phase space color charts where regions of chaotic
and regular motion were clearly identified, as well as for evaluating
the percentage of chaotic orbits.

\begin{figure}
\centerline{
\includegraphics[scale=0.42]{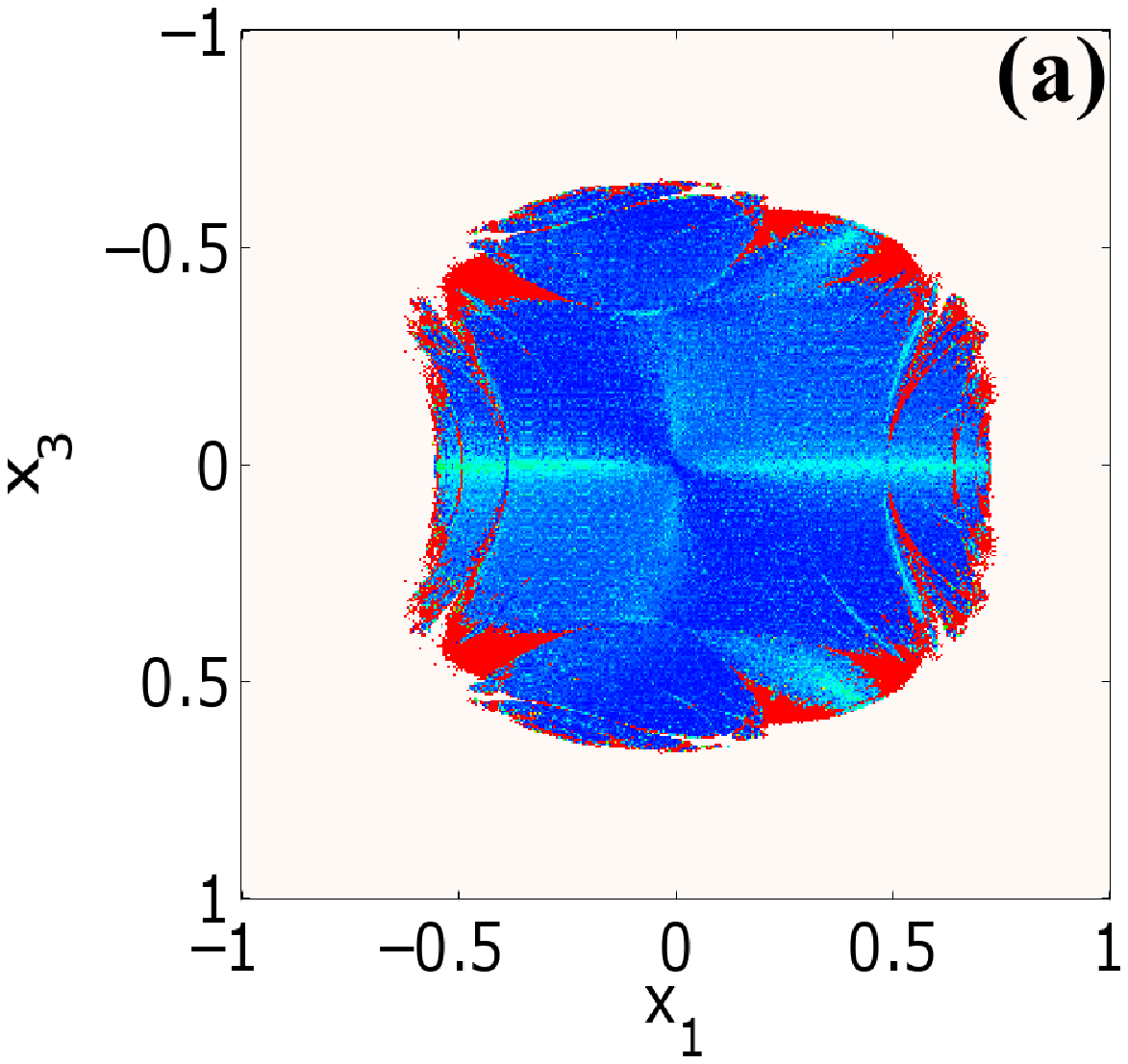}
\hspace{-0.4cm}
\includegraphics[scale=0.42]{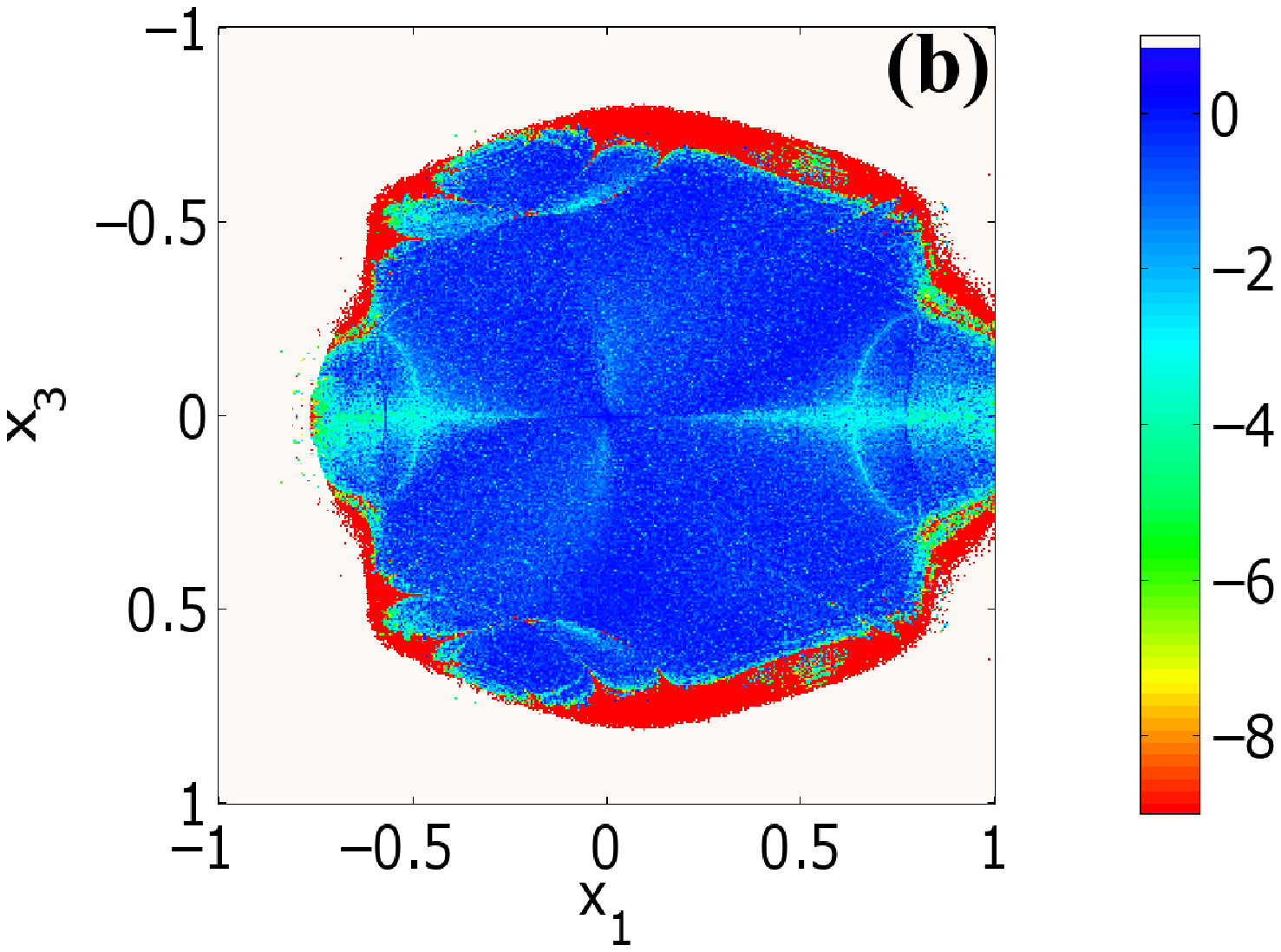} }
\caption{Regions of different SALI values of (a) the 4d uncontrolled
  accelerator map studied in \cite{BS_06} and (b) the controlled map
  constructed in \cite{BCSV_12}. The coordinates $x_1$, $x_3$
  respectively describe horizontal and vertical deflections of a
  charged particle from the ideal circular orbit passing from
  $x_1=x_3=0$ in some appropriate units (see \cite{BS_06} for more
  details). $16\, 000$ uniformly distributed initial conditions on the
  grid $(x_1,x_3) \in [-1,1]\times [-1,1]$ were evolved for $10^5$
  iterations of each map and colored according to the orbit's
  $\log(\mbox{SALI})$ value, using the color scales shown at the right
  of the panels. The white colored regions correspond to orbits that
  escape in less than $10^5$ iterations. Red points denote chaotic
  orbits, while regular ones are colored in blue. The increase of the
  stability region around the point $x_1=x_3=0$ is evident (after
  \cite{BCSPV_12}) }
\label{fig:accel_map}
\end{figure}

Later on, in \cite{BCSV_12,BCSPV_12} this map was used to test the
efficiency of chaos control techniques for increasing the stability
domain (the so-called `dynamic aperture') around the ideal circular
orbit of this simplified accelerator model. These techniques turned
out to be quite successful, as the addition of a rather simple control
term, which potentially could be approximated by real multipole
magnets, increased the the stability region of the map as can be seen
in Fig.~\ref{fig:accel_map}.

\subsubsection{A Hamiltonian Model of a Bose-Einstein condensate}
\label{sect:appl_studies_Ham_BEC}

Let us now turn our attention to a 2D Hamiltonian system describing
the interaction of three vortices in an atomic Bose-Einstein
condensate, which was studied in \cite{KKSK_14}.  By means of SALI
color plots the extent of chaos in this model was accurately measured
and its dependence on physically important parameters, like the energy
and the angular momentum of the vortices, were determined.

\begin{figure}
\centerline{
\begin{tabular}{cc}
\multicolumn{2}{c}{\includegraphics[scale=0.36]{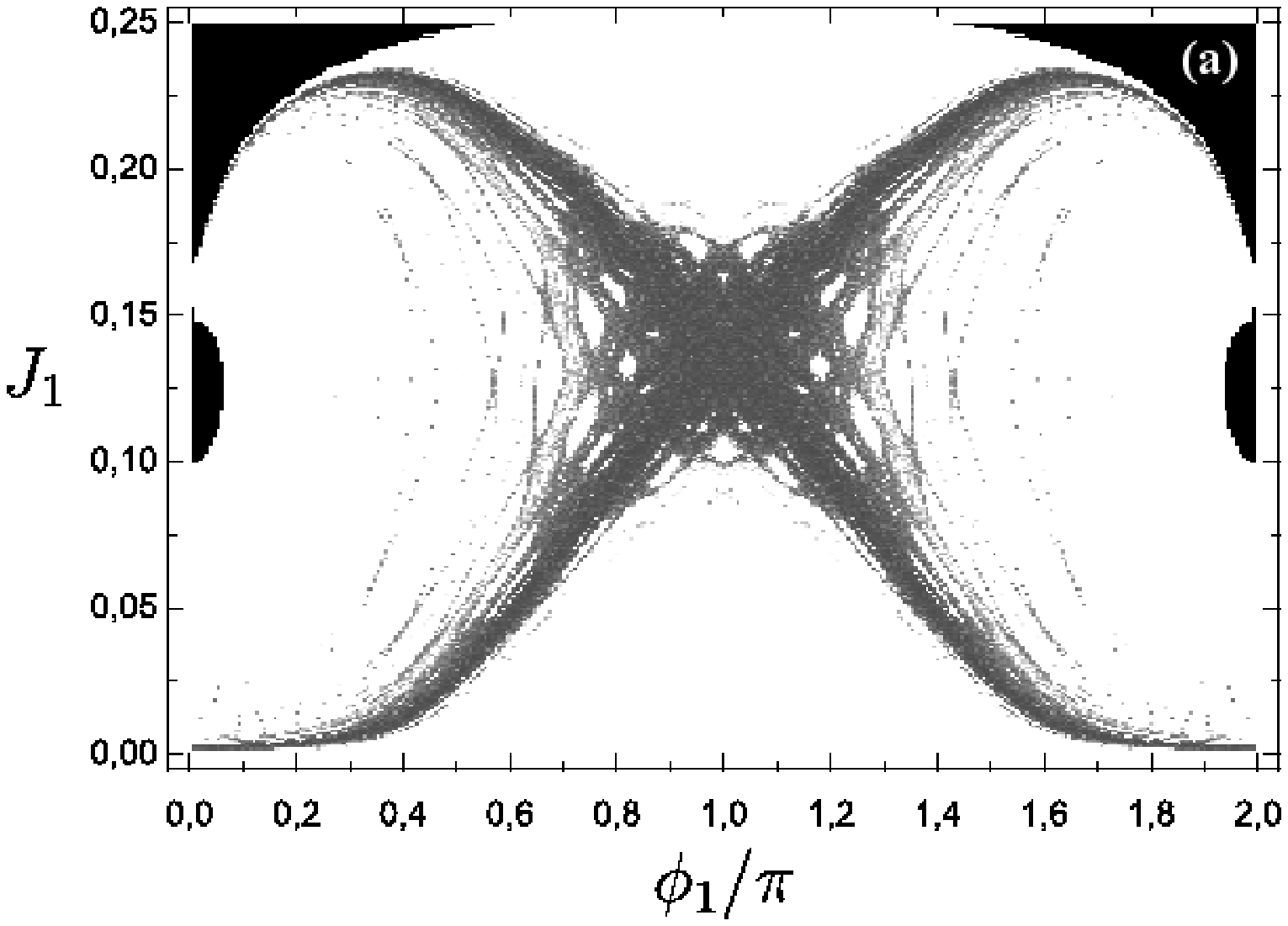} \,
\includegraphics[scale=.72]{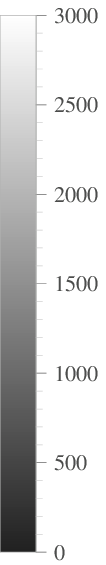}}
\\
\\
\hspace{0.cm}
\includegraphics[scale=0.36]{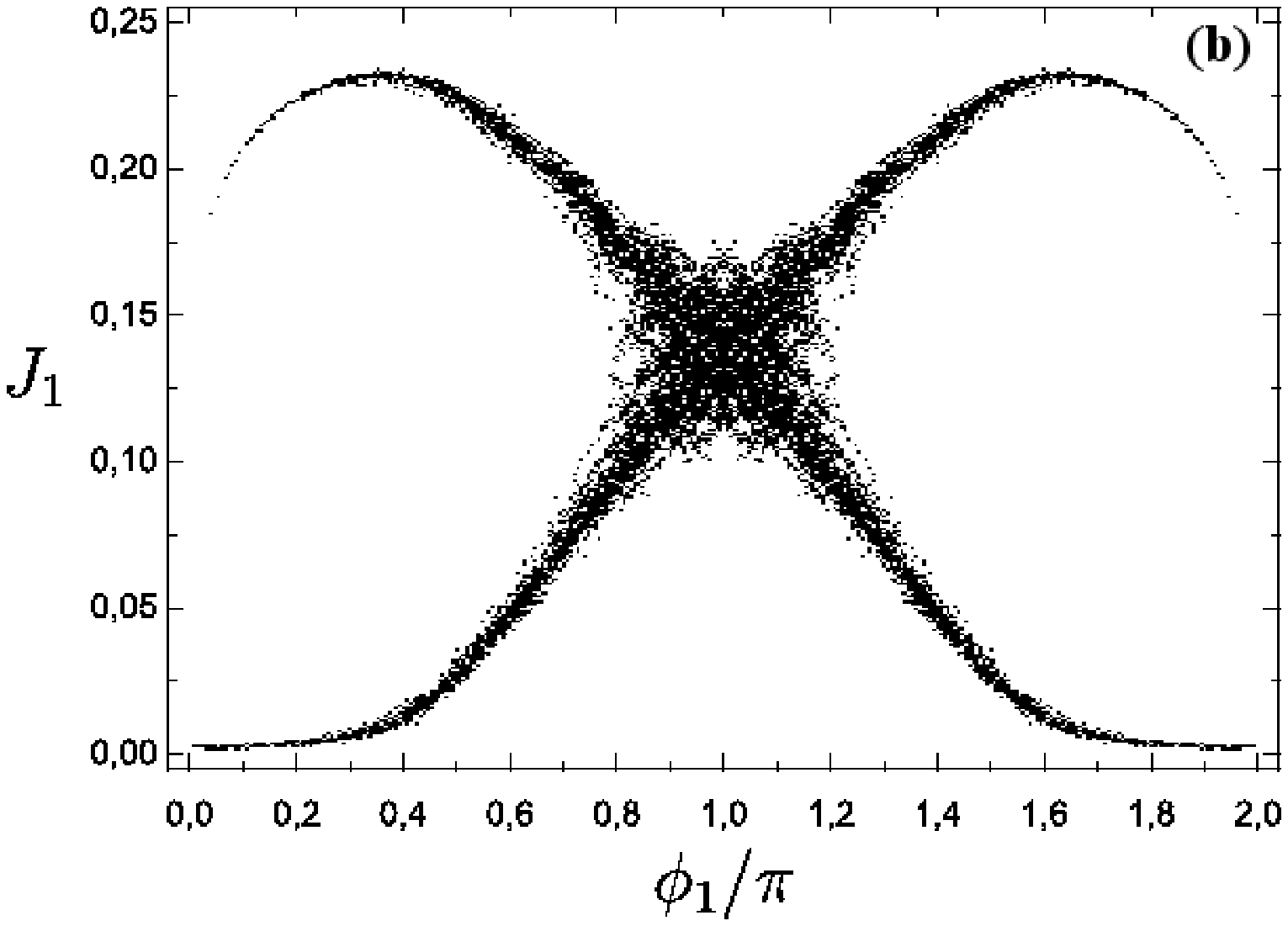} &
\hspace{0.cm}
\includegraphics[scale=0.36]{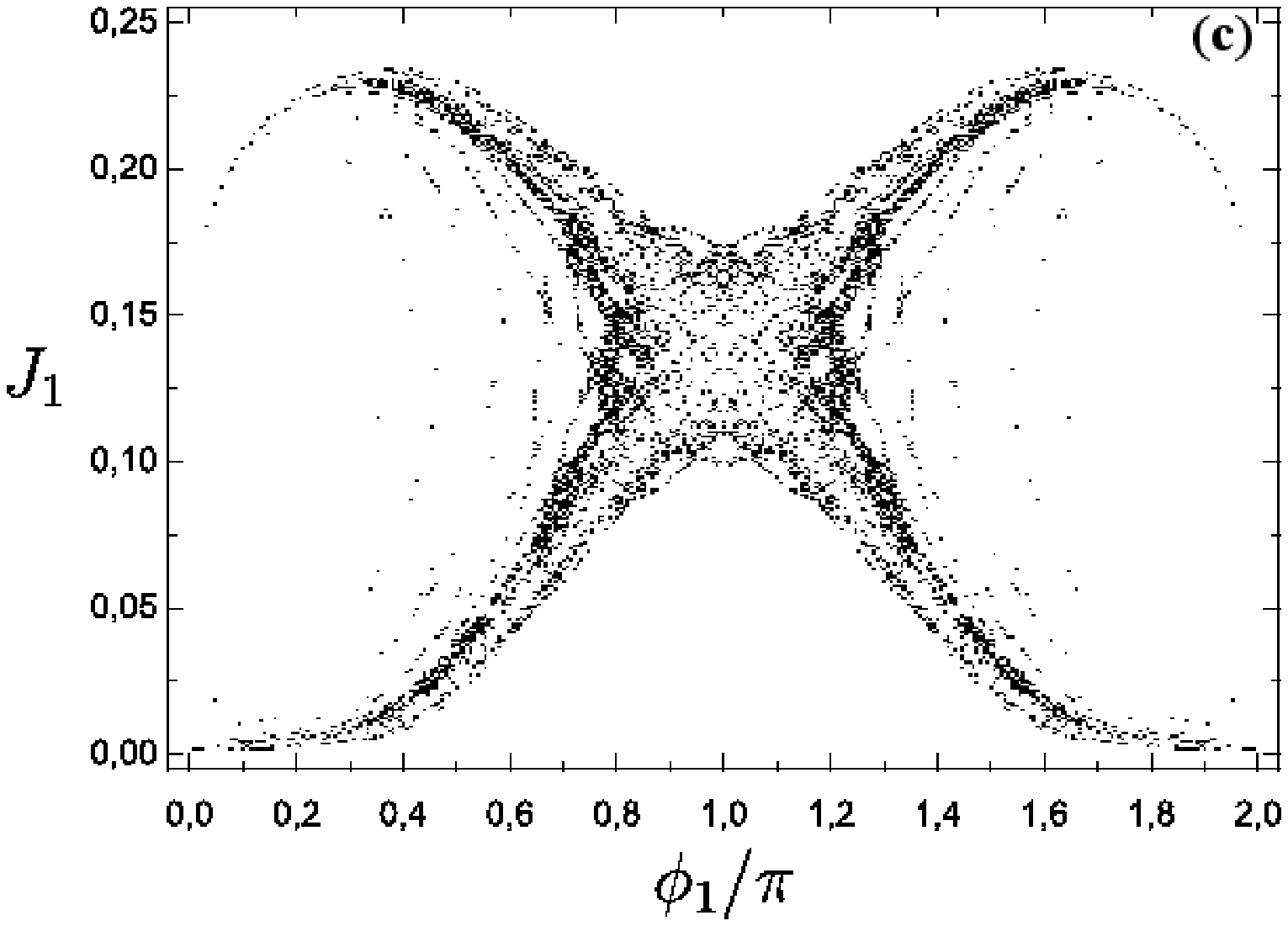}
\\
\\
\hspace{0.cm}
\includegraphics[scale=0.36]{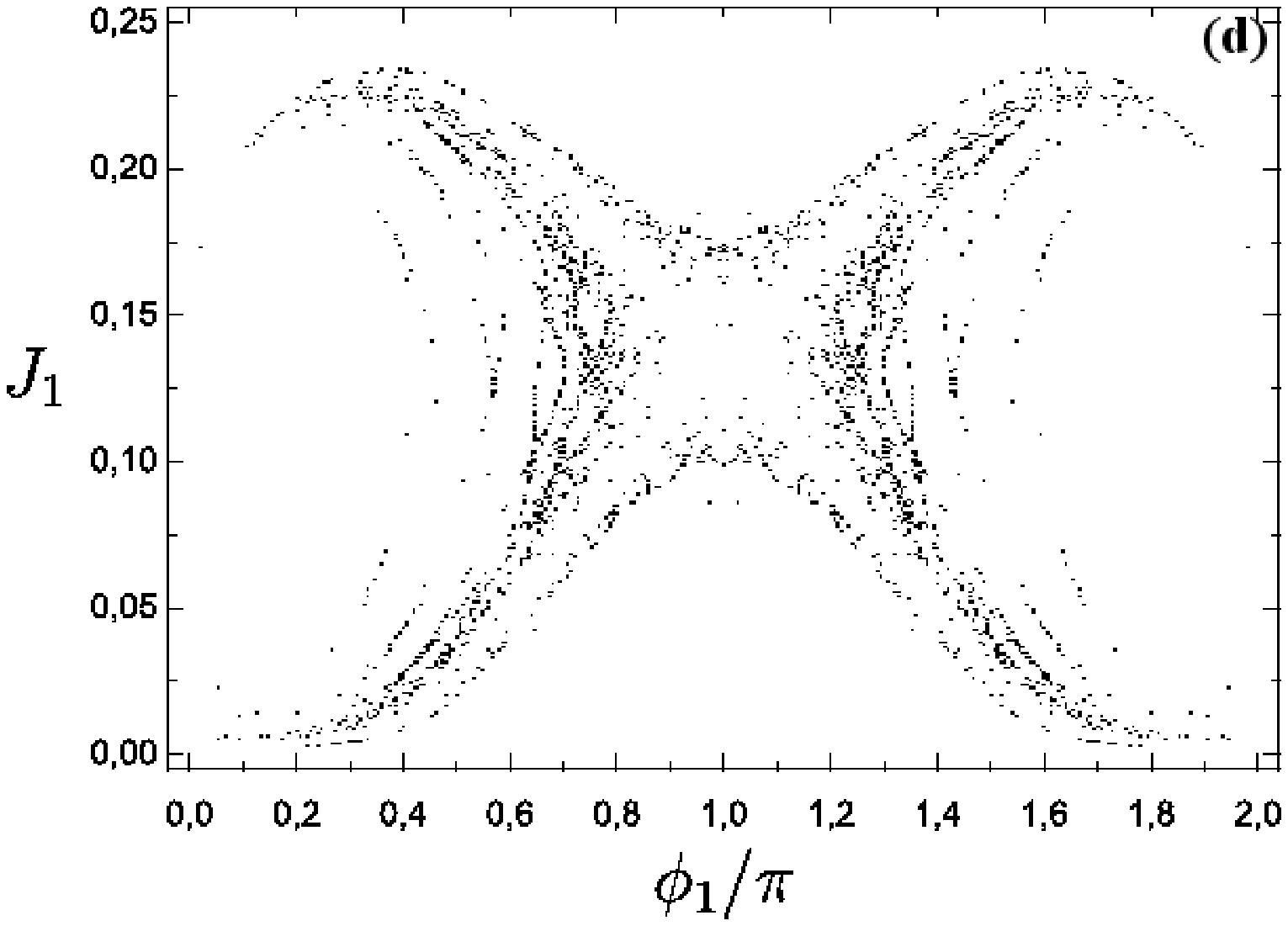} &
\hspace{0.cm}
\includegraphics[scale=0.36]{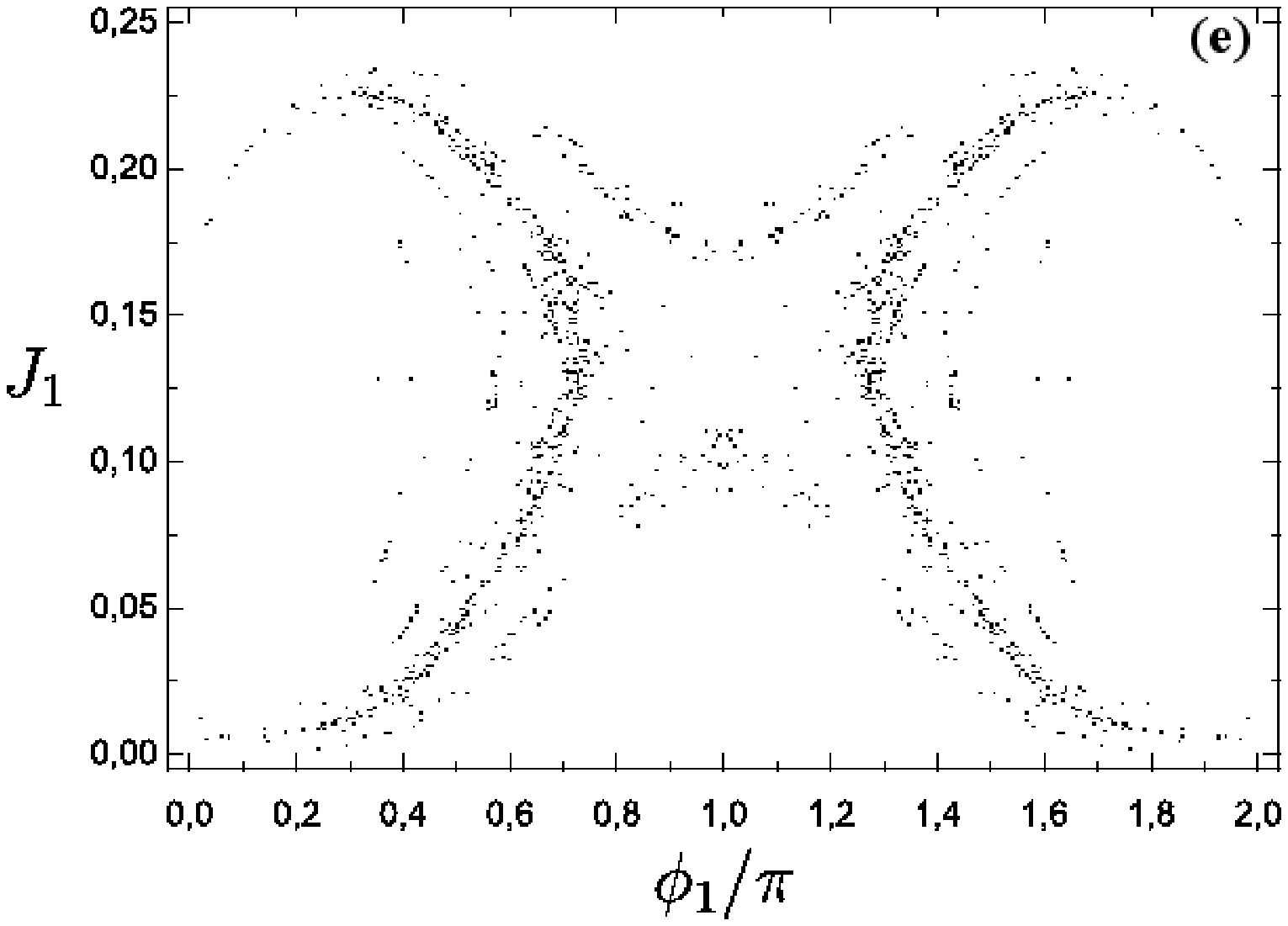}
\end{tabular}
}
\caption{ (a) Regions of different values of the time $t_{th}$ needed
  for the SALI to become less than $10^{-12}$ for a 2D Hamiltonian
  describing the interaction of three vortices in an atomic
  Bose-Einstein condensate. The explicit definition of the coordinates
  $J_1$ and $\phi_1/\pi$ can be found in \cite{KKSK_14} where this
  model was studied in detail. Each orbit is integrated up to
  $t=3\,000$ time units. White regions correspond to regular orbits,
  while black areas at the upper two corners, as well as in the middle
  of the vertical axes at both sides of the plot, denote not permitted
  initial conditions. The color scales shown at the right of the panel
  are used to color each point according to the orbit's $t_{th}$
  value. The initial conditions of (a) are decomposed in four
  different sets according to their $t_{th}$ value: (b) $140 \leq
  t_{th} \leq 500$, (c) $500 < t_{th} \leq 1\,000$, (d) $1\,000 <
  t_{th} \leq 1\,500$ and (e) $1\,500 < t_{th} \leq 2\,000$ (after
  \cite{KKSK_14}) }
\label{fig:appl_BEC}
\end{figure}

In real experiments, from which the study of this model was motivated,
the life time of Bose-Einstein condensates is limited. For this reason
the time in which the chaotic nature of orbits is uncovered played a
significant role in the analysis presented in \cite{KKSK_14}.
Actually, different `degrees of chaoticity' are revealed by
registering the time $t_{th}$ that the SALI of a chaotic orbit
requires in order to become $\leq 10^{-12}$
(Fig.~\ref{fig:appl_BEC}). This approach is similar to the one
presented in Sect.~\ref{sect:global_N-2N}, and allows the
identification of regions with different strengths of chaos.

The chaotic orbits of Fig.~\ref{fig:appl_BEC}(a) are decomposed in
Figs.~\ref{fig:appl_BEC}(b)--(e) in four different sets according to
their $t_{th}$ value: $t_{th} \in [140,500]$
(Fig.~\ref{fig:appl_BEC}(b)), $t_{th} \in (500, 1\,000]$
(Fig.~\ref{fig:appl_BEC}(c)), $t_{th} \in (1\,000, 1\,500]$
(Fig.~\ref{fig:appl_BEC}(d)) and $t_{th} \in (1\,500, 2\,000]$
(Fig.~\ref{fig:appl_BEC}(e)), where time is measured in some
appropriate units (see \cite{KKSK_14} for more details).  From these
results we see that, as the initial conditions move further away from
the center of the x-shaped region of Fig.~\ref{fig:appl_BEC}(a) the
orbits need more time to show their chaotic nature and consequently,
some of them can be considered as regular from a practical
(experimental) point of view. For instance, in real experiments one
would expect to detect chaotic motion in regions shown in
Fig.~\ref{fig:appl_BEC}(b) where orbits have relatively small $t_{th}$
values.  Thus, an analysis of this kind can provide practical
information about where one should look for chaotic behavior in actual
experimental set ups.

\subsubsection{Further Applications of the SALI and the GALI Methods}
\label{sect:appl_studies_further}

The SALI and the GALI methods have been successfully employed in
studies of various physical problems and mathematical toy models, as
well as for the investigation of fundamental aspects of nonlinear
dynamics (e.g.~see \cite{CEGM_14}). In what follows we briefly present
some of these studies

In \cite{MSAB2008} the SALI/GALI$_2$ method was used for the global
study of the standard map (\ref{eq:2d_map}).  By considering large
ensembles of initial conditions the percentage of chaotic motion was
accurately computed as a function of the map's parameter $K$.  This
work revealed the periodic re-appearance of small (even tiny) islands
of stability in the system's phase space for increasing values of $K$.
Subsequent investigations of the regular motion of the standard map in
\cite{ManRob2014} led to the clear distinction between typical islands
of stability and the so-called accelerator modes, i.e.~motion
resulting in an anomalous enhancement of the linear in time orbits'
diffusion.  Typically, this motion is highly superdiffusive and is
characterized by a diffusion exponent $\approx 2$.

In \cite{BMC2009} the GALI was used for the detection of chaotic
orbits in many dimensions, the prediction of slow diffusion, as well
as the determination of quasiperiodic motion on low dimensional tori
in the system (\ref{eq:Md_map}) of many coupled standard
maps. Additional applications of the SALI in studying maps can be
found in \cite{PABV2008}, where the index was used for shedding some
light in the properties of accelerator models, while in
\cite{SS2012} a coupled logistic type predator-prey model describing
population growths in biological systems was considered.  Further
studies of 2d and 4d maps based on the SALI method were performed in
\cite{FMT2012}.

Models of dynamical astronomy and galactic dynamics are considered to
be the spearhead of the chaos detection methods \cite{Cont_book}.
Actually, many of these methods have been used, or often even
constructed, to investigate the properties of such systems. Several
applications of the SALI to systems of this kind can be found in the
literature.  In \cite{SBD2007,SPB2008,BP2009} the stability properties
of orbits in a particular few-body problem, the so-called the Sitnikov
problem, were studied, while in \cite{Voy2008} the long term stability
of two-planet extrasolar systems initially trapped in the 3:1 mean
motion resonance was investigated.  The SALI was also used to study
the dynamics of the Caledonian symmetric four-body problem
\cite{SESS2013}, as well as the circular restricted three-body problem
\cite{R_14}.

In systems modeling the dynamics of galaxies special care should be
taken with respect to the determination of the star motion's nature,
because this has to be done as fast as possible and in physically
relevant time intervals (e.g.~smaller than the age of the universe).
Hence, in order to check the adequacy of a proposed galactic model, in
terms of being able to sustain structures resembling the ones seen in
observations of real galaxies, the detection of chaotic and regular
motion for rather small integration times is imperative.  The SALI and
the GALI methods have proved to be quite efficient tools for such
studies, as they allow the fast characterization of orbits. This
ability reduces significantly the required computational burden, as in
many cases the determination of the orbits' nature is achieved before
the predefined, final integration time.

In particular, the SALI method has been used successfully in studying
the chaotic motion and spiral structure in self-consistent models of
rotating galaxies \cite{VHC2007}, the dynamics of self-consistent
models of cuspy triaxial galaxies with dark matter haloes
\cite{CDLMV2007}, the orbital structure in $N$ body models of
barred-spiral galaxies \cite{HK2009}, the secular evolution of
elliptical galaxies with central masses \cite{K2008}, the chaotic
component of cuspy triaxial stellar systems \cite{CMN_14}, as well as
the chaoticity of non-axially symmetric galactic models
\cite{ZotCar2013} and of models with different types of dark matter
halo components \cite{Zot2014}.

The SALI was used in \cite{MSAB2008} for investigating the dynamics of
2D and 3D Hamiltonian models of rotating bared galaxies. This work was
extended in \cite{MA2011} by using the GALI for studying the global
dynamics of different galactic models of this type. In particular, the
effects of several parameters related to the shape and the mass of the
disk, the bulge and the bar components of the models, as well as the
rotation speed of the bar, on the amount of chaos appearing in the
system were determined. Moreover, the implementation of the GALI$_3$
in the 3D Hamiltonians allowed the detection of regular motion on low
(2d) dimensional tori, although these systems support, in general, 3d
orbits. The astronomical significance of these orbits was discussed in
detail in \cite{MA2011}.

Implementations of the SALI to nuclear physics systems can be found in
\cite{SCM2007,MSCHJD2007,SHC2009,MDSC2010,MDC2010} where the chaotic
behavior of boson models is investigated, as well as in \cite{ABB2010}
where the dynamics of a Hamiltonian model describing a confined
microplasma was studied. Recently the SALI and the GALI methods,
together with other chaos indicators, were reformulated in the
framework of general relativity, in order to become invariant under
coordinate transformation \cite{Luk2014}.

The SALI and the GALI have been also used to study the dynamics of
nonlinear lattice models. Applications of these indices to the
Fermi-Pasta-Ulam model can be found in
\cite{AB2006,CB2006,ABS_06,SBA_08,PP2008,CEB2010,AC2011,CE2013} where
the properties of regular motion on low dimensional tori, the long
term stability of orbits, as well as the interpretation of
Fermi-Pasta-Ulam recurrences were studied. In \cite{ManRuf2011} the
GALI method managed to capture the appearance of a second order phase
transition that the Hamiltonian Mean Field model exhibits at a certain
energy density. The index successfully verified also other
characteristics of the system, like the sharp transition from weak to
strong chaos. Further applications of the SALI method to other models
of nonlinear lattices can be found in \cite{PBS_04,ABS_06}.

In addition, the SALI was further used in studying the chaotic and
regular nature of orbits in non-Hamiltonian dynamical systems
\cite{HuaWu2011,ABDNT2013}, some of which model chaotic electronic
circuits \cite{HuaWu2012,HuaZhou2013,HuaCao2014}.

\subsection{Time Dependent Hamiltonians}
\label{sect:time_dep}

The applications presented so far concerned autonomous dynamical
systems. However, there are several phenomena in nature whose modeling
requires the invocation of parameters that vary in time. Whenever
these phenomena are described according to the Hamiltonian formalism,
the corresponding Hamiltonian function is not an integral of motion as
its value does not remain constant as time evolves.

The SALI and the GALI methods can be also used to determine the
chaotic or regular nature of orbits in time dependent systems as long
as, their phase space does not shrink ceaseless or expand unlimited,
with respect to its initial volume, during the considered times.  This
property allows us to utilize the time evolution of the volume defined
by the deviation vectors, as in the case of the time independent
models, and estimate accurately its possible decay for time intervals
where the total phase space volume has not changed significantly.

In conservative time independent Hamiltonians orbits can be periodic
(stable or unstable), regular (quasiperiodic) or chaotic and their
nature does not change in time. Sticky chaotic orbits may exhibit a
change in their orbital morphologies from almost quasiperiodic to
completely chaotic behaviors, but in reality their nature does not
change as they are weakly chaotic orbits.  On the other hand, in time
dependent models, individual orbits can display abrupt transitions
from regular to chaotic behavior, and vice versa, during their time
evolution. This is an intriguing characteristic of these systems which
should be captured by the used chaos indicator.  Such transitions
between chaotic and regular behaviors can be seen for example in $N$
body simulations of galactic models. For this reason, time dependent
analytic potentials trying to mimic the evolution of $N$ body galactic
systems, are expected to exhibit similar transitions.

An analytic time dependent bared galaxy model consisting of a bar, a
disk and a bulge component, whose masses vary linearly in time was
studied in \cite{MBS2013}. The time dependent nature of the model
influences drastically the location and the size of stability islands
in the system's phase space, leading to a continuous interplay between
chaotic and regular behaviors. The GALI was able to capture subtle
changes in the nature of individual orbits (or ensemble of orbits)
even for relatively small time intervals, verifying that it is an
ideal diagnostic tool for detecting dynamical transitions in time
dependent systems.

Although both 2D and 3D time dependent Hamiltonian models were studied
in \cite{MBS2013}, we  further discuss here only the 3D model in
order to illustrate the procedure followed for detecting the various
dynamical epochs in the evolution of an orbit. The main idea for doing
that is the re-initialization of the computation of the GALI$_k$, with
$2 \leq k \leq N$, whenever the index reaches a predefined low value
(which signifies chaotic behavior) by considering $k$ new, orthonormal
deviation vectors resetting GALI$_k=1$.

Let us see this procedure in more detail.  In \cite{MBS2013} the
evolution of the GALI$_3$ was followed for each studied orbit. The
three randomly chosen, initial deviations vectors set GALI$_3=1$ in
the beginning of the numerical simulation (t=0). These vectors were
evolved according to the dynamics induced by the 3D, time dependent
Hamiltonian up to the time $t=t_d$ that the GALI$_3$ became smaller
than $10^{-8}$ for the fist time. At that point the time $t=t_d$ was
registered and three new, random, orthonormal vectors were considered
resetting GALI$_3=1$. Afterwards, the evolution of these vectors was
followed until the next, possible occurrence of GALI$_3<
10^{-8}$. Then the same process was repeated.

Why was this procedure implemented? What is the reason behind this
strategy? In order to reveal this reason let us assume that an orbit initially
behaves in a chaotic way and later on it drifts to a regular behavior. The
volume formed by the deviation vectors will shrink exponentially fast, becoming very small during the initial chaotic epoch and will remain small throughout
the whole evolution in the regular epoch, unless one re-initializes the
deviation vectors and the volume they define. In this way the deviation vectors will be able to `feel' the new, current dynamics.

\begin{figure}
\centerline{
\begin{tabular}{c}
\hspace{-0.14cm}
\includegraphics[scale=1.1]{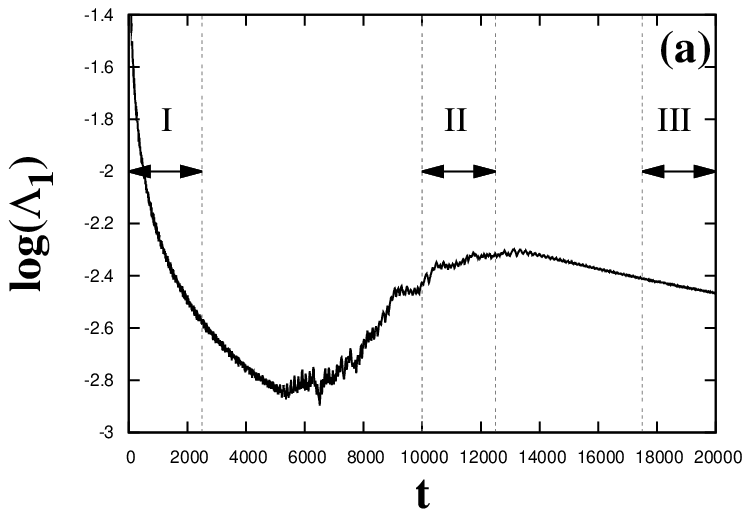}
\\
\hspace{0.13cm}
\includegraphics[scale=1.0475]{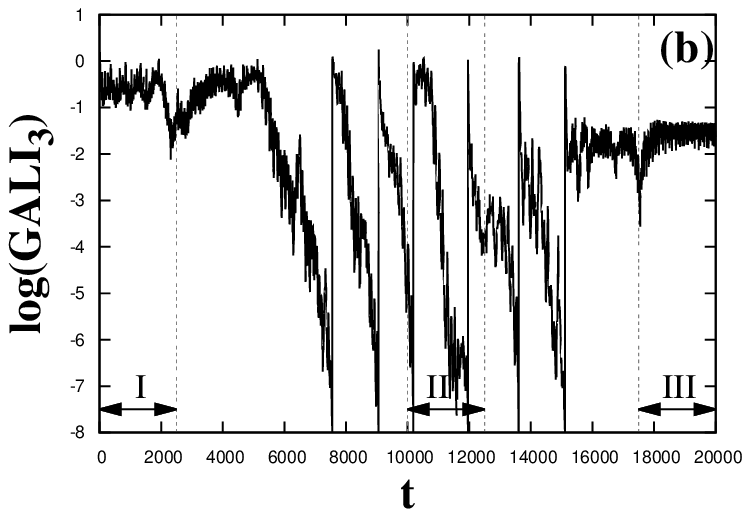}
\\
\hspace{-0.35cm}
\includegraphics[scale=1.125]{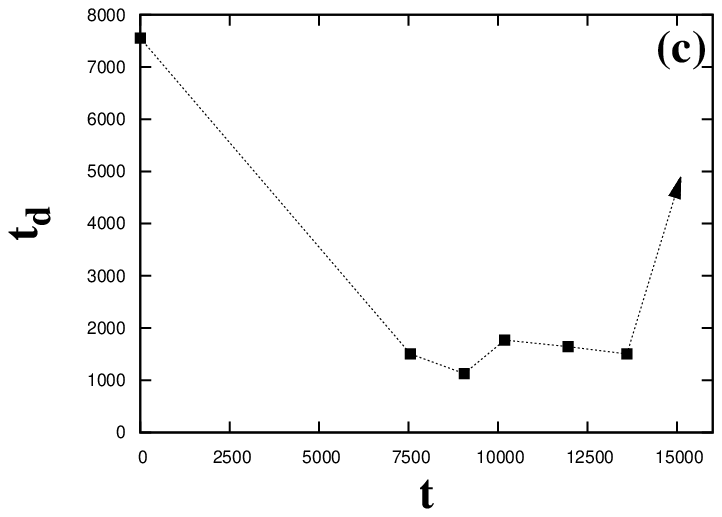}
\end{tabular}
}
\caption{ Time evolution of (a) the finite time mLE $\Lambda_1$, (b)
  the re-initialized GALI$_3$, and (c) the time $t_d$ needed for the
  re-initialized GALI$_3$ to decrease from GALI$_3=1$ to GALI$_3\leq
  10^{-8}$ for a particular orbit of the 3D time dependent galactic
  model studied in \cite{MBS2013}.  The orbit changes its dynamical
  nature from regular to chaotic and again to regular. Three
  characteristic epochs are located between the vertical dashed gray
  lines in (a) and (b) and are denoted by I (regular), II (chaotic)
  and III (regular).  The arrow at the right end of (c) indicates that
  after $t\gtrsim 15000$ the GALI$_3$ in (b) does not fall back to
  zero (until of course, the final integration time $t=20\, 000$),
  which is a clear indication that in this time interval the orbit is
  regular (after \cite{MBS2013}) }
\label{fig:appl_TD}
\end{figure}

An example case of this kind is shown in Fig.~\ref{fig:appl_TD}.  In
particular, in Fig.~\ref{fig:appl_TD}(a) we see that the evolution of the
finite time mLE $\Lambda_1$ is not able to provide valid
information about the different dynamical epochs that the studied orbit
experiences. This is due to the index's averaging nature which takes
into account the whole history of the evolution.  On the other hand, the
re-initialized GALI$_3$ (whose time evolution is shown in
Fig.~\ref{fig:appl_TD}(b)) clearly succeeds in depicting the transitions
between regular epochs, where it oscillates around positive values (such time
intervals are denoted by I and III in Figs.~\ref{fig:appl_TD}(a) and (b)), and
chaotic ones, where it exhibits repeated exponential decays to very small
values (epoch II). From the results of Fig.~\ref{fig:appl_TD}(a) it becomes
evident that the computation of the mLE cannot be used as a reliable criterion
for determining the chaotic or regular nature of the orbit in these three time
intervals.

Another way to visualize the results of Fig.~\ref{fig:appl_TD}(b) is through
the measurement of the time $t_d$ needed for the repeated
re-initializations of the GALI$_3$, or in other words, of the time needed for
the GALI$_3$ to decrease from GALI$_3=1$ to GALI$_3\leq 10^{-8}$.  In
Fig.~\ref{fig:appl_TD}(c) we present $t_d$ as a function of the evolution time
of the orbit. From the results of this figure we see that during the time
interval $7500 \lesssim t \lesssim 14000$ the value of $t_d$ is rather small,
indicating strong chaotic motion. For smaller times, $t \lesssim 7500$, the
GALI$_3$ takes a long time to become small, suggesting the presence of regular
motion or of (relatively) weaker chaotic motion. The upwardly pointing arrow,
after $t\gtrsim 15000$, shows that the GALI$_3$ no longer falls to zero, which
again indicates the appearance of a regular epoch.

After the first, successful application of the GALIs to time dependent
Hamiltonians in \cite{MBS2013}, the same approach was followed for the
study of a more sophisticated time dependent galactic model in
\cite{ManMach2014}. This analytic Hamiltonian model succeeded to
incorporate the evolution of the basic morphological features of an
actual $N$ body simulation, by allowing all the relevant parameters of
its dynamical components to vary in time.

\section{Summary}
\label{sect:disc}

In this chapter we presented how the SALI and the various GALIs can be
used to study the chaotic behavior of dynamical systems.

Following the history of the evolution of these indices, we initially
presented in Sect.~\ref{sect:SALI} the underlying idea behind the
introduction of the SALI: the index actually quantifies the possible
alignment of two initially distinct deviation vectors. The natural
generalization of this idea, by considering more than two deviation
vectors and checking if they become linearly dependent, led later on,
to the introduction of the GALI, as we explained in
Sect.~\ref{sect:GALI}. The close relation between the two indices was
also pointed out, as according to (\ref{eq:SALI-GALI2_equi}) the
GALI$_2$ and the SALI practically coincide
\[
    \mbox{GALI}_2\propto\mbox{SALI}.
\]

Avoiding the presentation of mathematical proofs (which the interested reader can find in the related references), we formulated in Sect.~\ref{sect:GALI} the laws that the indices follow for chaotic and regular orbits, providing also several numerical results which demonstrate their validity.

In particular, for $N$D Hamiltonian systems ($N\geq 2$) and $2N$d
symplectic maps ($N\geq 1$) the GALI$_k$ tends exponentially to zero
for chaotic orbits and unstable periodic orbits following
(\ref{eq:GALI_ch})
\[
\mbox{GALI}_k(t) \propto \exp \left\{-\left[ (\lambda_1-\lambda_2) +
  (\lambda_1-\lambda_3)+ \cdots+ (\lambda_1-\lambda_k)\right]t
\right\},
\]
while for regular motion on as $s$d torus, with $2\leq s \leq N$, the
evolution of the GALI$_k$ is given by (\ref{eq:GALI_reg_low_tor})
\[
\mbox{GALI}_k (t) \propto
  \left\{ \begin{array}{ll} \mbox{constant} & \mbox{if $2\leq k \leq
      s$} \\ \frac{1}{t^{k-s}} & \mbox{if $s< k \leq 2N-s$}
    \\ \frac{1}{t^{2(k-N)}} & \mbox{if $2N-s< k \leq 2N .$} \\
\end{array}\right.
\]
The latter formula is quite general as a) for $s=N$ it provides
(\ref{eq:GALI_reg}), which describes the behavior of the GALI$_k$ for
motion on an $N$d torus, i.e.~the most common situation of regular
motion in the $2N$d phase space of the system, b) for $k=2$, $s=1$ and
$N=1$ it gives (\ref{eq:GALI_reg_2dmap}), which describes the power
law decay of the GALI$_2$ in the case of a 2d map (the GALI$_2$ is
only possible GALI in this case), and c) for $s=1$ it becomes
(\ref{eq:GALI_stable_po_Ham}), which provides the power law decay of
the GALI$_k$ for stable periodic orbits of Hamiltonian systems (we
remind that in the case of stable periodic in maps all the GALIs
remain constant (\ref{eq:GALI_stable_po_map})).

In our presentation, we paid much attention to issues concerning the
actual computation of the indices.  In Sect.~\ref{sect:GALI_compute}
we explained in detail an efficient way to evaluate the GALI$_k$,
which is based on the SVD procedure (\ref{eq:GALI_SVD}), while in the
Appendix we provide pseudo-codes for the computation of the SALI and
the GALI. In Sect.~\ref{sect:low_search} we discussed a numerical
strategy for the detection of regular motion on low dimensional tori
(see Figs.~\ref{fig:GALI_search_low_1} and
\ref{fig:GALI_search_low_2}), while in Sect.~\ref{sect:GALI_po} we
showed how the evaluation of the GALI for an ensemble of orbits can
lead to the location of stable periodic orbits (see
Figs.~\ref{fig:Ham_spo_neighbor_scan} and
\ref{fig:map_spo_neighbor}). In addition, the effect of the choice of
the initial deviation vectors on the color plots depicting the global
dynamics of a system, was discussed in Sect.~\ref{sect:global_2-N},
where specific strategies to avoid the appearance of spurious
structures in these plots were presented (see
Fig.~\ref{fig:HH_global_GALI2}).

One of the main advantages of the SALI and the GALI methods is their
ability to discriminate between chaotic and regular motion very
efficiently. The GALI$_k$ with $2 \leq k \leq N$ tends exponentially
fast to zero for chaotic orbits, while it attains positive values for
regular ones. Due to these different behaviors these indices, and in
particular the GALI$_2$/SALI and the GALI$_N$, can reveal even tiny
details of the underlying dynamics, if one follows the procedure
presented in Sect.~\ref{sect:global_2-N}.  Implementing the numerical
strategies developed in Sect.~\ref{sect:global_N-2N} we can also use
the completely different time rates with which the GALI$_k$ with $N <
k \leq 2N$, tends to zero (exponentially fast for chaotic orbits and
power law decay for regular ones) in order to study the dynamics
globally. Finally, in Sect.~\ref{sect:time_dep} a particular numerical
method, the re-initialization of the GALI$_k$, proved to be the
suitable approach to reveal even brief changes in the dynamical nature
of orbits in time dependent Hamiltonians.

The SALI and the GALI have already proven their usefulness in chaos
studies as their many applications to a variety of dynamical systems
show (see Sect.~\ref{sect:appl_studies}).  Nevertheless, several other
chaos indicators have been developed over the years. A few, sporadic
comparisons between some of these methods have been performed in
studies of particular dynamical systems
(e.g.~\cite{S_01,SABV_04,BBB_09,R_14}). Recently, detailed and
systematic comparisons between many chaos indicators based on the
evolution of deviation vectors were conducted
\cite{MDCG2011,DMCG2012}, and the SALI method was added in the
software package \verb"LP-VIcode" \cite{CMD2014}, which includes
several of these indicators. The main outcome of these comparative
studies was that the use of more than one chaos indicators is useful,
if not imperative, for revealing the dynamics of a system.

\begin{acknowledgement}
  Many of the results described in this chapter were obtained in close
  collaboration with Prof.~T.~Bountis, Dr.~Ch.~Antonopoulos and
  Dr.~E.~Gerlach. This work was partially supported by the European
  Union (European Social Fund - ESF) and Greek national funds through
  the Operational Program ``Education and Lifelong Learning'' of the
  National Strategic Reference Framework (NSRF) - Research Funding
  Program: `THALES'. Ch.~S. would like to thank the Research Office of
  the University of Cape Town for the Research Development Grant which
  funded part of this study, as well as the Max Planck Institute for
  the Physics of Complex Systems in Dresden for its hospitality during
  his visit in December 2014 -- January 2015, when part of this work
  was carried out. In addition, Ch.~S. thanks T.~van Heerden for
  the careful reading of the manuscript and for his valuable comments.
  We are also grateful to the three anonymous referees whose constructive remarks helped us improve the content and the clarity of the chapter.
\end{acknowledgement}

\section*{Appendix: Pseudo--codes for the Computation of the SALI and the GALI$_k$}

We present here pseudo--codes for the numerical computation of the
SALI (Table \ref{tab:SALI}) and the GALI$_k$ (Table \ref{tab:GALI})
methods, according to the algorithms presented in
Sects.~\ref{sect:SALI} and \ref{sect:GALI_compute} respectively.

\begin{table}
  \caption{Numerical computation of the SALI. The algorithm for the computation of the SALI according to equation (\ref{eq:SALI}). The program
  computes the evolution of the SALI with respect to
  time $t$ up to a given upper value of time $t=T_M$ or until the
  index becomes smaller than a low  threshold value $S_{m}$. In the latter
  case the studied orbit is considered to be chaotic.}
\label{tab:SALI}
\begin{tabular}{p{2.3cm}p{9.0cm}}
\hline\noalign{\smallskip}
Input:  & 1. Hamilton equations of motion  and variational equations, or\\
        & \quad \quad equations of the map  and of the tangent map.\\
        & 2. Initial condition for the orbit $\vec{x}(0)$.\\
        & 3. Initial \textit{orthonormal} deviation vectors $\vec{w}_1(0)$, $\vec{w}_2(0)$.\\
        & 4. Renormalization time $\tau$.\\
        & 5. Maximum time: $T_M$ and small threshold value of the SALI: $S_{m}$.\\
\hline\noalign{\smallskip}
Step 1  & \textbf{Set} the stopping flag, $\textsl{SF} \gets 0$, the counter, $i \gets 1$, and the orbit \\
        & \quad \quad characterization variable, $\textsl{OC} \gets \mbox{`regular'}$.\\
Step 2  & \textbf{While} $(\textsl{SF}=0)$\/ \textbf{Do}\\
        & \quad \quad \textbf{Evolve} the orbit and the deviation vectors from time $t=(i-1)\tau$\\
        & \quad \quad \quad to $t=i \tau$, i.~e.~\textbf{Compute} $\vec{x}(i \tau)$ and $\vec{w}_1(i \tau)$, $\vec{w}_2(i \tau)$. \\
Step 3  & \quad \quad \textbf{Normalize} the two vectors, i.e.  \\
        & \quad \quad \quad \textbf{Set} $\vec{w}_1(i \tau) \gets \vec{w}_1(i \tau)/\| \vec{w}_1(i \tau)\|$ and  $\vec{w}_2(i \tau) \gets \vec{w}_2(i \tau)/\| \vec{w}_2(i \tau)\|$. \\
Step 4  & \quad \quad \textbf{Compute} and \textbf{Store} the current value of the SALI: \\
        & \quad \quad \quad $\mbox{SALI}(i \tau)=\min \left\{\| \vec{w}_1(i \tau) + \vec{w}_2(i \tau)
\|, \| \vec{w}_1(i \tau) - \vec{w}_2(i \tau)
\|\right\}$.\\
Step 5  & \quad \quad \textbf{Set}  the counter $i \gets i+1$. \\
Step 6  & \quad \quad \textbf{If} $[ \mbox{SALI}((i-1)\tau) < S_{m}]$ \textbf{Then} \\
        & \quad \quad \quad \quad \textbf{Set}  $\textsl{SF} \gets 1$ and $\textsl{OC} \gets \mbox{`chaotic'}$. \\
        & \quad \quad \textbf{End If}  \\
Step 7  & \quad \quad \textbf{If} $[ (i \tau > T_M) ]$ \textbf{Then} \\
        & \quad \quad \quad \quad \textbf{Set}  $\textsl{SF} \gets 1$. \\
        & \quad \quad \textbf{End If}  \\
        & \textbf{End While}\\
Step 8  & \textbf{Report} the time evolution of the SALI and the nature of the orbit.\\
\hline
\end{tabular}
\end{table}

\begin{table}
  \caption{\textbf{Numerical computation of the GALI$_k$.}
    The algorithm for the computation of
    the GALI$_k$ according to equation (\ref{eq:GALI_SVD}). The program
    computes the evolution of the GALI$_k$ with respect to
    time $t$ up to a given upper value of time $t=T_M$ or until the
    index becomes smaller than a low  threshold value $G_{m}$. In the latter
    case the studied orbit is considered to be chaotic.}
\label{tab:GALI}
\begin{tabular}{p{2.3cm}p{9.0cm}}
\hline\noalign{\smallskip}
Input:  & 1. Hamilton equations of motion  and variational equations, or\\
        & \quad \quad equations of the map  and of the tangent map.\\
        & 2. Order $k$ of the desired GALI.\\
        & 3. Initial condition for the orbit $\vec{x}(0)$.\\
        & 4. Initial \textit{orthonormal} deviation vectors $\vec{w}_1(0)$, $\vec{w}_2(0)$, $\ldots$, $\vec{w}_k(0)$.\\
        & 5. Renormalization time $\tau$.\\
        & 6. Maximum time: $T_M$ and small threshold value of the GALI: $G_{m}$.\\
\hline\noalign{\smallskip}
Step 1  & \textbf{Set} the stopping flag, $\textsl{SF} \gets 0$, the counter, $i \gets 1$, and the orbit \\
        & \quad \quad characterization variable, $\textsl{OC} \gets \mbox{`regular'}$.\\
Step 2  & \textbf{While} $(\textsl{SF}=0)$\/ \textbf{Do}\\
        & \quad \quad \textbf{Evolve} the orbit and the deviation vectors from time $t=(i-1)\tau$\\
        & \quad \quad \quad to $t=i \tau$, i.~e.~\textbf{Compute} $\vec{x}(i \tau)$ and $\vec{w}_1(i \tau)$, $\vec{w}_2(i \tau)$, $\ldots$, $\vec{w}_k(i \tau)$. \\
Step 3  & \quad \quad \textbf{Normalize} the vectors:  \\
        & \quad \quad \quad \textbf{Do} for $j=1$ to $k$ \\
        & \quad \quad \quad \quad \quad  \textbf{Set} $\vec{w}_j(i \tau) \gets \vec{w}_j(i \tau)/\| \vec{w}_j(i \tau)\|$.\\
        & \quad \quad \quad \textbf{End Do} \\
Step 4  & \quad \quad \textbf{Compute} and \textbf{Store} the current value of the GALI$_k$: \\
        & \quad \quad \quad \textbf{Create} matrix $\textbf{A}(i \tau) $ having as rows the deviation vectors $\vec{w}_1(i \tau)$,\\
        & \quad \quad \quad \quad  $\vec{w}_2(i \tau)$, $\ldots$, $\vec{w}_k(i \tau)$.\\
        & \quad \quad \quad \textbf{Compute} the singular values $z_1(i \tau)$, $z_2(i \tau)$, $\ldots$, $z_k(i \tau)$ of \\
        & \quad \quad \quad \quad matrix $\textbf{A}^T(i \tau)$ by applying the SVD algorithm.\\
        & \quad \quad \quad $\mbox{GALI}_k (i \tau)=\prod_{j=1}^k z_j(i \tau)$.\\
Step 5  & \quad \quad \textbf{Set}  the counter $i \gets i+1$. \\
Step 6  & \quad \quad \textbf{If} $[ \mbox{GALI}_k((i-1)\tau) < G_{m}]$ \textbf{Then} \\
        & \quad \quad \quad \quad \textbf{Set}  $\textsl{SF} \gets 1$ and $\textsl{OC} \gets \mbox{`chaotic'}$. \\
        & \quad \quad \textbf{End If}  \\
Step 7  & \quad \quad \textbf{If} $[ (i \tau > T_M) ]$ \textbf{Then} \\
        & \quad \quad \quad \quad \textbf{Set}  $\textsl{SF} \gets 1$. \\
        & \quad \quad \textbf{End If}  \\
        & \textbf{End While}\\
Step 8  & \textbf{Report} the time evolution of the GALI$_k$ and the nature of the orbit.\\
\hline
\end{tabular}
\end{table}



\begin{thebibliography}{99.}%
%
%

\bibitem{AB2006} Antonopoulos,~Ch., Bountis,~T.: Detecting order and
  chaos by the Linear Dependence Index (LDI) method. ROMAI
  J. \textbf{2}(2), 1--13 (2006)

\bibitem{AC2011} Antonopoulos,~Ch., Christodoulidi,~H.: Weak chaos
  detection in the Fermi-Pasta-Ulam-$\alpha$ system using $q$-Gaussian
  statistics. Int.~J.~Bifurcation Chaos \textbf{21}, 2285 (2011)

\bibitem{AMS_05} Antonopoulos,~Ch., Manos,~A., Skokos,~Ch.: SALI: an
  efficient indicator of chaos with application to 2 and 3 degrees of
  freedom Hamiltonian systems. In Tsahalis,~D.T. (ed.) From Scientific
  Computing to Computational Engineering. Proceedings of the 1st
  International Conference, Patras Univ. Press, Vol. III,
  pp.~1082-1088 (2005)

\bibitem{ABS_06} Antonopoulos,~Ch., Bountis,~T., Skokos,~Ch.: Chaotic
  dynamics of $N$-degree of freedom Hamiltonian
  systems. Int.~J.~Bifurcation Chaos \textbf{16}, 1777--1793 (2006)

\bibitem{ABB2010} Antonopoulos,~Ch., Basios,~V., Bountis,~T.: Weak
  chaos and the `Melting Transition' in a confined microplasma system.
  Phys. Rev. E \textbf{81}, 016211 (2010)

\bibitem{ABDNT2013} Antonopoulos,~C., Basios,~V., Demongeot,~J,
  Nardone,~P., Thomas,~ R.: Linear and nonlinear arabesques: A study
  of closed chains of negative 2-element circuits. Int.~J.~Bifurcation
  Chaos \textbf{23}, 1330033 (2013)

\bibitem{B05} Bario,~R.: Sensitivity tools vs.~Poincar\'e sections. Chaos,
    Solitons and Fractals  \textbf{25} 711--726 (2005)

\bibitem{B06} Bario,~R.:  Painting chaos: a gallery of sensitivity plots
of classical problems. Int.~J.~Bifurcation Chaos \textbf{16} 2777--2798 (2006)

\bibitem{BBB_09} Barrio,~R., Borczyk,~W., Breiter,~S.: Spurious
  structures in chaos indicators maps. Chaos, Solitons and Fractals
  \textbf{40} 1697--1714 (2009)

\bibitem{BG_79} Benettin,~G., Galgani,~L.: Lyapunov characteristic
  exponents and stochasticity. In: Laval,~G., Gr\'{e}sillon,~D. (eds.)
  Intrinsic Stochasticity in Plasmas, pp.~93--114, Edit.~Phys.~Orsay
  (1979)

\bibitem{BGS_76} Benettin,~G., Galgani,~L., Strelcyn J.M.: Kolmogorov
  entropy and numerical experiments Phys.~Rev.~A \textbf{14},
  2338--2344 (1976)

\bibitem{BGGS_78} Benettin,~G., Galgani,~L., Giorgilli,~A.,
  Strelcyn,~J.M.: Tous les nombres caract\'eristiques sont
  effectivement calculables.  C.~R.~Acad.~Sc.~Paris S\'er.~A
  \textbf{286}, 431--433 (1978)

\bibitem{BFS_79} Benettin,~G. Froeschl\'{e},~C., Scheidecker,~J.P.:
  Kolmogorov entropy of a dynamical system with an increasing number
  of degrees of freedom. Phys.~Rev.~A \textbf{19}, 2454--2460 (1979)

\bibitem{BGGS_80a} Benettin,~G., Galgani,~L., Giorgilli,~A.,
  Strelcyn,~J.M.: Lyapunov characteristic exponents for smooth
  dynamical systems and for Hamiltonian systems; A method for
  computing all of them. Part 1: theory. Meccanica (March) 9--20
  (1980)

\bibitem{BGGS_80b} Benettin,~G., Galgani,~L., Giorgilli,~A.,
  Strelcyn,~J.M.: Lyapunov characteristic exponents for smooth
  dynamical systems and for Hamiltonian systems; A method for
  computing all of them. Part 2: Numerical application. Meccanica
  (March) 21--30 (1980)

\bibitem{BCSV_12} Boreux,~J., Carletti,~T., Skokos,~Ch., Vittot,~M.:
  Hamiltonian control used to improve the beam stability in particle
  accelerator models. Commun. Nonlinear
  Sci. Num. Simulat. \textbf{17}, 1725--1738 (2012)

\bibitem{BCSPV_12} Boreux,~J., Carletti,~T., Skokos,~Ch.,
  Papaphilippou,~Y., Vittot,~M.: Efficient control of accelerator
  maps. Int.~J.~Bifurcation Chaos \textbf{22}(9), 1250219 (2012)

\bibitem{BP2009} Bountis,~T., Papadakis,~K.E.: The stability of
  vertical motion in the $N$-body circular Sitnikov
  problem. Cel. Mech. Dyn. Astr. \textbf{104}, 205-225 (2009)

\bibitem{BS_06} Bountis,~T., Skokos,~Ch.: Application of the SALI
  chaos detection method to accelerator mappings. Nucl. Instr. Meth.
  Phys. Res. – Sect. A \textbf{561}, 173--179 (2006)

\bibitem{BS_12} Bountis,~T.C., Skokos,~Ch.: Complex Hamiltonian
  Dynamics. Springer-Verlag, Berlin (2012)

\bibitem{BMC2009} Bountis,~T., Manos,~T., Christodoulidi,~H.:
  Application of the GALI Method to localization dynamics in nonlinear
  systems. J. Comp. Appl. Math. \textbf{227}, 17--26 (2009)

\bibitem{B_69} Broucke, R.~A.: Periodic orbits in the elliptic
  restricted three--body problem. NASA, Jet Propulsion Laboratory,
  Tech.~Rep.~32-1360 (1969)

\bibitem{CDLMV2007} Capuzzo-Dolcetta,~R., Leccese,~L., Merritt,~D.,
  Vicari,~A.: Self-consistent models of cuspy triaxial galaxies with
  dark matter haloes. Astroph. J. \textbf{666}, 165--180 (2007)

\bibitem{CMD2014} Carpintero,~D.D., Maffione,~N., Darriba,~L.:
  LP--VIcode: A program to compute a suite of variational chaos
  indicators. Astronomy and Computing \textbf{5}, 19--27 (2014)

\bibitem{CMN_14} Carpintero,~D.D., Muzzio,~J. C., Navone,~H.D.: Models
  of cuspy triaxial stellar systems – III. The effect of velocity
  anisotropy on chaoticity. Mon. Not. R. Astron. Soc. \textbf{438},
  2871--2881 (2014)

\bibitem{CCF_80} Casati,~G., Chirikov,~B.V., Ford,~J.: Marginal local
  instability of quasi-periodic motion. Phys.~Let.~A \textbf{77},
  91--94 (1980)

\bibitem{CB2006} Christodoulidi,~H., Bountis,~T.: Low-dimensional
  quasiperiodic motion in Hamiltonian systems. ROMAI J \textbf{2}(2),
  37–-44 (2006)

\bibitem{CE2013} Christodoulidi,~H., Efthymiopoulos,~Ch.:
  Low-dimensional $q$-tori in FPU lattices: dynamics and localization
  properties. Physica D \textbf{261}, 92 (2013)

\bibitem{CEB2010} Christodoulidi,~H., Efthymiopoulos,~Ch.,
  Bountis,~T.: Energy localization on $q$-tori, long-term stability,
  and the interpretation of Fermi-Pasta-Ulam recurrences. Phys. Rev. E
  \textbf{81}, 016210 (2010)

\bibitem{CEGM_14} Cincotta,~P.M., Efthymiopoulos,~C., Giordano,~C.M.,
  Mestre,~M.F.: Chirikov and Nekhoroshev diffusion estimates: bridging
  the two sides of the river.  Physica D \textbf{266}, 49--64 (2014)

\bibitem{Cont_book} Contopoulos,~G.: Order and Chaos in Dynamical
  Astronomy. Springer-Verlag, Berlin, Heidelberg (2002)

\bibitem{CGG_78} Contopoulos,~G., Galgani,~L., Giorgilli,~A.: On the
  number of isolating integrals in Hamiltonian systems. Phys.~Rev.~A
  \textbf{18}, 1183--1189 (1978)

\bibitem{DMCG2012} Darriba,~L.A., Maffione,~N.P., Cincotta,~P.M.,
  Giordano,~C.M.: Comparative study of variational chaos indicators
  and ODEs' numerical integrators. Int.~J.~Bifurcation Chaos
  \textbf{22}, 1230033 (2012)

\bibitem{DMS00} Dullin,~H.~R., Meiss,~J.~D., Sterling,~D.: Generic twistless bifurcations. Nonlinearity  \textbf{13} 203--224 (2000)

\bibitem{H_75} Hadjidemetriou,~J.: The stability of periodic orbits in
  the three-body problem. Celest. Mech. \textbf{12}, 255--276 (1975)

\bibitem{H_83} Haken,~H.: At least one Lyapunov exponent vanishes if
  the trajectory of an attractor does not contain a fixed
  point. Phys.~Let.~A \textbf{94}, 71--72 (1983)

\bibitem{HK2009} Harsoula,~M., Kalapotharakos,~C.: Orbital structure
  in N-body models of barred-spiral
  galaxies. Mon. Not. R. Astron. Soc.  \textbf{394}, 1605--1619 (2009)

\bibitem{HH_64} H\'{e}non,~M., Heiles,~C.: The applicability of the
  third integral of motion: Some numerical
  experiments. Astron. J. \textbf{69}, 73--79 (1964)

\bibitem{HD_98} Howard,~J.~E., Dullin,~H.~R.: Linear stability of
  natural symplectic maps. Phys. Lett. A \textbf{246}, 273--283 (1998)

\bibitem{HM_87} Howard,~J.~E., MacKay,~R.~S.: Linear stability of
  symplectic maps. J. Math. Phys. \textbf{28}, 1036--1051 (1987)

\bibitem{HuaCao2014} Huang,~G., Cao,~Z.: Numerical analysis and
  circuit realization of the modified L\"{U} chaotic
  system. Syst. Sci. $\&$ Control Eng. \textbf{2}, 74--79 (2014)

\bibitem{HuaWu2011} Huang,~G-Q, Wu X.: Analysis of permanent-magnet
  synchronous motor chaos system. Lecture Notes in Computer Science
  \textbf{7002}, 257--263 (2011)

\bibitem{HuaWu2012} Huang,~G.Q., Wu,~X.: Analysis of new
  four-dimensional chaotic circuits with experimental and numerical
  methods. Int.~J.~Bifurcation Chaos \textbf{22}, 1250042 (2012)

\bibitem{HuaZhou2013} Huang,~G., Zhou,~Y.: Circuit simulation of the
  modified Lorenz system. J. Inform. Comput. Sci. \textbf{10},
  4763--4772 (2013)

\bibitem{FMT2012} Faranda,~D., Mestre,~M.F., Turchetti,~G.: Analysis
  of round off errors with reversibility test as a dynamical
  indicator. Int.~J.~Bifurcation Chaos \textbf{22}, 1250215 (2012)

\bibitem{FPU_55} Fermi,~E., Pasta,~J. and Ulam,~S.: Studies of
  nonlinear problems. I. Los Alamos Rep LA-1940 (1955)

\bibitem{FGL_97} Froeschl\'{e},~C., Gonczi,~R., Lega,~E.: The fast
  Lyapunov indicator: a simple tool to detect weak chaos. Application
  to the structure of the main asteroidal belt. Planet.~Space
  Sci. \textbf{45}, 881--886 (1997)

\bibitem{GES_12} Gerlach,~E., Eggl,~S., Skokos,~Ch.: Efficient
  integration of the variational equations of multi--dimensional
  Hamiltonian systems: Application to the Fermi--Pasta--Ulam
  lattice. Int.~J.~Bifurcation Chaos \textbf{22}, 1250216 (2012)

\bibitem{GM_04} Gottwald,~G.A., Melbourne,~I.: A new test for chaos in
  deterministic systems. Proc.~Roy.~Soc.~London A \textbf{460},
  603--611 (2004)

\bibitem{K2008} Kalapotharakos,~C.: The rate of secular evolution in
  elliptical galaxies with central
  masses. Mon. Not. R. Astron. Soc. \textbf{389}, 1709--1721 (2008)

\bibitem{KG_88} Kantz,~H., Grassberger,~P.: Internal Arnold diffusion
  and chaos thresholds in coupled symplectic maps. J.~Phys.~A
  \textbf{21}, L127--133 (1988)

\bibitem{KKSK_14} Kyriakopoulos,~N., Koukouloyannis,~V., Skokos,~Ch.,
  Kevrekidis,~P.: Chaotic behavior of three interacting vortices in a
  confined Bose-Einstein condensate. Chaos \textbf{24}, 024410 (2014)

\bibitem{LL_92} Lichtenberg,~A.~J., Lieberman,~M.~A.: Regular and
  Chaotic Dynamics (2nd edition) Springer-Verlag, Berlin (1992)

\bibitem{Luk2014} Lukes-Gerakopoulos,~G.: Adjusting chaotic indicators
  to curved spacetimes. Phys. Rev. D \textbf{89}, 043002 (2014)

\bibitem{Lyapunov_1892} Lyapunov,~A.M.: The general problem of the
  stability of motion. Taylor and Francis, London (1992) (English
  translation from the French: Liapounoff,~A.: Probl\`{e}me
  g\'{e}n\'{e}ral de la stabilit\'{e} du
  mouvement. Annal.~Fac.~Sci.~Toulouse \textbf{9}, 203--474
  (1907). The French text was reprinted in Annals Math.~Studies
  Vol.~17 Princeton Univ.~Press (1947). The original was published in
  Russian by the Mathematical Society of Kharkov in 1892)

\bibitem{MSCHJD2007} Macek,~M., Str\'{a}nsk\'{y},~P., Cejnar,~P.,
  Heinze,~S., Jolie,~J., Dobe\v{s},~J.: Classical and quantum
  properties of the semiregular arc inside the Casten
  triangle. Phys. Rev. C \textbf{75}, 064318 (2007)

\bibitem{MDSC2010} Macek,~M, Dobe\v{s},~J., Str\'{a}nsk\'{y},~P.,
  Cejnar,~P.: Regularity-induced separation of intrinsic and
  collective dynamics. Phys. Rev. Let. \textbf{105}, 072503 (2010)

\bibitem{MDC2010} Macek,~M, Dobe\v{s},~J., Cejnar,~P.: Occurrence of
  high-lying rotational bands in the interacting boson
  model. Phys. Rev. C \textbf{82}, 014308 (2010)

\bibitem{MDCG2011} Maffione,~N.P., Darriba,~L.A., Cincotta,~P.M.,
  Giordano,~C.M.: A comparison of different indicators of chaos based
  on the deviation vectors: application to symplectic
  mappings. Cel. Mech. Dyn. Astron. \textbf{111}, 285--307 (2011)

\bibitem{MA2011} Manos,~T., Athanassoula,~E.: Regular and chaotic
  orbits in barred galaxies - I. Applying the SALI/GALI method to
  explore their distribution in several
  models. Mon. Not. R. Astron. Soc. \textbf{415}, 629--642 (2011)

\bibitem{ManMach2014} Manos,~T., Machado,~R.E.G.: Chaos and dynamical
  trends in barred galaxies: bridging the gap between N-body
  simulations and time-dependent analytical
  models. Mon. Not. R. Astron. Soc., \textbf{438}, 2201--2217 (2014)

\bibitem{ManRob2014} Manos,~T., Robnik,~M.: Survey on the role of
  accelerator modes for the anomalous diffusion: The case of the
  standard map. Phys. Rev. E \textbf{89}, 022905 (2014)

\bibitem{ManRuf2011} Manos,~T., Ruffo,~S.: Scaling with system size of
  the Lyapunov exponents for the Hamiltonian Mean Field
  model. Transp. Theory Stat. Phys. \textbf{40}, 360--381 (2011)

\bibitem{MSAB2008} Manos,~T., Skokos,~Ch., Athanassoula,~E.,
  Bountis,~T.: Studying the global dynamics of conservative dynamical
  systems using the SALI chaos detection method, Nonlinear Phenomena
  in Complex Systems \textbf{11}(2), 171--176 (2008)

\bibitem{MSB_08} Manos,~T., Skokos,~Ch., Bountis,~T.: Application of
  the Generalized Alignment Index (GALI) method to the dynamics of
  multi-dimensional symplectic maps. In: Chandre C., Leoncini X. and
  Zaslavsky G. (eds.) Chaos, Complexity and Transport: Theory and
  Applications.  Proceedings of the CCT 07, World Scientific,
  pp.~356-364 (2008)

\bibitem{MSB_09} Manos,~T., Skokos,~Ch., Bountis,~T.: Global dynamics
  of coupled standard maps. In: Contopoulos G. and Patsis
  P. A. (eds.), Chaos in Astronomy, Astrophysics and Space Science
  Proceedings, Springer-Verlag, pp.~367-371 (2009)

\bibitem{MSA_12} Manos,~T., Skokos,~Ch., Antonopoulos,~Ch.: Probing
  the local dynamics of periodic orbits by the generalized alignment
  index (GALI) method. Int.~J.~Bifurcation Chaos \textbf{22}, 1250218
  (2012)

\bibitem{MBS2013} Manos,~T., Bountis,~T., Skokos,~Ch.: Interplay
  between chaotic and regular motion in a time-dependent barred galaxy
  model. J. Phys. A: Math. Theor., \textbf{46}, 254017 (2013)


\bibitem{NS_77} Nagashima,~T., Shimada,~I.: On the C--system--like
  property of the Lorenz system. Prog.~Theor.~Phys. \textbf{58},
  1318--1320 (1977)

\bibitem{O_68} Oseledec,~V.I.: A multiplicative ergodic
  theorem. Ljapunov characteristic numbers for dynamical
  systems. Trans.~Moscow Math.~Soc. \textbf{19}, 197--231 (1968)

\bibitem{PP2008} Paleari,~S., Penati,~T.: Numerical Methods and
  Results in the FPU Problem. Lect. Notes Phys. \textbf{728}, 239--282
  (2008)

\bibitem{PBS_04} Panagopoulos,~P., Bountis,~T.C., Skokos,~Ch.:
  Existence and stability of localized oscillations in 1-dimensional
  lattices with soft spring and hard spring potentials. J. Vibration
  \& Acoustics \textbf{126}, 520--527 (2004)

\bibitem{PABV2008} Petalas,,~Y.G., Antonopoulos,~C.G., Bountis,~T.C.,
  Vrahatis,~M. N.: Evolutionary methods for the approximation of the
  stability domain and frequency optimization of conservative
  maps. Int.~J.~Bifurcation Chaos, \textbf{18}, 2249--2264 (2008)

\bibitem{NumRec} Press,~W.H., Teukolsky,~S.A., Vetterling,~W.T.,
  Flannery,~B.P.: Numerical Recipes in Fortran 77. Second Edition.
  The Art of Scientific Computing. Cambridge University Press,
  Cambridge (1992)

\bibitem{R_14} Racoveanu,~O.: Comparison of chaos detection methods in
  the circular restricted three-body
  problem. Astron. Nachr. \textbf{335}, 877--885 (2014)

\bibitem{SS2012} Saha,~L.M., Sahni,~N.: Chaotic evaluations in a
  modified coupled logistic type predator-prey
  model. App. Math. Sci. \textbf{6}(139), 6927--6942 (2012)

\bibitem{SESF_04} S\'andor,~Zs., \'Erdi,~B., Sz\'ell,~A., Funk,~B.:
  The relative Lyapunov indicator: an efficient method of chaos
  detection. Cel.~Mech.~Dyn.~Astron. \textbf{90}, 127--138 (2004)

\bibitem{SN_79} Shimada,~I., Nagashima,~T.: A numerical approach to
  ergodic problem of dissipative dynamical
  systems. Prog.~Theor.~Phys. \textbf{61}, 1605--1615 (1979)

\bibitem{S_01} Skokos,~Ch.: Alignment indices: a new, simple method
  for determining the ordered or chaotic nature of orbits. J.~Phys.~A
  \textbf{34}, 10029--10043 (2001)

\bibitem{S_01b} Skokos,~Ch.: On the stability of periodic orbits of
  high dimensional autonomous Hamiltonian systems. Physica D
  \textbf{159}, 155--179 (2001)

\bibitem{S_10} Skokos,~Ch.: The Lyapunov Characteristic Exponents and
  their computation. Lect.~Notes Phys. \textbf{790}, 63--135 (2010)

\bibitem{SABV_03} Skokos,~Ch., Antonopoulos,~Ch., Bountis,~T.C.,
  Vrahatis, M.N.: How does the smaller alignment index (SALI)
  distinguish order from chaos? Prog.~Theor.~Phys.~Supp. \textbf{150},
  439--443 (2003)

\bibitem{SABV_04} Skokos,~Ch., Antonopoulos,~Ch., Bountis,~T.C.,
  Vrahatis, M.N.: Detecting order and chaos in Hamiltonian systems by
  the SALI method. J.~Phys.~A \textbf{37}, 6269--6284 (2004)

\bibitem{SBA_07} Skokos,~Ch., Bountis,~T.C., Antonopoulos,~Ch.:
  Geometrical properties of local dynamics in Hamiltonian systems: The
  generalized alignment index (GALI) method. Physica D \textbf{231},
  30--54 (2007)

\bibitem{SBA_08} Skokos,~Ch., Bountis,~T.C., Antonopoulos,~Ch.:
  Detecting chaos, determining the dimensions of tori and predicting
  slow diffusion in Fermi-Pasta-Ulam lattices by the generalized
  alignment index method.  Eur.~Phys.~J.~Spec.~Top. \textbf{165},
  5--14 (2008)

\bibitem{SCM2007} Str\'{a}nsk\'{y},~P., Cejnar,~P., Macek,~M.: Order
  and chaos in the Geometric Collective
  Model. Phys. Atom. Nucl. \textbf{70}(9), 1572--1576 (2007)

\bibitem{SHC2009} Str\'{a}nsk\'{y},~P., Hru\v{s}ka,~P. $\&$
  Cejnar,~P.: Quantum chaos in the nuclear collective model:
  Classical-quantum correspondence. Phys. Rev. E \textbf{79}, 046202
  (2009)

\bibitem{SBD2007} Soulis,~P., Bountis,~T., Dvorak,~R.: Stability of
  motion in the Sitnikov 3-body
  problem. Cel. Mech. Dyn. Astr. \textbf{99}, 129--148 (2007)

\bibitem{SPB2008} Soulis,~P.S., Papadakis,~K.E., Bountis,~T.: Periodic
  orbits and bifurcations in the Sitnikov four-body
  problem. Cel. Mech. Dyn. Astr. \textbf{100}, 251--266 (2008)

\bibitem{SESS2013} Sz\'{e}ll,~A., \'{E}rdi,~B., S\'{a}ndor,~Z.,
  Steves,~B.: Chaotic and stable behavior in the Caledonian Symmetric
  Four-Body problem. Mon. Not. R. Astron. Soc. \textbf{347}, 380--388
  (2004)

\bibitem{VCE_98} Voglis,~N., Contopoulos,~G., Efthymiopoulos,~C.:
  Method for distinguishing between ordered and chaotic orbits in
  four-dimensional maps. Phys.~Rev.~E \textbf{57}, 372--377 (1998)

\bibitem{VCE_99} Voglis,~N., Contopoulos,~G., Efthymiopoulos,~C.:
  Detection of ordered and chaotic motion using the dynamical
  spectra. Celest.~Mech.~Dyn.~Astr. \textbf{73}, 211--220 (1999)

\bibitem{VHC2007} Voglis,~N., Harsoula,~M., Contopoulos,~G.: Orbital
  structure in barred
  galaxies. Mon. Not. R. Astron. Soc. \textbf{381}, 757--770 (2007)

\bibitem{Voy2008} Voyatzis,~G.: Chaos, order, and periodic orbits in
  3:1 resonant planetary dynamics. Astroph. J. \textbf{675}, 802--816
  (2008)

\bibitem{WSSV_85} Wolf,~A., Swift,~J.B., Swinney,~H.L., Vastano,~J.A.:
  Determining Lyapunov exponents from a time series. Physica D
  \textbf{16}, 285--317 (1985)

\bibitem{Zot2014} Zotos,~E.E.: Classifying orbits in galaxy models
  with a prolate or an oblate dark matter halo
  component. Astron. Astroph. \textbf{563}, A19 (2014)

\bibitem{ZotCar2013} Zotos,~E.E., Caranicolas,~N.D.: Order and chaos
  in a new 3D dynamical model describing motion in non-axially
  symmetric galaxies. Nonlinear Dyn. \textbf{74}, 1203--1221 (2013)


\end{thebibliography}
\end{document}